\documentclass[times,review,nopreprintline]{elsarticle}

\usepackage{jasr}
\usepackage{framed,multirow}

\usepackage{amssymb}
\usepackage{latexsym}

\usepackage{bm}
\usepackage{caption}
\usepackage{subcaption}
\usepackage{mathtools}
\usepackage{adjustbox}
\usepackage[scr=boondox]   
           {mathalpha}

\usepackage{algorithm}
\usepackage{algpseudocode}
\usepackage{graphicx}
\usepackage{amsmath}

\usepackage{appendix}

\usepackage[switch]{lineno}

\usepackage{url}
\usepackage{xcolor}
\definecolor{newcolor}{rgb}{.8,.349,.1}

\usepackage[citebordercolor=white]{hyperref}

\usepackage{cleveref}

\journal{Advances in Space Research}
\begin{document}
\verso{Given-name Surname \textit{etal}}

\begin{frontmatter}

\title{Single track orbit determination analysis for low Earth orbit with approximated J2 dynamics}%

\author[1]{Jose M. \snm{Montilla}\fnref{fn0}}
\fntext[fn0]{Substitute Professor, Aerospace Engineering and Fluid Mechanics Department,  
  jmontillag@us.es;}
\author[2]{Jan A. \snm{Siminski}\fnref{fn2}}
\author[1]{Rafael \snm{Vazquez}\fnref{fn1}}
\fntext[fn1]{Professor, Aerospace Engineering and Fluid Mechanics Department, rvazquez1@us.es;}
\fntext[fn2]{Space Debris Specialist, Space Debris Office, ESOC, jan.siminski@esa.int;}

\affiliation[1]{organization={Universidad de Sevilla, Escuela T\'ecnica Superior de Ingenier\'ia},
                addressline={Camino de los Descubrimientos s.n},
                city={Sevilla},
                postcode={41092},
                country={Spain}}

\affiliation[2]{organization={ESA/ESOC},
                addressline={Robert-Bosch-Str 5},
                city={Darmstadt},
                postcode={64293},
                country={Germany}}

\received{23 Dec 2023}
\finalform{23 Dec 2023}
\accepted{23 Dec 2023}
\availableonline{15 May 2013}
\communicated{S. Sarkar}

\begin{abstract}
In the domain of Space Situational Awareness (SSA), the challenges pertaining to orbit determination and catalog correlation are notably pronounced, partly attributable to the escalating presence of non-cooperative satellites engaging in unspecified maneuvers at irregular intervals. This study introduces an initial orbit determination methodology reliant upon data obtained from a single surveillance radar, such as the Spanish Space Surveillance and Tracking Surveillance Radar (S3TSR). The need for fast algorithms within an operational context is considered here as the main design driver. The result is a least-squares fitting procedure that incorporates an analytically formulated approximation of the dynamics under the $\text{J}_2$ perturbation, valid for short-term propagation. The algorithm makes use of all available observables, including range-rate, which makes it distinct from other similar methods. The method is compared in a battery of synthetic tests against a classical range and angles fitting method (GTDS) to study the effect of the track length and density of measurements on the full state estimation. The presented methodology is quite versatile, and it is leveraged to improve the estimation quality by adding information of the object\textquotesingle s orbital plane obtained from predictions. The resulting method has been named OPOD.
\end{abstract}


\end{frontmatter}

\nolinenumbers

\section{Introduction}
\label{sec1}

Over the last sixty years, human activities in space have led to a significant increase in orbiting objects~\citep{esaReport}, a concern that has grown more pressing with the emergence of mega-constellations for communication, like Starlink and OneWeb. This mounting congestion in Earth\textquotesingle s low orbit emphasizes the urgent need for nations and space agencies to reinforce tracking and cataloging efforts. Within this context, there arises a critical necessity for robust and efficient Orbit determination (OD) algorithms. These algorithms play a pivotal role in satellite tracking and cataloging procedures, swiftly identifying correlated tracks to maintain an accurate and up-to-date satellite database. Their reliability is essential in bolstering Space Situational Awareness (SSA) and ensuring the safety and sustainability of orbital operations. The development of such algorithms for correlation and maneuver detection represents a crucial step in mitigating the escalating risks posed by the growing complexity of space activities.

OD as a discipline is almost as old as orbital mechanics itself. Indeed, Johannes Kepler utilized Tycho Brahe\textquotesingle~s meticulous observations of Mars to establish the planet\textquotesingle~s elliptical orbit and spatial orientation, formulating in the process his three laws of planetary motion. The discipline evolved during the years, with notable contributions from prominent figures such as Newton, Euler, Lambert or Gauss~\citep{hawking2004illustrated}. Classically, a common strategy employed in numerous orbital determination scenarios involves an Initial Orbit Determination (IOD) phase, generating an estimate which is latter refined. A summary of classical methods can be found in \cite{schutz2004statistical}. These IOD procedures encompass Laplace\textquotesingle~s and Gauss\textquotesingle methods~\citep{escobal1970methods}, double r-iteration~\citep{escobal1970methods}, or Gooding\textquotesingle~s method~\citep{gooding1993new}. All these IOD methodologies necessitate measurements encompassing at least six independent parameters to generate a six-parameter orbit. For instance, the determination of an elliptical orbit typically demands three angle pairs~\citep{gooding1993new}. Introducing constraints on the type of orbits, such as removing certain parameters, may potentially reduce the required number of independent measurement parameters (a circular orbit requires two angle pairs). The quality of the initial estimation is influenced by the spacing of the measurements, benefiting if they originate from different observation arcs, which are referred here as tracks\footnote{A series of sequential observations delayed by seconds, which in turn are composed by a set of measurements (of different nature) at a common epoch.}. 

More generally, even during the IOD phase introducing more information can be leveraged to improve the quality of the estimation. One example of this is the well known range and angles method of the Goddard Trajectory Determination System (GTDS)~\citep{long1989goddard}, used in this work as a reference for single track IOD. Also related to the method introduced here, in \cite{vallado1998accurate} the authors apply a classical differential correction method for IOD with dense observation arcs and numerical propagation models. The present work on the other hand is focused in taking advantage of analytical methods for scenarios with low density radar tracks. In \cite{huang2021short} the IOD classical approach is expanded by also including precise range data from another GEO object, thus improving the achievable accuracy in determining the semi-major axis (SMA). The idea of doing the track initialization with angles and angle-rates data is explored in \cite{demars2012initial}, which is later used for data association by means of admissible regions and Mahalanobis Distance (MD). The sub-problem of very short track IOD is treated in \cite{zhang2019initial}, where the inclusion of extra radial measurements is leveraged to determine the velocity. The use of data from two different stations is considered in \cite{shang2018initial} with analytical methods and incorporating phase-derived velocity and acceleration. 

Subsequently, the initial estimate derived from the IOD is refined using a batch processor, such as a square-root information filter~\citep{schutz2004statistical}, before transitioning to sequential estimation techniques like an Extended Kalman Filter (EKF)~\citep{julier1997new} or an Unscented Kalman Filter (UKF)~\citep{van2004sigma}. Filtering techniques are very prominent for the integration of new information into past estimations, and also for the purpose of maneuver detection~\citep{goff2015orbit}. For example, \cite{psiaki2022gaussian} relies solely in angular information obtained from ground station, and manages to reduce the number of Gaussian mixture components with the use of Modified Equinoctial elements. \cite{pastor2021initial} presents a review of the state of the art for IOD+OD methods, emphasizing the lack of an universal solution, as the applicability and performance of a given methodology is generally a compromise between speed and accuracy (dependent of the available data). \cite{pastor2021initial} also proposes a single step IOD+OD method for full sate orbital estimation (including covariance) for optical measurements as a batch least-squares. An important aspect of the present work has been to understand the capability of the proposed method to achieve a reliable uncertainty characterization. In a similar note, in \cite{cano2023covariance} the authors address the dynamics\textquotesingle~modeling errors to properly achieve uncertainty realism with a batch least-square OD algorithm. Other methods for dealing with correct uncertainty characterization in IOD have been proposed by \cite{armellin2018probabilistic,armellin2016dealing}, based on the use of Taylor differential algebra (DA) schemes for the mapping of uncertainties to orbital elements space, and applied to various orbital regimes and measurement types.

This study aims to create reliable IOD methods that could be later used in the automatic procedures of cataloging uncorrelated tracks and posterior maneuver detection in the context of the S3TSR~\citep{gomez2019initial}. This radar was installed and validated by Indra with the funding of CDTI (Centro de Desarrollo Tecnol\'ogico e Industrial) under the technical and contractual management of ESA. Maintaining an up-to-date satellite database relying solely in the data acquired by a single radar is a challenging scenario, and that is the assumption considered in this work. Building upon the authors\textquotesingle~prior investigation~\citep{montilla2023manoeuvre,montilla2023conference}, the primary concern is to effectively utilize all the information obtained from the observations of one radar track only. To achieve this, the entire set of measurements provided by the surveillance radar is employed, encompassing not only the range and range-rate, but also the not so precise line-of-sight information (azimuth and elevation). Consequently, an IOD process is necessary for the radar-acquired data, resulting in an estimation of the full state of the satellite. The main driver and ultimate motivation of this work is maneuver detection--specifically, the capability to accurately determine whether an object has performed an undisclosed maneuver since the last known state was determined. Non-disclosed maneuvers are the main roadblock in the SSA environment, rendering past estimations less relevant an complicating correlation tasks. The use of the developed IOD algorithms for maneuver detection is however not pursued here and will be the subject of a subsequent publication.

In the context of this study, the emphasis on radar-based orbit determination has been to keep all aspects pertaining to it as analytical as possible. Initially the use of attributables was considered \citep{montilla2023manoeuvre}. The attributable has the advantage of using a polynomial model for the least-squares algorithm, which makes the method simple and fast, and the performance is excellent for long tracks. Choosing the state representation when doing the fitting is not straightforward though. For instance, the distinct qualitative evolution of the observables in measurement space make this a bad choice, as each is better approximated by a different order polynomial. The inertial position, or even position in relation to the radar frame of reference are smooth curves that can be approximated by the same order polynomial in the sense of a least-squares fitting~\citep{reihs2021application}. On the other hand, attributables use no information of the dynamics that the fitted trajectory is following, which is known a priory with some certainty and could increase the accuracy of the estimation. It also requires to choose the order of the fitting, which is dependent on the track length for a given method.

The use of attributables is thus replaced by a procedure that includes information of the dynamics, similar to how the GTDS range and angles method uses Keplerian dynamics for the estimation \citep{siminski2016techniques,long1989goddard}. The main driving aspect for seeking an alternative to the well-proven GTDS algorithm is the need to leverage all available observables, including range-rate data. This has resulted in the creation of a fully analytical approximate method that incorporates the $\text{J}_2$ perturbation, detailed within this text. Interestingly, the unexpected requirement to incorporate Earth\textquotesingle s flattening in \emph{short} track orbital determination emerged as a notable outcome of this study, extensively explored within the ensuing results. For the purpose of long term non-Keplerian propagation other semi-analytical~\citep{amato2019non} or even analytical methods~\citep{martinusi2015analytic} could be leveraged, but the approximate nature of these averaged methods is not suited for the short track IOD application sought here. On the other hand, exact numerical methods for uncertainty propagation~\citep{hernando2023near} do not fulfill the computational efficiency requirement in the context of large scale real-time track association~\citep{pastor2021initial}. The alternative proposed here presents a trade-off between error and speed that is perfectly suited for short propagations. The \emph{generalized equinoctial orbital elements} in \cite{giulio2021generalization} have been used in this paper due to the explicit inclusion of potential derived perturbations in the equations of motion. Coupled with the natural slow evolution of these orbital representation, the direct application of a Taylor expansion for short term propagation becomes an interesting option. This study produces the analytical time derivatives of these orbital elements, up to the fourth order, and calculates the error state transition matrix through the Taylor expansion of the solution. Subsequently, this is harnessed to formulate a differential correction algorithm utilized in IOD with the $\text{J}_2$ zonal harmonic included as a perturbation, along with an extended application involving past orbital data.

Thus, the primary contribution concerns the full state estimation algorithm. A basic linear least-squares method is employed to fit radar observables directly. Within it, the state and uncertainty computation is being carried out with an analytical (approximated) J$_2$ propagator that has been developed for this purpose. As an extension, this approach is improved by integrating information about the anticipated orbital plane of the object, a method denoted as OPOD. In the Low Earth Orbit (LEO) regime, this orbital plane is shown to be consistently predicted based on a prior estimation, even accounting for the possibility of a maneuver. The novel methodologies introduced in this work are compared to a well-established orbit determination method, showcasing significant enhancements in synthetic testing for IOD. This progress is achieved while maintaining the analytical nature of the procedure, a crucial design constraint for the operational applicability of this approach.

Next, an overview of the paper\textquotesingle s organization is presented. In Section \ref{sec:estimation} the full state estimation algorithm is detailed first as a general linear least-square method, so that it can be particularized to the three main analytical methods compared latter. This section includes an extra methodology that leverages past orbital information to improve the estimation. Section \ref{sec:simulated_results} compiles detailed synthetic testing of the estimation methods, which are then applied to several scenarios that introduce variability in the track length and the frequency of measurements. The document is closed with some conclusions in Section \ref{sec:conlusions}. The \ref{sec:appendix} has been included with most of the formulation needed in the presented work.

\section{Estimation from a single radar track}\label{sec:estimation}

This section contains the main contribution of this work, namely the algorithms for IOD that use a $\text{J}_2$ dynamical model for fitting any kind of satellite state-derived variable, and leveraged for fitting radar observables directly. After stating the IOD problem under consideration and setting the notation in Section \ref{sec:statement_notation}, Section \ref{sec:gen_lin_ls} reviews a general linear least-squares algorithm applicable to the stated problem for any kind of observable and propagation model. Then Section \ref{sec_gtds_fitting} particularizes the general algorithm to that of the GTDS, where position measurements (from range and angles observations) are fitted with Keplerian dynamics. In Section \ref{sec:kep_fit_radar_observables}, radar observables alone are fitted also with a simple Kepler model. However, for a more accurate approximation of the derivatives in the differential correction algorithm, a Taylor expansion is employed. In Section \ref{sec_j2_fit_observables} the generalized equinoctial orbital elements are leveraged for an approximation of the $\text{J}_2$ dynamics through a Taylor expansion, directly applicable for the short-track IOD problem under consideration. Finally, the inclusion of orbital plane information is proposed in Section \ref{sec:orb_plane_fitting} as an enhancement of the orbit determination algorithm, an immediate modification of the general method that simply requires the adequate observation function and derivatives.

\subsection{Problem statement and notation}\label{sec:statement_notation}

\begin{figure}
    \centering
    \includegraphics[width=0.8\textwidth]{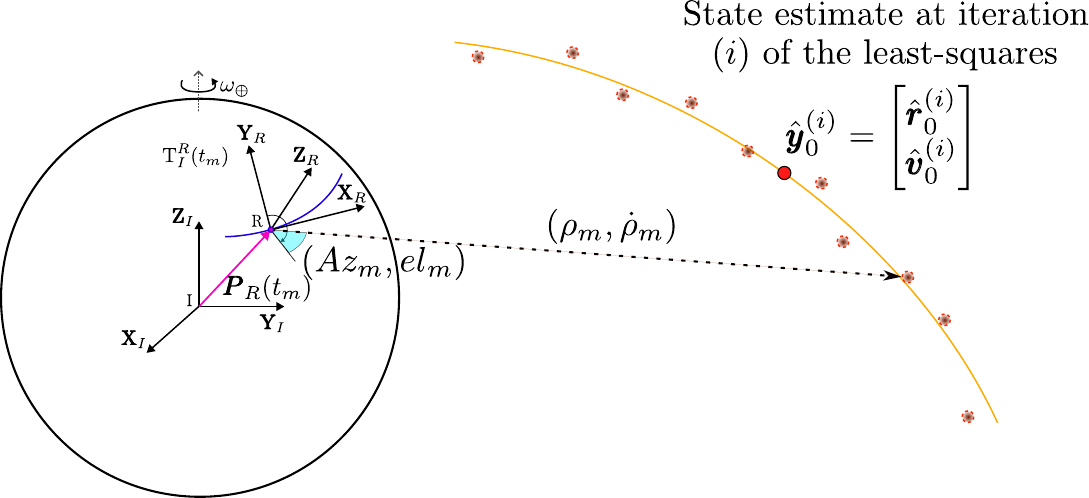}
    \caption{The estimation problem from a single radar track is represented here as an iterative fitting of the radar observables. The observations (plots) correspond to a set of range, azimuth, elevation and range-rate values $\left(\rho_m , {Az}_m , {el}_m, \dot{\rho}_m \right)$, all of which create a track of $n$ consecutive observations from the same Resident Space Object. Given the inertial frame of reference (I), the measurements are captured from a radar station with frame (R) of known position ($\bm{P}_R$), velocity ($\bm{V}_R$) and orientation with respect to the inertial one ($\text{T}_{I}^{R}$), for any given instant $t_m$.} 
    \label{fig:est_frame}
\end{figure}


The aim of the methodology presented here is to obtain an estimation of the full state of the observed satellite from the information of a single radar track, see Figure \ref{fig:est_frame} for a representation of the scenario. This involves defining a radar station with known inertial position ($\bm{R}_r$), velocity ($\bm{V}_r$) and antenna orientation ($\text{T}_{I}^{R}$) for any required instant. The radar has a designated Field of Regard (FoR) characterized by a specific revisit time for the measurement action. When a Resident Space Object (RSO) goes through the sensor\textquotesingle s FoR it generates a track composed by series of observations or plots, which in turn are a set of measurements taken at a common epoch $t_m$. Here the sub-index $m$ is used for any of the measurement instants of a radar track, where $m\in\left[ 1,n \right]$ and $n$ is the number of individual plots. Typically, radar observables for LEO objects are range $\rho_m$, azimuth ${Az}_m$, elevation ${el}_m$ and range-rate $\dot{\rho}_m$, all of which are leveraged here for the purpose of IOD. Each measurement has error, characterized by a known standard deviation and included in a covariance matrix $\text{C}_{z_m}$ for the set, where some correlations between the different observables could be considered as well. 

It is assumed that the observation function $\bm r_{m} = \text{OBS}_{fun}\left( \rho_m , {Az}_m , {el}_m \right)$ allows to consider the inertial position as a \emph{measurement} instead. The covariance of such position measurement $\text{C}_{r_m}$ has to be calculated from the original radar uncertainty ($\text{C}_{z_m}$). A linear transformation from measurement space could be applied, following \cite{siminski2016techniques}. On the other hand, the chosen alternative is to rely on the Unscented Transform (UT), as defined in \cite{goff2015orbit}, and use the $\text{OBS}_{fun}$ function directly to map the sampled sigma points around the mean value $\left( \rho_m , {Az}_m , {el}_m \right)$. The sigma points are then reconstructed in inertial space to obtain the covariance matrix $\text{C}_{r_m}$. This choice results in a more conservative representation of the uncertainty than the linear method.

At the middle instant between $t_1$ and $t_n$ the epoch of the radar track is defined as $t_0=t_1 + (t_n-t_1)/2$, where the estimation of the satellite\textquotesingle s state $ \hat{\bm{y}}_0 = \left[ \hat{\bm{r}}_0 \, \hat{\bm{v}}_0 \right] $ and its covariance $\text{C}_y$ are computed with a fitting methodology. Note that $\hat{\bm{r}}_0$ and $\hat{\bm{v}}_0$ are the estimated inertial position and velocity. This approach requires the calculation of predicted measurements, $\hat{z}_m$ in general, from the estimate $\hat{\bm{y}}_0$, which in turns necessitates the propagation of the orbital state to the instant $t_m$. The propagation step is abstracted for now within the function $\hat{\bm{y}}_m = \left[ \hat{\bm{r}}_m \, \hat{\bm{v}}_m \right] = \mathcal{P} \left( \hat{\bm{y}}_{0} , \Delta t_m \right)$, where $\Delta t_m = t_m - t_0$ is the duration of the propagation from the reference instant $t_0$, which is omitted~\footnote{The propagation generally requires to specify the reference epoch $t_0$, but potential-based dynamics that go up to the second zonal harmonic do not depend on the reference instant, so only the total propagation time $\Delta t_m$ is included in the formulation for brevity.}. For the complete set of tools needed the function $ z = h \left( \bm{y}, t \right) $ is defined as the conversion of an inertial state $\bm{y}$ at an instant $t$ to any observable $z$ (depending on the type of fitting). The function $h$ allows to compute the aforementioned predicted measurement $ \hat{z}_m = h \left( \hat{\bm{y}}_m, t_m \right)$.

\cite{siminski2016techniques} utilizes a reference method for fitting position measurements (so that $\hat{\bm{r}}_m = h \left( \hat{\bm{y}}_m, t_m \right)$) in a very efficient way, by simply assuming the motion of the spacecraft is Keplerian in nature and iteratively fitting the trajectory spawned from the reference state $\hat{\bm{y}}_0^{(i)}$. Following the notation of the source, the $\left(i\right)$ super-index denotes the $i^{th}$ iteration of the differential correction process, so that the square sum of the position residuals ($\sum_{m=1}^n  \lvert \bm r_{m} - \hat{\bm{r}}_m^{(i)} \lvert^2 $) are minimized at every iteration until convergence occurs. This method has the advantage of including information that the radar is not providing, that is, the dynamics of the motion. As long as the length of the radar track is short enough this Keplerian approximation should be safe to assume, and it is exactly what the GTDS method applies \citep{siminski2016techniques,long1989goddard}. The result is a fast analytical algorithm that outputs a full state and its covariance, which in turn depends on the uncertainty of the measurements. The only disadvantage of the GTDS method is that range-rate information cannot be included, as it is exclusively a position fit.

Here, a general weighted least-square algorithm for fitting any type of measurement is first introduced in Section \ref{sec:gen_lin_ls}. This algorithm is subsequently adapted to the specific approach employed in \cite{siminski2016techniques} for position information only and a Kepler model for the dynamics (which includes some extra simplifications), denoted simply as the GTDS algorithm and reviewed in Section \ref{sec_gtds_fitting}. The direct fitting of the radar observables with a Keplerian model is presented in Section \ref{sec:kep_fit_radar_observables} as an alternative to the GTDS method that incorporates range-rate data. The versatility of the general method permits the incorporation of alternative dynamic models. Section \ref{sec_j2_fit_observables} presents the proposed fitting methodology where the dynamics are given by an (approximated) analytical $\text{J}_2$ propagator, which is later improved by adding the orbital plane as a virtual measurement obtained from a past orbital estimation, see Section \ref{sec:orb_plane_fitting}.


\subsection{Iterative linearized least-squares fitting algorithm}\label{sec:gen_lin_ls}

To maintain a broad scope, the fitting algorithm is derived from a general scalar observable $z_m$ of any kind at time $t_m$. This measurement has some error (with known standard deviation $\sigma_m$), so the predicted measurement $\hat{z}_m = h \left( \hat{\bm{y}}_m, t_m \right)$ from the estimated full state $\hat{\bm{y}}_{0}$ is not equal in general, even without model inaccuracies in $\hat{\bm{y}}_m=\mathcal{P} \left( \hat{\bm{y}}_{0} , \Delta t_m \right)$. As a result it can be stated that $z_m = \hat{z}_m + \xi_m$, where $\xi_m$ is the residual of the observed value compared to the predicted one. The goal is to find the value of $\hat{\bm{y}}_{0}$ that minimizes the sum of $\xi_m^2$ for all $n$ measurements. More generally, it is necessary to consider some normalizing factor, as the different measurements may use different units, and thus not be comparable. Therefore, the objective is to minimize the sum of $(w_m\xi_m)^2$, where $w_m = 1/\sigma_m$, as expressed in Equation \ref{eq:min_general}. A prerequisite for this to properly define a least-squares problem is that the count of individual measurements to fit must exceed the dimension of the state ($n>6$).
\begin{equation} \label{eq:min_general}
    J_z = \min_{\hat{\bm{y}}_0}\sum_{m=1}^n (w_m\xi_m)^2.
\end{equation}
This approach leads to a non-linear minimization problem that can be expensive to solve with no simplifications. Typically, it is advisable to opt for an iterative approach, where the result after $i$ iterations, $\hat{\bm{y}}_{0}^{(i)}$, is a close approximation to the solution of the general minimization problem. Assuming a good enough first estimate of $\hat{\bm{y}}_0$ is known, an approximation of the the predicted observable $\hat{z}_m$ can be calculated as a linearization around the value $\hat{z}_m^{(i)}$ for a variation of the estimation at the current iteration $\Delta \hat{\bm{y}}_0^{(i)}$ (so that $ \hat{\bm{y}}_0^{(i+1)} = \hat{\bm{y}}_0^{(i)} + \Delta \hat{\bm{y}}_0^{(i)}$), see Equation \ref{eq:aprox_z}.
\begin{equation}\label{eq:aprox_z}
    h(\hat{\bm{y}}_m,t_m) \approx h \left( \mathcal{P} \left( \hat{\bm{y}}_{0}^{(i)} , \Delta t_m \right),t_m \right) + \frac{\partial h}{\partial \hat{\bm{y}}_0}\biggr\rfloor_{(\hat{\bm{y}}_0^{(i)},t_m)} \Delta \hat{\bm{y}}_0^{(i)} = h(\hat{\bm{y}}_m^{(i)},t_m) + \frac{\partial h}{\partial \bm y} \biggr\rfloor_{(\hat{\bm{y}}_m^{(i)},t_m)} \frac{\partial \mathcal{P}}{\partial \hat{\bm{y}}_0 } \biggr\rfloor_{(\hat{\bm{y}}_0^{(i)},\Delta t_m)} \Delta \hat{\bm{y}}_0^{(i)}.
\end{equation}

Considering all the measurements $\bm{z} = \left[ z_1,\cdots,z_n \right]^\intercal$, and the predicted measurements $\hat{\bm{z}}^{(i)} = \left[ \hat{z}_1^{(i)},\cdots,\hat{z}_n^{(i)} \right]^\intercal$ from $\hat{\bm{y}}_{0}^{(i)}$, the error $\Delta \bm{z}^{(i)}=\bm{z} - \hat{\bm{z}}^{(i)}$ is defined. Now, the derivatives of the predicted observed value for the last iteration $\partial h/\partial \hat{\bm{y}}_0\rfloor_{(\hat{\bm{y}}_0^{(i)},t_m)}$ are stacked in the matrix $\text{A}$ to get the linear equation $\Delta \bm{z}^{(i)} = \text{A} \, \Delta \bm{y}_0^{(i)} + \bm{\xi} $, with $\bm{\xi}$ being the the residuals vector for the iteration. With this expression, the same minimization problem as in Equation \ref{eq:min_general} can be posed and solved as a linear least-squares problem, see Equation \ref{eq:llsp}, where the matrix $\text{C}_y = \left( \text{A}^\intercal \text{W} \text{A}\right)^{-1}$ is the covariance of the estimation.
\begin{equation}\label{eq:llsp}
\begin{split}
    \Delta \bm{z}^{(i)} = & \text{A} \, \Delta \bm{y}_0^{(i)} + \bm{\xi} \\
    J_z  = & \min_{\Delta \hat{\bm{y}}_0^{(i)}} \: \bm \xi^\intercal \text{W} \bm \xi = \min_{\Delta \hat{\bm{y}}_0^{(i)} } \: (\Delta \bm z^{(i)} - \text{A} \, \Delta \hat{\bm{y}}_0^{(i)} )^\intercal \text{W} ( \Delta \bm z^{(i)} - \text{A} \, \Delta \hat{\bm{y}}_0^{(i)} ), \\
    \Delta \bm y_0^{(i)}  = & \left( \text{A}^\intercal \text{W} \text{A}\right)^{-1} ( \text{A}^\intercal \text{W} \, \Delta \bm z^{(i)} ).
\end{split}
\end{equation}

In a similar note to the general least-squares problem, see Equation \ref{eq:min_general}, a normalization of the residuals is required. The matrix $\text{W} = \text{C}_{z}^{-1}$ with the inverted measurements covariance is chosen. In this general formulation $\text{C}_{z} = \text{diag}\left( \sigma_1^2, \cdots , \sigma_n^2 \right)$, although the addition of correlations is also possible. This differential correction approach is very convenient, as now solving the full optimization problem becomes equivalent to several iterations of a linear least-squares problem. The general convergence of this approach to the true global minimum is not assured, primarily owing to the potential presence of nearby local minima. However, provided the initial estimation is sufficiently close, it presents a favorable trade-off between speed and accuracy as analyzed by \cite{vallado2001fundamentals}. Given the output of every iteration as $\Delta \hat{\bm{y}}_0^{(i)}=\left[\Delta \hat{\bm{r}}_0^{(i)} \, \Delta \hat{\bm{v}}_0^{(i)} \right]$, the stopping condition should be a function of the norm of position displacement $\vert \Delta \hat{\bm r}^{(i)}_0 \vert$, so that when the change is under some threshold the last iteration is the solution. The matrix of predicted observations derivatives $\text{A}$ has to be recomputed for every iteration.

There has been no mention so far of the propagation method inside $\mathcal{P}(\hat{\bm{y}}_{0},\Delta t_m)$, or the relation between the spacecraft state and the measurements nature in $h(\bm y,t)$. Now three different fully analytical methods are considered. The analytical aspect is important, as the goal is to obtain an algorithm comparable to the GTDS method speed-wise, which is only possible for fast propagation functions.

\subsection{Keplerian fitting of non-weighted Cartesian position - GTDS} \label{sec_gtds_fitting}

The GTDS range and angles method is reviewed in this section as a particular case of the general algorithm of Section \ref{sec:gen_lin_ls}. In this method the fit is in inertial position, where measurements are represented by $\bm r_m$ at each instant $t_m$, so the vector with all is $\bm z = \left[ \bm r_1^\intercal,\bm r_2^\intercal,\cdots,\bm r_n^\intercal \right]^\intercal$. 

The predicted measurements $\hat{\bm{r}}_m$ are thus an explicit function of the propagated estimation $\hat{\bm{y}}_m$, so that $\hat{\bm{r}}_m = h \left( \hat{\bm{y}}_m, t_m \right)$. The propagation function $\mathcal{P}\left( \hat{\bm{y}}_{0},\Delta t \right)$ assures Keplerian dynamics. Various methods are available for performing this analytical propagation; however, in this case, the formulation of the universal variable $\chi$ as outlined in \cite{vallado2001fundamentals} has been selected. This is a general method, valid for all orbit types, which allows to compute the values of the $f$ and $g$ \emph{Kepler functions} that linearly combine the initial conditions and get the propagated state, see Equation \ref{eq:kep_pos_f_g}.
\begin{equation}\label{eq:kep_pos_f_g}
    \hat{\bm{y}}_m = \mathcal{P}\left( \hat{\bm{y}}_{0},\Delta t_m \right) = \begin{bmatrix}
        \hat{\bm{r}}(\hat{\bm{y}}_{0},\Delta t_m) \\
        \hat{\bm{v}}(\hat{\bm{y}}_{0},\Delta t_m)
    \end{bmatrix} = \begin{bmatrix}
        f(\hat{\bm{r}}_0,\hat{\bm{v}}_0,\Delta t_m) \bm \hat{\bm{r}}_0 + g(\hat{\bm{r}}_0,\hat{\bm{v}}_0,\Delta t_m) \hat{\bm{v}}_0 \\
        \dot{f}(\hat{\bm{r}}_0,\hat{\bm{v}}_0,\Delta t_m) \hat{\bm{r}}_0 + \dot{g}(\hat{\bm{r}}_0,\hat{\bm{v}}_0,\Delta t_m) \hat{\bm{v}}_0
    \end{bmatrix}.
\end{equation}

The details to compute $f$ and $g$ values are in \emph{Algorithm 8} of \cite{vallado2001fundamentals}. In order to apply the iterative method the predicted position $\hat{\bm{r}}(\hat{\bm{y}}_{0},\Delta t)$, in Equation \ref{eq:kep_pos_f_g}, has to be linearized around $\hat{\bm r}_0^{(i)}$ and $\hat{\bm v}_0^{(i)}$ for a variation $\Delta \hat{\bm r}_0^{(i)}$ and $\Delta \hat{\bm v}_0^{(i)}$ respectively, as shown in Equation \ref{eq:kep_pos_f_g_linear}. In the sequel, the dependence over $\Delta t$ has been dropped.
\begin{equation}
\begin{split}\label{eq:kep_pos_f_g_linear}
    \hat{\bm{r}}(\hat{\bm{r}}_0,\hat{\bm{v}}_0) \approx & \, \hat{\bm{r}}(\hat{\bm{r}}_0^{(i)},\hat{\bm{v}}_0^{(i)}) + \frac{\partial \hat{\bm{r}} }{\partial \hat{\bm{r}}_0 }\biggr\rfloor^{(i)} \Delta \hat{\bm{r}}_0^{(i)} + \frac{ \partial \hat{\bm{r}}}{\partial \hat{\bm{v}}_0 }\biggr\rfloor^{(i)} \Delta \hat{\bm{v}}_0^{(i)}, \\
    \frac{\partial \hat{\bm{r}}}{\partial \hat{\bm{r}}_0}\biggr\rfloor^{(i)}  = & f^{(i)} \frac{\partial \hat{\bm{r}}_0}{\partial \hat{\bm{r}}_0} + \hat{\bm{r}}_0^{(i)} \frac{\partial f}{\partial \hat{\bm{r}}_0}\biggr\rfloor^{(i)} + \hat{\bm{v}}_0^{(i)} \frac{\partial g}{\partial \hat{\bm{r}}_0}\biggr\rfloor^{(i)}, \\
    \frac{\partial \hat{\bm{r}}}{\partial \hat{\bm{v}}_0}\biggr\rfloor^{(i)}  = & g^{(i)} \frac{\partial \hat{\bm{v}}_0}{\partial \hat{\bm{v}}_0} + \hat{\bm{r}}_0^{(i)} \frac{\partial f}{\partial \hat{\bm{v}}_0}\biggr\rfloor^{(i)} + \hat{\bm{v}}_0^{(i)} \frac{\partial g}{\partial \hat{\bm{v}}_0}\biggr\rfloor^{(i)}.
\end{split}
\end{equation}

In order to compute the derivatives that appear in Equation \ref{eq:kep_pos_f_g_linear}, the derivatives of the $f$ and $g$ functions are apparently necessary. That said, at this point the GTDS method applies the simplification of $\partial f/\partial \bm r_0 = \partial f/\partial \bm v_0 = \partial g/\partial \bm r_0 = \partial g/\partial \bm v_0 = \bm{0}^\intercal $, which essentially bypasses the need of any diferentiation. The derivatives in Equation \ref{eq:kep_pos_f_g_linear} are thus reduced to $\partial \hat{\bm{r}} / \partial \hat{\bm{r}}_0 \rfloor^{(i)} \approx f^{(i)} \text{I}_{3} $ and $\partial \hat{\bm{r}} / \partial \hat{\bm{v}}_0 \rfloor^{(i)} \approx g^{(i)} \text{I}_{3} $. The matrix $\text{I}_{3}$ is the identity matrix of order 3. Applying this technique to the linearization of the predicted position results in: 
\begin{equation}\label{eq:kep_pos_f_g_linear_aprox}
\begin{aligned}
    \hat{\bm{r}}(\hat{\bm{r}}_0,\hat{\bm{v}}_0)  \approx &  \, \hat{\bm{r}}(\hat{\bm{r}}_0^{(i)},\hat{\bm{v}}_0^{(i)}) + \frac{\partial \hat{\bm{r}} }{\partial \hat{\bm{r}}_0}\biggr\rfloor^{(i)} \Delta \hat{\bm{r}}_0^{(i)} + \frac{\partial \hat{\bm{r}}}{\partial \hat{\bm{v}}_0}\biggr\rfloor^{(i)} \Delta \hat{\bm{v}}_0^{(i)} \\
     \approx & f^{(i)} \hat{\bm{r}}_0^{(i)}  + g^{(i)} \hat{\bm{v}}_0^{(i)} + f^{(i)} \Delta \hat{\bm{r}}_0^{(i)} +  g^{(i)} \Delta \hat{\bm{v}}_0^{(i)} \\
     = & f^{(i)}(\hat{\bm{r}}_0^{(i)} + \Delta \hat{\bm{r}}_0^{(i)}) + g^{(i)}(\hat{\bm{v}}_0^{(i)} + \Delta \hat{\bm{v}}_0^{(i)}) = f^{(i)} \hat{\bm{r}}_0^{(i+1)} + g^{(i)} \hat{\bm{v}}_0^{(i+1)}.
\end{aligned}
\end{equation}

After the linearization and later simplification the predicted measurement at $t_m$ can be expressed as a linear function of the state of the next iteration:
\begin{equation}\label{eq:mess_lin_pos}
\begin{split}
    \hat{\bm{r}}_m = \hat{\bm{r}}(\hat{\bm{r}}_0,\hat{\bm{v}}_0,\Delta t_m) \approx & f(\hat{\bm{r}}_0^{(i)},\hat{\bm{v}}_0^{(i)},\Delta t_m) \hat{\bm{r}}_0^{(i+1)} + g(\hat{\bm{r}}_0^{(i)},\hat{\bm{v}}_0^{(i)},\Delta t_m) \hat{\bm{v}}_0^{(i+1)} \\ 
    = & f_m^{(i)} \text{I}_3 \hat{\bm{r}}_0^{(i+1)} + g_m^{(i)} \text{I}_3 \hat{\bm{v}}_0^{(i+1)} \\ 
    = & \begin{bmatrix} f_m & 0 & 0 & g_m & 0 & 0 \\
                                    0 & f_m & 0 & 0 & g_m & 0 \\
                                    0 & 0 & f_m & 0 & 0 & g_m \\ \end{bmatrix}^{(i)} \begin{bmatrix}
                                        \hat{\bm{r}}_0^{(i+1)} \\
                                        \hat{\bm{v}}_0^{(i+1)}
                                    \end{bmatrix}.
\end{split}
\end{equation}

By utilizing the form in Eq. (\ref{eq:mess_lin_pos}) it is possible to express the vector of (approximated) predicted positions ($\hat{\bm{z}}$) by linearly combining the matrix A, see Equation (\ref{eq:matrixA_pos}), with the estimation $\hat{\bm{y}}_0^{(i+1)}$.  The sub-indices in the elements of A refer to the measurement instant, and the $(i)$ super-index refers to this matrix being computed at every iteration. 

\begin{equation}\label{eq:matrixA_pos}
    \text{A}  = \begin{bmatrix} f_1 & 0 & 0 & g_1 & 0 & 0 \\
                                    0 & f_1 & 0 & 0 & g_1 & 0 \\
                                    0 & 0 & f_1 & 0 & 0 & g_1 \\
                                    \vdots & \vdots & \vdots & \vdots & \vdots & \vdots \\
                                    f_n & 0 & 0 & g_n & 0 & 0 \\
                                    0 & f_n & 0 & 0 & g_n & 0 \\
                                    0 & 0 & f_n & 0 & 0 & g_n \\
        \end{bmatrix}^{(i)}.
\end{equation}

The position residuals in $\bm z = \text{A}  \hat{\bm{y}}_0^{(i+1)} + \bm \xi$ are thus a explicit function of the estimation $\hat{\bm{y}}_0^{(i+1)}$. By solving the linear least-squares in Eq. (\ref{eq:llsp}) with $W$ being the identity matrix (non-weighted), the solution in Equation (\ref{eq:GTDS_min}) is that of the GTDS method. Emphasis has to be put on the choice of not weighting position residuals despite the availability of an uncertainty characterization. The result is a fit that minimizes errors in all directions equally. The covariance of the estimation is computed from the matrix $\text{H}=\left( \text{A}^\intercal \text{A}\right)^{-1} \text{A}^\intercal$ as a linear approximation after the estimation is computed (in the last iteration). For a more compact form to the solution of Eq. (\ref{eq:GTDS_min}) see \cite{siminski2016techniques}.
\begin{equation} \label{eq:GTDS_min}
\begin{split}
        \bm y_0^{(i+1)} & = \left( {\text{A}}^\intercal \text{A}\right)^{-1} {\text{A}}^\intercal \bm z = \text{H} \, \bm z , \\
        \text{C}_y & = \text{H} \text{C}_r \text{H}^\intercal, \, \text{where }\text{C}_r=\text{diag}\left( \text{C}_{r_1} , \cdots , \text{C}_{r_n} \right).
\end{split}
\end{equation}

\subsection{Keplerian fitting of radar observables} \label{sec:kep_fit_radar_observables}

This method is proposed as a variation of GTDS where measurements are fitted directly with Keplerian dynamics, which allows to incorporate the range-rate information into the fitting process. The measurement vector comprises all observables directly, represented as $\bm z = \left[ \rho_1, Az_1, {el}_1, \dot\rho_1, \cdots , \rho_n, Az_n, {el}_n, \dot\rho_n \right]^\intercal$. The $\left[ \rho, {Az}, {el}, \dot{\rho} \right] = h(\bm y,t)$ function for these measurements is in Equation (\ref{eq:rad_measurements}), where $\bm y  = \left[ \bm r \, \bm v\right]$, $t$ is omitted for brevity, and the relative inertial position $\bm \rho = \bm r - \bm P_R$ and velocity $\bm u = \bm v - \bm V_R$ are defined for a compact notation. The unit vectors forming the base of a Cartesian frame are denoted as $\left( \bm{e}_1, \bm{e}_2, \bm{e}_3 \right)$. It is important to note that incorporating range-rate necessitates the utilization of both predicted velocity and position. In the case of the Keplerian model, this entails computing the $\dot f$ and $\dot g$ functions, as shown in Equation (\ref{eq:kep_pos_f_g}) (see \cite{vallado2001fundamentals} \emph{Algorithm 8}).

\begin{equation}\label{eq:rad_measurements}
\begin{aligned}
    \rho & = \left[ \left( \bm r - \bm P_R \right)^\intercal \left( \bm r - \bm P_R \right) \right]^{1/2} =  \left[  \bm \rho^\intercal \bm \rho \right]^{1/2}, \\    
    Az & = \text{atan2}{\left( { \bm{e}_1^\intercal \bm \rho\rfloor_{R}},{ \bm{e}_2^\intercal \bm \rho\rfloor_{R}} \right)} = \text{atan2}{\left( {\bm{e}_1^\intercal \text{T}_{I}^{R} \bm \rho},{\bm{e}_2^\intercal \text{T}_{I}^{R} \bm \rho } \right)}, \\
    el & = \arcsin{\left( \frac{\bm{e}_3^\intercal \bm \rho\rfloor_{R}}{\rho} \right)}  = \arcsin{\left( \frac{\bm{e}_3^\intercal \text{T}_{I}^{R} \bm \rho}{\rho} \right)},  \\
    \dot\rho & = \frac{1}{\rho} \left( \bm r- \bm P_R \right)^\intercal \left( \bm v - \bm V_R \right) = \frac{\bm \rho^\intercal \bm u}{\rho}.
\end{aligned}
\end{equation}

The computation of $\partial h/\partial \hat{\bm{y}}_0\rfloor_{(\hat{\bm{y}}_0^{(i)},t_m)}= \partial h/\partial {\bm{y}}\rfloor_{(\hat{\bm{y}}_m^{(i)},t_m)} \partial {\mathcal{P}}/\partial \hat{\bm{y}}_0\rfloor_{(\hat{\bm{y}}_0^{(i)}, \Delta  t_m)}$ is divided in two parts. For one, the observation function derivatives, denoted as $\partial h / \partial \bm y$, are detailed in the \ref{appendix:meas_derivatives} for the radar observables. The predicted full state derivatives ($ \partial {\mathcal{P}}/\partial \hat{\bm{y}}_0$) are also required. The position part is present already in Section \ref{sec_gtds_fitting}, see Equation (\ref{eq:kep_pos_f_g_linear}). Nevertheless, the GTDS method approximates these derivatives by neglecting the Kepler function derivatives contribution, a simplification no longer valid for the fitting of radar observables. The predicted measurements are now a much more non-linear function of the estimation, so in order to improve convergence the approximation of GTDS is not used. Furthermore, the predicted velocity derivatives are also required, see Equation (\ref{eq:kep_vel_derivatives}). In conclusion, the present method requires the computation of all the Kepler function derivatives, or a better approximation, and the rest of the section delves on this.
\begin{equation}\label{eq:kep_vel_derivatives}
\begin{split}
    \frac{\partial \hat{\bm{v}}}{\partial \hat{\bm{r}}_0}\biggr\rfloor^{(i)} & = \dot f^{(i)} \frac{\partial \hat{\bm{r}}_0}{\partial \hat{\bm{r}}_0} + \hat{\bm{r}}_0^{(i)} \frac{\partial \dot f}{\partial \hat{\bm{r}}_0}\biggr\rfloor^{(i)} + \hat{\bm{v}}_0^{(i)} \frac{\partial \dot g}{\partial \hat{\bm{r}}_0}\biggr\rfloor^{(i)}, \\
    \frac{\partial \hat{\bm{v}}}{\partial \hat{\bm{v}}_0}\biggr\rfloor^{(i)} & = \dot g^{(i)} \frac{\partial \hat{\bm{v}}_0}{\partial \hat{\bm{v}}_0} + \hat{\bm{r}}_0^{(i)} \frac{\partial \dot f}{\partial \hat{\bm{v}}_0}\biggr\rfloor^{(i)} + \hat{\bm{v}}_0^{(i)} \frac{\partial \dot g}{\partial \hat{\bm{v}}_0}\biggr\rfloor^{(i)}.
\end{split}
\end{equation}

An analytical approximation for the derivatives of $f$, $g$, $\dot f$ and $\dot g$ has been developed for this work that applies a Taylor expansion around the values $\hat{\bm{r}}_0$ and $\hat{\bm{v}}_0$ at the reference epoch, see Equation (\ref{eq:taylor_expansion}). The notation used for partial derivatives here is a slight modification of Newton\textquotesingle s notation, where $\partial s / \partial t = \dot{s} =  \overset{\scriptscriptstyle(1)}{s}$. This is a deliberate choice that allows for a compact expression of a high order time derivative differentiated with respect to another variable $x$. See the \ref{appendix:f_g_derivatives} for a justification and example of use.
\begin{equation}\label{eq:taylor_expansion}
\begin{split}
    \hat{\bm{r}}(\hat{\bm{r}}_0,\hat{\bm{v}}_0,\Delta t) & = \hat{\bm{r}}_0 + \hat{\bm{v}}_0 \Delta t + \ddot{\hat{\bm{r}}}(\hat{\bm{r}}_0,\hat{\bm{v}}_0)\frac{\Delta t^2}{2!} + \dddot{\hat{\bm{r}}}(\hat{\bm{r}}_0,\hat{\bm{v}}_0)\frac{\Delta t^3}{3!}+  \overset{\scriptscriptstyle(4)}{\hat{\bm{r}}}(\hat{\bm{r}}_0,\hat{\bm{v}}_0)\frac{\Delta t^4}{4!} + \cdots \\
    \hat{\bm{v}}(\hat{\bm{r}}_0,\hat{\bm{v}}_0,\Delta t) & = \hat{\bm{v}}_0 + \ddot{\hat{\bm{r}}}(\hat{\bm{r}}_0,\hat{\bm{v}}_0) \Delta t + \dddot{\hat{\bm{r}}}(\hat{\bm{r}}_0,\hat{\bm{v}}_0)\frac{\Delta t^2}{2!} + \overset{\scriptscriptstyle(4)}{\hat{\bm{r}}}(\hat{\bm{r}}_0,\hat{\bm{v}}_0)\frac{\Delta t^3}{3!} + \cdots
\end{split}
\end{equation}

Obtaining the Kepler functions is a matter of formulating the expansion in Equation (\ref{eq:taylor_expansion}) as a linear combination of $\hat{\bm{r}}_0$ and $\hat{\bm{v}}_0$, as the multiplying coefficients are the Kepler functions by definition. For this purpose, the derivatives $\overset{\scriptscriptstyle(k)}{\hat{\bm{r}}}$ of the series can be obtained from the equations of motion $ \ddot{\bm r} = -\mu \bm r / r^3 = -u \bm r$,  expressed as a linear combination of $\bm r$ and $\bm v = \dot{\bm r}$ evaluated at the reference epoch, as outlined in Equation (\ref{eq:coeff_taylor1}). 
\begin{align}\label{eq:coeff_taylor1}
    \ddot{ \hat{\bm{r}} }_0 & =  \frac{-\mu}{ r^3} \bm r = -u \hat{\bm{r}}_0, & \overset{\dots}{\hat{\bm{r}}}_0  & =  -\dot{u} \hat{\bm{r}}_0 - u \hat{\bm{v}}_0, & \overset{\scriptscriptstyle(4)}{\hat{\bm{r}}}_0 & =  -\ddot{u} \hat{\bm{r}}_0  + (u^2 - 2 \dot{u}) \hat{\bm{v}}_0.
\end{align}

The value of $u$ and its time derivatives are a function of the scalar $r = (\bm r^\intercal \bm r)^{1/2}$ and its derivatives, see Equation (\ref{eq:align_components}).
\begin{align}\label{eq:align_components}
    \nonumber u & = \frac{\mu}{r^3}, & r & =  (\bm r^\intercal \bm r)^{1/2}, \\ 
    \dot{u} & = -3 u \frac{\dot{r}}{r}, & \dot{r} & =\frac{\bm r^\intercal \bm v}{r} , \\
    \nonumber \ddot{u} & = -3 \left[ \dot{u}\frac{\dot{r}}{r} + u\left(\frac{\ddot{r}}{r} - \frac{\dot{r}^2}{r^2} \right)  \right], &  \ddot{r} &  = \frac{v^2 - u r^2 - \dot{r}^2 }{r} .
\end{align}

Finally, by introducing the elements of Equation (\ref{eq:coeff_taylor1}) into Equation (\ref{eq:taylor_expansion}) it is straightforward to obtain the solution as a linear combination of the initial conditions, which allows to extract the Taylor expansions of $f$, $g$, $\dot f$ and $\dot g$, see Equation \ref{eq:fg_expansion}.

\begin{equation} \label{eq:fg_expansion}
    \begin{aligned}
        f = & 1 - \frac{u}{2}\Delta t^2 - \frac{\dot{u}}{6}\Delta t^3 - \frac{\ddot{u}-u^2}{24}\Delta t^4 + \cdots \\
        g = & \Delta t - \frac{u}{6}\Delta t^3- \frac{\dot{u}}{12} \Delta t^4 + \cdots \\
        \dot{f} = & -u \Delta t - \frac{\dot{u}}{2}\Delta t^2 + \frac{u^2 - \ddot{u}}{6}\Delta t^3 + \cdots \\
        \dot{g} = & 1 - \frac{u}{2}\Delta t^2 - \frac{\dot{u}}{3}\Delta t^3 + \cdots \\
    \end{aligned}
\end{equation}

However, these analytical approximations of the Kepler functions are employed exclusively to compute the derivatives required in \cref{eq:kep_pos_f_g_linear,eq:kep_vel_derivatives}. The detailed expressions have been omitted to avoid an excessive extension, but these are straightforward to derive from Eq. (\ref{eq:align_components}). Thus, the element $\partial \mathcal{P} / \partial  \hat{\bm{y}}_0 = \left[\begin{smallmatrix}\partial \hat{\bm{r}}/\partial \hat{\bm{r}}_0 & \partial \hat{\bm{r}}/\partial \hat{\bm{v}}_0 \\ \partial \hat{\bm{v}}/\partial \hat{\bm{r}}_0 & \partial \hat{\bm{v}}/\partial \hat{\bm{v}}_0\end{smallmatrix}\right]$ in Equation (\ref{eq:aprox_z}) can be computed at each $\Delta t_m$ with the necessary degree of approximation.

Now that the derivatives of the predicted measurement $\partial h/\partial \hat{\bm{y}}_0$ can be computed, all the elements to solve the linear least-squares problem in Equation (\ref{eq:llsp}) are available. In particular, the errors in $\Delta \bm z^{(i)}$ are computed directly with the propagated estimates of the last iteration  $\hat{\bm{y}}_m^{(i)}$, using Eq. (\ref{eq:kep_pos_f_g}), evaluated in the observation function in Eq. (\ref{eq:rad_measurements}). This is the second analytical method that is used for comparison in the results.
 




\subsection{$\text{J}_2$ fitting of radar observables}\label{sec_j2_fit_observables}

Finally, a third algorithm considering the $\text{J}_2$ perturbation has been developed. Several analytical methods exist for propagation that include the perturbation of an oblate planet, see \cite{martinusi2015analytic}, always at the cost of losing precision, in the form of mean orbital elements and with validity only for a small range of eccentricities. For the purpose of this work an exact solution is considered, even if valid only for a short time span. The approach employs the generalized equinoctial orbital elements described in \cite{giulio2021generalization}, henceforth GEqOE. 

\subsubsection{GEqOE formulation}
Given a perturbation force derivable from a potential energy $\mathscr{U}$, the GEqOE have the perturbing potential embedded in their definition, thus generalizing the classical equinoctial elements \citep{broucke1972equinoctial}. The formulation produces the generalized semi-major axis ($a$) and Laplace vector ($\mu \bm g$) that define a non-osculating ellipse in the orbital plane. Projecting $\bm g$ along the in-plane axes define the elements $p_1$ and $p_2$ (generalized versions of the $h$ and $k$ equinoctial elements). Kepler\textquotesingle s equation is written as a function of the generalized mean longitude $\mathcal{L}$. The generalized mean motion $\nu$ is derived from the total energy, which includes $\mathscr{U}$. Finally, the elements $q_1$ and $q_2$ (analogous to the classical $p$ and $q$) complete the set of six GEqOE $\bm \chi = \left[ \nu , p_1 , p_2 , q_1 , q_2 ,\mathcal{L} \right]$. The first order time derivatives of these elements are given as an explicit function of the perturbing potential $\mathscr{U}$ and the dissipative force $\bm P$. The general formulation is leveraged here to derive a Taylor expansion of the solution in a simplified case of these perturbations.

Equation \ref{eq:J2_geqoe} provides a specific instance of the GEqOE time derivatives with just the $\text{J}_2$ component of the potential perturbation $\mathscr{U}$. Notice that the evolution of the generalized mean motion $\nu$ is reduced to a constant value, which comes from particularizing $\mathscr{U}$ to an stationary field in an inertial frame\footnote{This is a simplification compatible with the inclusion of zonal harmonics only, where the gravitational potential has axial symmetry.}, and having no dissipating forces. The inclusion of drag is left out of the proposed algorithm due to the short propagation time spans of the present application.
\begin{equation}\label{eq:J2_geqoe}
    \begin{split}
        \dot{\nu} & = 0, \\
        \dot{p}_1 & = p_2 \left( \frac{h-c}{r^2} - I \hat{z} \right) - \frac{1}{c} \left( \frac{X}{a} + 2 p_2 \right) \mathscr{U}, \\
        \dot{p}_2 & = p_1 \left( I \hat{z} - \frac{h-c}{r^2}  \right) + \frac{1}{c} \left( \frac{Y}{a} + 2 p_1 \right) \mathscr{U}, \\
        \dot{q}_1 & = -I \frac{Y}{r}, \\
        \dot{q}_2 & = -I \frac{X}{r}, \\
        \dot{\mathcal{L}} & = \nu + \frac{h-c}{r^2} - I \hat{z} - \frac{1}{c} \left[ \frac{1}{\alpha} + \alpha \left( 1 - \frac{r}{a} \right) \right]\mathscr{U}. \\
    \end{split}
\end{equation}
Equation (\ref{eq:geqoe_parts}) includes some needed definitions throughout the this section, and Algorithms (\ref{alg:eci2geqoe}-\ref{alg:geqoe2eci}) in the \ref{appendix:eci2geqoe} provide the conversion of inertial position and velocity to (and from) the set of GEqOE. 
\begin{equation} \label{eq:geqoe_parts}
    \begin{aligned}
        \mathscr{U} & = - \frac{A}{r^3} (1-3 \hat{z}^2), &
        A & = \frac{\mu J_2 R_{\oplus}^2}{2}, \\
        \hat{z} & = \frac{z}{r} = \frac{ 2 \left(Y q_2 - X q_1 \right) }{r (1+q_1^2+q_2^2)}, &
        I & = \frac{3 A}{h r^3} \hat{z} (1-q_1^2-q_2^2), \\
        h & = \sqrt{c^2 - 2 r^2 \mathscr{U}}, &
        c & = \left( \frac{\mu^2}{\nu} \right)^{1/3} \sqrt{1 - p_1^2 - p_2^2}.
    \end{aligned}
\end{equation}
The numerical integration of the ordinary differential equation (ODE) system in Eq. (\ref{eq:J2_geqoe}) requires the computation of the $X$ and $Y$ values. Given the initial values of the GEqOE the first step is to solve the generalized Kepler\textquotesingle s Equation (\ref{eq:gen_kepler}). 
\begin{equation} \label{eq:gen_kepler}
    \mathcal{L} =  \mathcal{K} + p_1 \cos{\mathcal{K}} - p_2 \sin{\mathcal{K}}.
\end{equation}
The output $\mathcal{K}$ is then used to compute $r$, $X$ and $Y$ from Eq. (\ref{eq:r_X_Y}), where $a=\left( \mu / \nu^2 \right)^{1/3}$ and $\alpha = 1 / \sqrt{1-p_1^2 - p_2^2}$. These are the steps needed to numerically solve the GEqOE dynamics. Instead, the goal of this section is to derive an analytical solution that can be evaluated once to obtain the propagated states and derivatives for the fitting problem in Eq. (\ref{eq:llsp}).
\begin{equation}\label{eq:r_X_Y}
\begin{split}
    r  = & a \left( 1 - p_1 \sin{\mathcal{K}} - p_2 \cos{\mathcal{K}}, \right) \\
    X  = & a  \left[ \alpha p_1 p_2 \sin{\mathcal{K}} + (1-\alpha p_1^2)\cos{\mathcal{K}}  - p_2 \right], \\
    Y = &  a  \left[ \alpha p_1 p_2 \cos{\mathcal{K}} + (1-\alpha p_2^2)\sin{\mathcal{K}}  - p_1 \right]. \\
\end{split}
\end{equation}

In \cite{giulio2021generalization} the authors report the natural slower evolution of the GEqOE, which results in a much reduced numerical integration error when compared with other orbital elements representation. Applying a Taylor expansion in this particular state representation is thus much preferred to the alternatives. Differentiation with respect the initial GEqOE values $\bm \chi_0$ can then be applied to obtain the matrix of derivatives $\frac{\partial \bm \chi}{\partial \bm \chi_0}$. The methodology employed to achieve this is explained next.

\subsubsection{Taylor expansion in GEqOE} \label{sec:j2_expansion_theory}

A special notation is be used from now on in order to allow for a systematic differentiation of the ODEs in Eq. (\ref{eq:J2_geqoe}). For one, the function $\mathscr{f}_s=1/s$ is defined as the inverse of the scalar $s$. This is useful, as there are many parts of the equations of motion that involve division, and it is most convenient in order to apply the chain rule with the multiplying element $\mathscr{f}_s$ instead. For example, see Equation (\ref{eq:rdot}) for the time derivative of $r$ expressed with the true longitude $L$ \citep{giulio2021generalization}, and rewrite it as $\dot{r} = \mu \mathscr{f}_c r_{pl}$. 
\begin{equation} \label{eq:rdot}
    \dot{r} = \frac{\mu}{c} \left( p_2 \sin{L} - p_1 \cos{L} \right).
\end{equation}
Now the time derivative $\ddot{r}$ can be put down as $\ddot{r} = \mu \left( \dot{\mathscr{f}}_c r_{pl} + \mathscr{f}_c \dot{r}_{pl} \right)$. It is only a matter of having the elements needed in order to compute $\dot{\mathscr{f}}_s=-\dot{s}/s^2$, the derivative of $s$ itself. Thus, a function that computes time derivatives of $\mathscr{f}_s$ has been implemented which takes as input the vector $\left[s,\dot{s}, \ddot{s} ,\cdots \right]$ and computes the derivative up to the desired order, see the \ref{appendix:inverse_time_der} for the detailed expressions.
\begin{equation}\label{eq:XdotYdot}
    \begin{split}
        \cos{L}  = & \frac{X}{r}, \\
        \sin{L}  = & \frac{Y}{r}, \\
        \dot{X}  = & \, \dot{r} \cos{L} - \frac{h}{r} \sin{L}, \\
        \dot{Y}  = & \, \dot{r} \sin{L} + \frac{h}{r} \cos{L}.
    \end{split}
\end{equation} 

Another useful tool is the function $\mathscr{g}_{s}=s\dot{s}$, which appears when a derivative of $s^2$ is performed. See the \ref{appendix:s_dots_fun} for the expressions of $\mathscr{g}_{s}$ time derivatives.

Computing the first order derivatives of the Taylor expansion reduces to evaluating the ODEs. For the next orders, the expressions in \cref{eq:rdot,eq:XdotYdot} become useful. The methodology is simplified further by applying an extra intermediate step. The differentiation of the ODEs in Eq. (\ref{eq:J2_geqoe}) is going to need in turn the derivatives of the elements in Eq. (\ref{eq:geqoe_parts}), and this becomes tedious when going beyond the second order Taylor expansion. The solution has been to split them in parts that are easy to derive and work with. See for example Eq. (\ref{eq:zhat_der}), where this has been done for the variable $\hat{z}$ and its many components. The time derivative of $\mathscr{f}_{\mathcal{D}}$ is computed from $\dot{\mathcal{D}}$.
\begin{equation}\label{eq:zhat_der}
    \begin{aligned} 
        \hat{z} & =  \frac{ 2 \left(Y q_2 - X q_1 \right) }{r (1+q_1^2+q_2^2)} = \frac{ \mathcal{C} }{ \mathcal{D} } = \mathcal{C}\mathscr{f}_{\mathcal{D}},  & \dot{\hat{z}} & = \dot{\mathcal{C}}\mathscr{f}_{\mathcal{D}} + \mathcal{C}\dot{\mathscr{f}_{\mathcal{D}}}, \\
        \mathcal{C} & = 2 \left(Y q_2 - X q_1 \right), & \dot{\mathcal{C}} &  = 2 \left(\dot{Y} q_2 + Y \dot{q}_2 - \dot{X} q_1 - X \dot{q}_1\right), \\
        \mathcal{D} & = r (1+q_1^2+q_2^2) = r q_s, & \dot{\mathcal{D}} & = \dot{r} q_s + r \dot{q}_s, \\
        q_s & = (1+q_1^2+q_2^2), & \dot{q}_s & = 2 \left( q_1 \dot{q}_1 + q_2 \dot{q}_2 \right).
    \end{aligned}
\end{equation}

With this in mind, the expressions of the equations of motion become those in Eq. (\ref{eq:J2_geqoe_simplified}), where $ \mathscr{s}_L =\sin{L} $, $\mathscr{c}_L = \cos{L}$ and $\Gamma = \mathscr{f}_{\alpha} + \alpha \left(1 - r/a \right)$. The full derivation up to fourth order and all the symbols definitions can be found in the \ref{appendix:j2_derivatives}.
\begin{equation}\label{eq:J2_geqoe_simplified}
    \begin{aligned}
        \dot{\nu} & = 0, & \dot{p}_1 & = p_2 \left( \mathscr{d} - w_h \right) - \mathscr{f}_{c} \xi_1 \mathscr{U}, & \dot{p}_2 & = p_1 \left( w_h  - \mathscr{d}  \right) + \mathscr{f}_{c} \xi_2 \mathscr{U}, \\
        \dot{q}_1 & = -I \mathscr{s}_L, & \dot{q}_2 & = -I \mathscr{c}_L, & \dot{\mathcal{L}} & = \nu + \mathscr{d} - w_h - \mathscr{f}_{c} \Gamma \mathscr{U}. \\
    \end{aligned}
\end{equation}
Once all the derivatives have been computed at the epoch (middle of the track), a Taylor expansion can be used to propagate to all $\Delta t_m$ with a simple polynomial evaluation. The result for each evaluation is the state vector $ \bm \chi \left( \bm \chi_0, \Delta t_m \right) = \left[ \nu, p_1, p_2, q_1, q_2, \mathcal{L} \right]^{\intercal} $, which can then be converted to Cartesian using the inverse transformation in Algorithm \ref{alg:geqoe2eci}. 


The next step is to perform the analytical derivative of $\bm \chi \left( \bm \chi_0 , \Delta t_m \right)$ with respect to $\bm \chi_0$, so that $\frac{ \partial \bm \chi}{\partial \bm \chi_0}$ can be obtained. This is not a trivial task, as the expressions of the coefficients of the expansion, see the \ref{appendix:j2_derivatives}, are extensive. Furthermore, many contain the $\mathscr{f}_s$ and $\mathscr{g}_s$ functions (and its time derivatives), which need to be differentiated with respect to a generic orbital element, and the \ref{appendix:f_g_derivatives} includes the analytical formulation of these. In order to avoid an excessive extension this work lacks the complete derivation of $\frac{ \partial \bm \chi}{\partial \bm \chi_0}$. The analytical derivatives of the coefficients in the \ref{appendix:j2_derivatives} have been validated by the use of Taylor differential algebra with Hipparchus. Again, this is only for validation, as the fully analytical implementation performs better.


The output $\frac{ \partial \bm \chi}{\partial \bm \chi_0}$ is the error transition matrix from and to GEqOE, but the fitting algorithm works with Cartesian states and derivatives. In order to perform this conversion the jacobians of both transformations, RV2GEQOE $\left( \frac{ \partial \bm \chi}{\partial \bm X} \right)$ and GEQOE2RV $\left( \frac{ \partial \bm X}{\partial \bm \chi} \right)$, are required. These are available in analytical form in \cite{giulio2021generalization}, and the implementation has been validated by the application of automatic differentiation over Algorithms (\ref{alg:eci2geqoe}-\ref{alg:geqoe2eci}). Now, in order to obtain $\frac{ \partial \bm X}{\partial \bm X_0}$, it is simply a matter of multiplying by the corresponding jacobian evaluated at the correct instant, see Equation (\ref{eq:j2_pX_pX0_approx}).
\begin{equation}\label{eq:j2_pX_pX0_approx}
    \frac{ \partial \bm X}{\partial \bm X_0} = \frac{ \partial \bm X}{\partial \bm \chi}\biggr\rfloor_{t} \left( \frac{ \partial \bm \chi}{\partial \bm \chi_0} \right) \frac{ \partial \bm \chi}{\partial \bm X}\biggr\rfloor_{t_0}.
\end{equation}
Notice that the evaluation of $\frac{ \partial \bm X}{\partial \bm \chi}\biggr\rfloor_{t}$ is done at the state computed by the Taylor expansion itself. The correctness of the analytical approximation of $\frac{ \partial \bm X}{\partial \bm X_0}$ has been checked by numerically integrating the variational equations with the $\text{J}_2$ perturbation added. This completes the analytical method with the addition of the $\text{J}_2$ perturbation.


\subsection{Adding extra information: Orbital Plane-based Orbit Determination - OPOD}\label{sec:orb_plane_fitting}

The main setback in doing precise initial orbital state determination with the information available in a single radar track is precisely the scarcity of data. This will later affect the sensitivity of derived metrics for maneuver detection. This section presents an alternative approach in order to improve the quality and reliability of the estimation. The idea behind it is that in LEO maneuvers are usually performed to maintain altitude, and thus do not act on the orbital plane itself. This is in part due to how expensive it is to change the inclination $i$ of the orbit, or its right ascension of the ascending node $\Omega$ (RAAN). It is assumed that the object of the track has a known past orbit, which makes it possible to predict the orbital plane with relatively high precision. Here it is assumed that the process noise (drag uncertainty primarily) acts in a direction that does not affect \emph{directly} the $i$ and $\Omega$ prediction. The predictability of both is assumed for the moment, in particular with the presence of maneuvers, allowing to perform Orbital Plane-based Orbit Determination (OPOD).

The measurement function $h(\bm{y},t)$ for these new observables, as well as the analytical derivative with respect to the Cartesian state $\partial h / \partial \bm{y}$ are deduced next. Consider the specific angular momentum as the cross product of position and velocity both given in an inertial frame (I), $\bm h = \bm r \times \bm v$. The inclination \emph{i}  can be computed as a function of $\bm r$ and $\bm v$ from Equation (\ref{app_eq_i}).
\begin{align}\label{app_eq_i}
    h & = \left(\bm h^\intercal \bm h \right)^{1/2}, & h_z & = \bm{e}_3^\intercal \bm h, & i = \arccos{\left( \frac{h_z}{h} \right)}.
\end{align}
The right ascension of the ascending node $\Omega$ is computed from Eq. (\ref{app_eq_omega}).
\begin{align}\label{app_eq_omega}
    \bm n^{*} & = \bm{e}_3 \times \bm h, & n^{*} & = \left(\bm n^{*\intercal} \bm n^{*}  \right)^{1/2},  & \bm n & = \frac{\bm n^{*}}{n^{*}}, & \Omega & = \arccos{\left(\bm{e}_1^\intercal \bm n  \right)}.
\end{align}
For the derivation of $i$ and $\Omega$ with respect to position and velocity the cross product of a $3\times n$ matrix $M$ by a $3 \times 1$ vector $\bm d$ is defined as the cross product of each column in $\text{M}$ by $\bm d$. Consequently, the \emph{i-eth} column of $\text{D}=\text{M} \times \bm d$ meets $\text{D}_{i} = \text{M}_i \times \bm d$. In other words, the cross product of matrix and vector acts column-wise for the next derivation. The differentiation with respect to the state for the inclination results in the following:
\begin{align*}
    \frac{\partial \bm h}{\partial \bm r}  = & \, \frac{\partial \bm r}{\partial \bm r} \times \bm v + \bm r \times \frac{\partial \bm v}{\partial \bm r} = I \times \bm v, \\
    \frac{\partial \bm h}{\partial \bm v}  = & \, \frac{\partial \bm r}{\partial \bm v} \times \bm v + \bm r \times \frac{\partial \bm v}{\partial \bm v} = -I \times \bm r, \\
    \frac{\partial \bm h}{\partial \bm y}  = & \, \left[\frac{\partial \bm h}{\partial \bm r} , \frac{\partial \bm h}{\partial \bm v} \right], \\
    \frac{\partial h_z}{\partial \bm y}    = & \,  \bm{e}_3^\intercal \frac{\partial \bm h}{\partial \bm y}, \\ 
    \frac{\partial h}{\partial \bm y}      = & \, \frac{\bm h^\intercal}{h}\frac{\partial \bm h}{\partial \bm y}, \\ 
    \frac{\partial i}{\partial \bm y}      = & \, - \frac{1}{\sqrt{1 - \cos^2i }}\left[ \frac{\partial h_z}{\partial \bm y} - \cos{i} \frac{\partial h}{\partial \bm y}  \right]\frac{1}{h}.
\end{align*}

For the right ascension of the ascending node on the other hand:
\begin{align*}
    \frac{\partial \bm n^{*}}{\partial \bm y}  = & \, \bm{e}_3 \times \frac{\partial h}{\partial \bm y} =  - \frac{\partial h}{\partial \bm y} \times \bm{e}_3, \\ \frac{\partial n^{*}}{\partial \bm y}  = & \, \frac{ \bm n^{*\intercal}}{n^{*}}\frac{\partial \bm n^{*}}{\partial \bm y}, \\
    \frac{\partial \bm n}{\partial \bm y}  = & \, \left[\frac{\partial \bm n^{*}}{\partial \bm y} - \bm n \frac{\partial n^{*}}{\partial \bm y} \right] \frac{1}{n^{*}}, \\ \frac{\partial \Omega}{\partial \bm y} = & \, - \frac{1}{\sqrt{1+\cos^2\Omega}} \bm{e}_1^\intercal \frac{\partial \bm n}{\partial \bm y}.
\end{align*}

The addition of these virtual measurements can be applied to either the Keplerian or $\text{J}_2$ fitting of radar measurements, but not the GTDS method as it only works with position information.

\section{Simulated results}\label{sec:simulated_results}


For this section, synthetic simulations of the satellite\textquotesingle s trajectory and consequent radar measurements are leveraged to derive results on the IOD methods\textquotesingle~performance. The satellite dynamic\textquotesingle~s are simulated using high-fidelity models that account for the shape, attitude and solar array orientation, and include all perturbations relevant for the time spans of the simulations. The radar simulation is a simplified model that uses a fixed noise characterization for all range values. The implementation of the dynamical models and radar simulation make use of the Orekit library \citep{maisonobe2010orekit}, as explained in Section \ref{sec:dyn_radar_charaterization_res}. After that, Section \ref{sec:prel_testing_non_plane_alg} presents the necessary preliminary testing of the radar-only IOD methods, establishing a baseline for the performance and its dependency with the track length. Then, Section \ref{sec:prel_testing_plane_astror} performs the same preliminary testing on the orbital-plane enhanced method, denoted as OPOD, showing the effect of adding orbital plane information on the estimation, and the dependency with the assumed uncertainty in the presence of maneuvers. Section \ref{sec:scenarios_k2_results} presents the synthetic testing methodology and the particular scenarios that generate all the track data used in Section \ref{sec:k2_metric_def}. In this latest section a $k^2$ metric is defined to test the correct uncertainty characterization of the estimation errors.

\subsection{Dynamical modeling and radar characterization}\label{sec:dyn_radar_charaterization_res}

For a classical IOD method reliant upon sensor data alone it would be enough to simply simulate the ``real'' satellite trajectories and generate samples of radar tracks. The testing methodology has been made a bit more complex to accommodate for the OPOD algorithm as well (reliant upon predicted orbital plane), thus needing two distinct dynamics. Table \ref{tab:dynmodels} includes the dynamical models needed for all numerical simulations performed in this work. Notice that the main difference between \emph{Real} and \emph{Prediction} dynamics resides in the type of atmosphere model and the degree and order of the harmonics included. Both the NRLMSISE-00 \citep{picone2002nrlmsise} and DTM-2000 \citep{bruinsma2003dtm} are empirical atmospheres that feed on tabulated solar weather data, particularly mean and instantaneous solar flux, and geomagnetic indices dependent on the solar activity. Still, these are different in order to faithfully consider the ignorance of the real atmosphere at any time. 

On the other hand, satellite models are distinguished between high-fidelity (HF), see Table \ref{tab:real}, and low-fidelity (LF), see Table \ref{tab:sim}. In this case the differences are more pronounced, as the LF satellites use a simple spherical drag model, where in the HF case their orientation is relevant, and always fixed to the LVLH frame\footnote{X axis aligned with position, Z axis aligned with orbital momentum}. Suffice to say, for the simulation of real satellite trajectories a HF model is paired with Real dynamics, while propagations intended for predicting the orbital plane use a LF model in conjunction with Prediction dynamics. In contrast, within the IOD process propagations use the analytical models developed in Section \ref{sec:estimation}.
\begin{table*}[h]
    \caption{The two dynamical models defined, one for the real trajectory and other for the prediction, the third one is used in a special scenario for checking purposes. SRP stands for Solar Radiation pressure.}
   \label{tab:dynmodels}
    \centering 
   \begin{tabular}{c | c | r | r | r | r} 
      \hline 
      Model & Earth harmonics & Atmosphere & Sun & Moon & SRP \\
      \hline 
      Real     & $[150,150]$ & DTM-2000     & yes & yes & yes \\
      Prediction   & $[40,40]$   &  NRLMSISE-00 & yes & yes & yes \\
      Test   & $[2,2]$   &  None & no & no & no \\
      \hline
   \end{tabular}
\end{table*}

The name of the satellites are just a reference to the real-world equivalent in terms of the orbit type. The Sentinel-1A resides in a $700$ Km orbit, while Swarm-C is at an altitude of $430$ Km. Starlink-1 and 2 are lower inclination orbits at around $540$ Km. Table \ref{tab:sto} includes the initial conditions that are used for the synthetic scenarios of each. The aim of including these satellites is to generate variety in the length of the radar track. 

Out of all the HF satellites defined in Table \ref{tab:real} only Sentinel-1A and Swarm-C have a corresponding LF model for prediction computations (see Table \ref{tab:sim}). Thus, the algorithm that includes predicted orbital plane data (OPOD) is tested on these satellites alone. The other three radar-only IOD methods (Sections \ref{sec_gtds_fitting}, \ref{sec:kep_fit_radar_observables} and \ref{sec_j2_fit_observables}) are tested on all satellites.

\begin{table*}[h]
    \caption{Parameters defining the realistic satellites Sentinel-1A, Swarm-C and Starlink-1/2 using Orekit\textquotesingle s \emph{BoxAndSolarArraySpacecraft} force model class. These parameters are: mass, dimensions of the box-shaped body, solar array area, solar array rotation axis in the body frame, drag coefficient, absorption coefficient and reflection coefficient. The satellite body is always aligned with the LVLH frame of reference.}
   \label{tab:real}
        \centering 
   \begin{tabular}{c | r | r | r | r | r | r | r | r | r} 
      \hline 
      HF Satellite & $m$ ($Kg$) & $x_l$ ($m$) & $y_l$ ($m$)& $z_l$ ($m$)& $A_s$ ($m^2$)& $A_x$ [-]& $C_D$ [-]& $a_c$ [-]& $r_c$ [-]\\
      \hline 
      Sentinel-1A  &  2270  & 1.02 & 1.34 & 3.2 & 25.46 & $y$ & 1.5 & 0.8 & 0.5 \\
      Swarm-C   & 440       & 1.3     &  8 & 0.9 & 0.1 & $y$ & 0.95 & 0.9 & 0.5 \\
      Starlink-1   & 1250   & 2.8     &  2.3 & 0.3 & 25.2 & $z$ & 0.9 & 0.9 & 0.5 \\
      Starlink-2   & 800    & 2.5     &  2 & 0.2 & 20 & $z$ & 0.92 & 0.9 & 0.5 \\
      \hline
   \end{tabular}
\end{table*}

\begin{table*}[h]
    \caption{Parameters defining the low fidelity satellites Sentinel-1A$_{1}$ and Swarm-C$_{1}$ using Orekit\textquotesingle s isotropic drag and radiation force model classes. These parameters are: mass, drag cross section, drag coefficient, solar radiation cross section and reflection coefficient.}
   \label{tab:sim}
        \centering 
   \begin{tabular}{c | r | r | r | r | r} 
      \hline 
      LF Satellite & $m$ ($Kg$) & $S$ ($m^2$) & $C_D$ [-]& $SRP_s$ ($m^2$)& $r_c$ [-] \\
      \hline 
      Sentinel-1A$_{1}$  &  2200  & 9.5   & 2     & 10    & 1.4 \\
      Swarm-C$_{1}$   & 420       & 4     & 2.56  & 4     & 1.4 \\
      \hline
   \end{tabular}
\end{table*}

\begin{table*}[h]
    \caption{Initial condition (I.C.) for the reference orbit defined for the scenarios in Table \ref{tab:scenarios}, in Cartesian coordinates (Earth EME2000 inertial frame).  The first is a Sentinel-1A type orbit. The other is used in the Swarm-C scenario. The initial epoch $t_0$ is at 00:00:00 UT.}
   \label{tab:sto}
        \centering 
   \begin{tabular}{c | c | c | c | c | c | c | c} 
      \hline 
      I.C. & $x$ ($km$) & $y$ ($km$) & $z$ ($km$) & $v_x$ ($km/s$) & $v_y$ ($km/s$) & $v_z$ ($km/s$) & $t_0$  \\
      \hline 
      S1A-1    &  1459.975  & 436.989 & -6916.264  & -3.8952 & -6.282 & -1.219  & 01/05/2022   \\
      SWC-1    & 254.597  & -241.108 & 6792.396  & 7.151 & 2.655 & -0.1728  & 06/07/2020     \\
      SRL-1    & 4970.721  & 3941.433 & 602.664  & -4.458 & 4.973 & 4.244  & 27/08/2023     \\
      SRL-2    & 5030.115  & 3838.607 & 754.660  & -4.040 & 4.249 & 5.315  & 27/08/2023     \\
      \hline
   \end{tabular}
\end{table*}

\begin{figure}[H]
    \centering
    \begin{subfigure}{0.42\textwidth} 
      \centering
      \includegraphics[width=\textwidth]{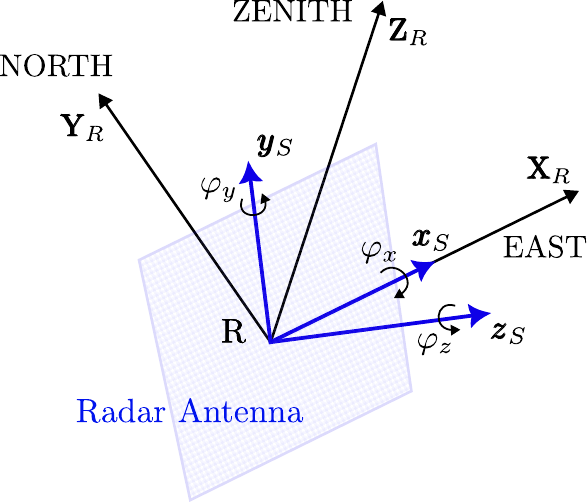}
      \caption{}
      \label{fig:radar_antena_a}
    \end{subfigure}
    \begin{subfigure}{0.42\textwidth}
      \centering
      \includegraphics[width=\textwidth]{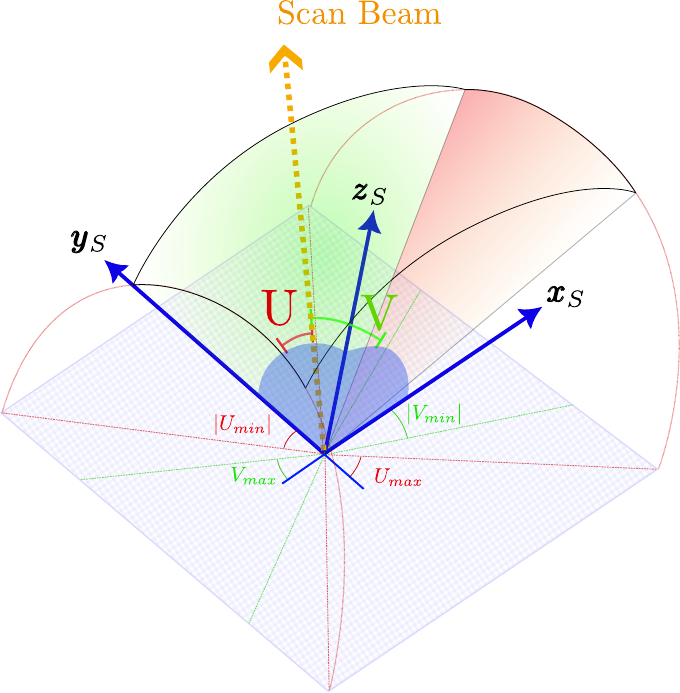}
      \caption{}
      \label{fig:radar_antena_b}
    \end{subfigure}
    \caption{The orientation of the radar antenna is given by 3 consecutive rotations around its local $x$, $y$ and $z$ axes, in that order. Initially it is parallel to the $XY$ plane of the topocentric frame of reference with the same origin, see (\ref{fig:radar_antena_a}). The field of regard (FoR) is expressed in terms of the scanning angles U and V, which are the angles between the scanning beam and the local $yz$ and $xz$ planes respectively. Each is limited in the positive and negative side, and define a volume shaped as the outside of two concentric cones that meet at the vertex. When the limits of both angles are considered the result is the volume outside 4 cones that intersect with the two perpendicular to themselves, see (\ref{fig:radar_antena_b}).}
    \label{fig:radar_antena}
\end{figure}

The dynamics part of the synthetic testing is carried out with Orekit \citep{maisonobe2010orekit}, a well proven and tested space dynamics library. This is also true for the measurement generation part. Orekit includes classes that can handle the type of sensor needed, taking into account these are two-way measurements (reflection time is part of the measurement generation). The particular characterization of the simulated field of regard (FoR) is part of the caption in Figure \ref{fig:radar_antena_b}. Technically, this surveillance volume is defined in a limited range from the origin (minimum and maximum), but that is overlooked for this work, as the orbits considered are within it at all times. The generation of the measurements includes the addition of noise over the real values. This noise is a normal distribution in measurement space, has constant covariance for all measurements (a simplified noise model), and includes no biases. The limited orbital regimes considered here justify this straightforward model, in contrast to \citep{siminski2016techniques} where the radar survey is applied to a wide extent of orbital heights. It is important to highlight that the uncertainty characterization is supposed to be known perfectly when performing the estimation. 


All the parameters characterizing a radar are specified in Table \ref{tab:radar_char} for three different stations that are used in the synthetic testing. It must be highlighted that these three radar stations are used separately, that is, in different scenarios (the data each generates is never used in conjunction with the others). Using more than one radar in the synthetic data generation is meant to enrich the track length variability and data density over different scenarios.


\begin{table}[ht]
	\fontsize{8}{8}\selectfont
    \caption{Complete characterization of the radar stations used for the synthetic testing. The angles $\lambda$ and $\phi$ are the geodetic coordinates in the WGS84 ellipsoid, and $h$ is the altitude over it. The observables have a standard deviation $\sigma$, and the angular measurements have a correlation given by $\xi_{Az,el}$. The revisit time $r_t$ is the time between measurements. Radar 1 has been modeled as an approximation of the S3T Surveillance Radar~\citep{hernandez2021operational}, but the FoR parameters (subject to changes during its real operation) have been omitted due to confidentiality reasons.}
   \label{tab:radar_char}
    \centering 
   \begin{tabular}{c | c | c | c | c | c | c | c | c | c | c | c} 
      \hline 
      Name & $\left( \lambda,\phi \right)$ ($^\circ$) & $h$ (m) & $\left(\varphi_x,\varphi_y,\varphi_z\right)$ ($^\circ$) & $\left(\text{U}_{min},\text{U}_{max}\right)$ ($^\circ$) & $\left(\text{V}_{min},\text{V}_{max}\right)$ ($^\circ$) & $\sigma_{\rho}$ (m) & $\sigma_{\dot{\rho}}$ (m/s) & $\sigma_{Az}$ ($^\circ$) & $\sigma_{el}$ ($^\circ$) & $\xi_{Az,el}$ & $r_t$ (s)  \\
      \hline 
      Radar 1 & $\left( -5.59,37.17 \right)$ & 142.32 & - & - & - & $7$ & $0.4$ & $0.3$ & $0.2$ & $0.043$ & 7  \\
      Radar 2 & $\left( 22.3,78.49 \right)$ & 22.14 & $\left( 40,0,0 \right)$ & $\left( -30,30 \right)$ & $\left( -20,20 \right)$ & $6.5$ & $0.35$ & $0.25$ & $0.15$ & $0.043$ & 4  \\
      Radar 3 & $\left( 175.68,-37.86 \right)$ & 36.2 & $\left( 30,0,0 \right)$ & $\left( -20,20 \right)$ & $\left( -45,45 \right)$ & $6.5$ & $0.35$ & $0.25$ & $0.15$ & $0.043$ & 4  \\
      \hline
   \end{tabular}
\end{table}

\subsection{Preliminary testing of the IOD algorithms}\label{sec:prel_testing_non_plane_alg}

With all aspects of the simulation in place, and the fitting methods that are to be compared, see Section \ref{sec:estimation}, it is necessary to first apply and compare them in a controlled case. Understanding the limitations and characteristics of each does help to better comprehend the estimation statistics later on, see Section \ref{sec:stats_IOD_radar_only}. The approach consists on taking a single radar track, and apply the three IOD algorithms over it (radar data alone is used). In order to study the effect of the track length on the estimation qualities (absolute error and uncertainty), a long track is chosen, and it is made shorter by taking out measurements from the extremes (two at a time), repeating the fitting each time.

The longest track generated by Radar 3 when simulating Sentinel-1A during 4 days has been chosen. It consists of 72 measurements and spans 284 seconds from the initial to the final measurement. This is the only track used for the test.


\subsubsection{Fitting of noiseless measurements}

\begin{figure}[h]
    \centering
    \includegraphics[width=0.75\textwidth]{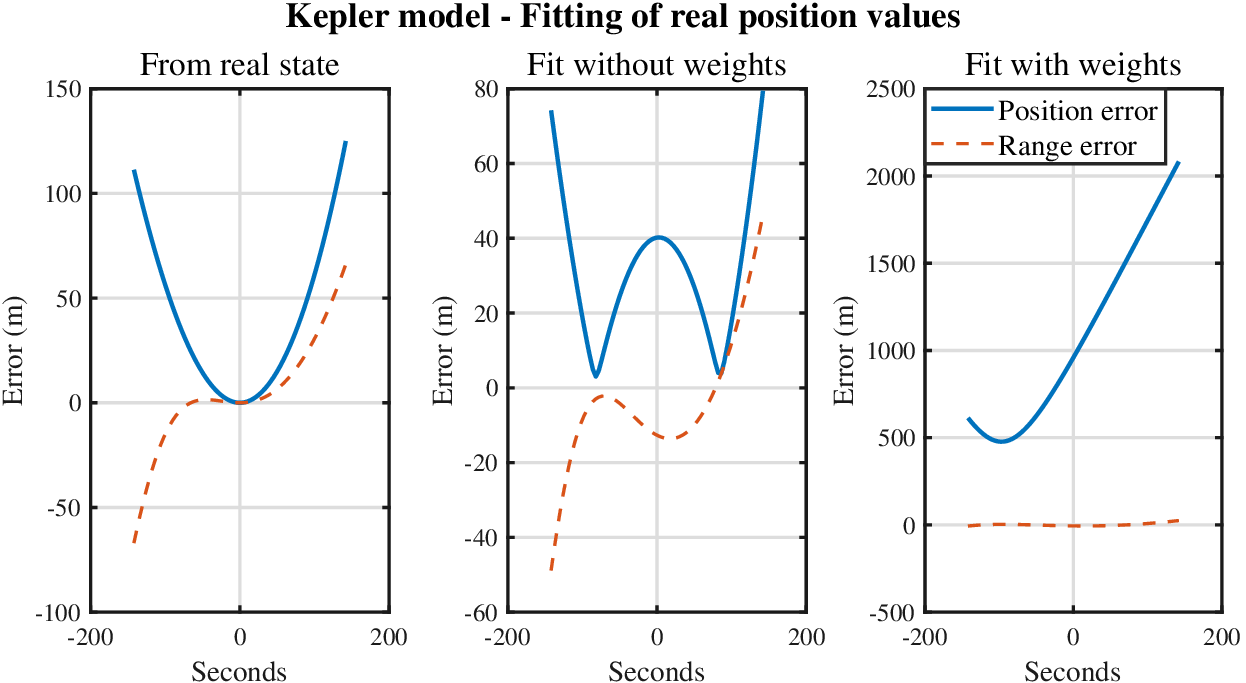}
    \caption{Position and range errors using a simple Kepler model for a very long radar track ($284$ seconds sample track from Radar 3). The left plot is a propagation from the true state at the epoch (middle of the track). The center plot is the estimation resulting from applying the GTDS algorithm (Section \ref{sec_gtds_fitting}), but using the \emph{real} position values of the Sentinel-1A high fidelity simulation as measurements. The right plot is a fitting, also with real position values, but applying weights based on the measurements uncertainty of this radar (conversion from radar to inertial covariance with Unscented Transform).}
    \label{fig:kepler_fit}
\end{figure}

First, it is relevant to consider not the noisy measurements, but the real values of these, in order to emphasize the base limitations of each method. Propagating the real position and velocity at the epoch of the test track with Kepler dynamics results in the errors of Figure \ref{fig:kepler_fit} (Left). Position errors are on the order of $100$ meters, and more importantly, range errors go over $50$ m. This is a trajectory that coincides with the real one at the epoch, and yet the Kepler model incurs in range errors one order of magnitude greater than the uncertainty of the radar measurements. 

Now, applying the GTDS algorithm over the real position measurements results in the estimation with errors of Figure \ref{fig:kepler_fit} (Center). The value of the square sum of errors ($\bm \xi^\intercal \bm \xi$), is reduced here to its minimum value. No weights are applied, so on average the position errors are lower along the complete trajectory, and not just at the epoch. Range errors on the other hand are not necessarily lower, as this fitting has no preferential direction for the minimization of error.

The opposite is happening in Figure \ref{fig:kepler_fit} (Right), where the position fitting is done by minimizing $\bm \xi^\intercal \text{W} \bm \xi$, with $\text{W}=\text{C}_r^{-1}$. Position covariance computed with an UT has greater variance in the direction of range (radar-satellite vector) than the original uncertainty in measurement space. Even then, it is still a very thin disk, so the covariance is much less homogeneous than the non-weighted case (same as an uniform weight in all directions). Notice that this fit does no correspond to any of the methods highlighted in Section \ref{sec:estimation}, but serves as a step prior to fitting observables directly. The error of the estimation is now on the order of the kilometer, while still being a fitting of exact position values. The only change is the increased weight of the error in the range direction, forcing the residuals (in that direction) to be much lower. The residuals on the perpendicular directions are not so relevant, and is even less important for measurements further apart from the radar, as the disk-shaped covariance gets wider. The problem resides in the pairing of Kepler dynamics with a strongly directional weighting. Figure \ref{fig:kepler_fit} (Left) showed how this model is not capable of keeping low range residuals for such a long pass, so the fitting algorithm has to change the trajectory considerably. The result is a kilometer wide gap at the epoch for the resulting estimation (but with range errors on the order of $10$ m).



\begin{figure}[h]
    \centering
    \includegraphics[width=0.75\textwidth]{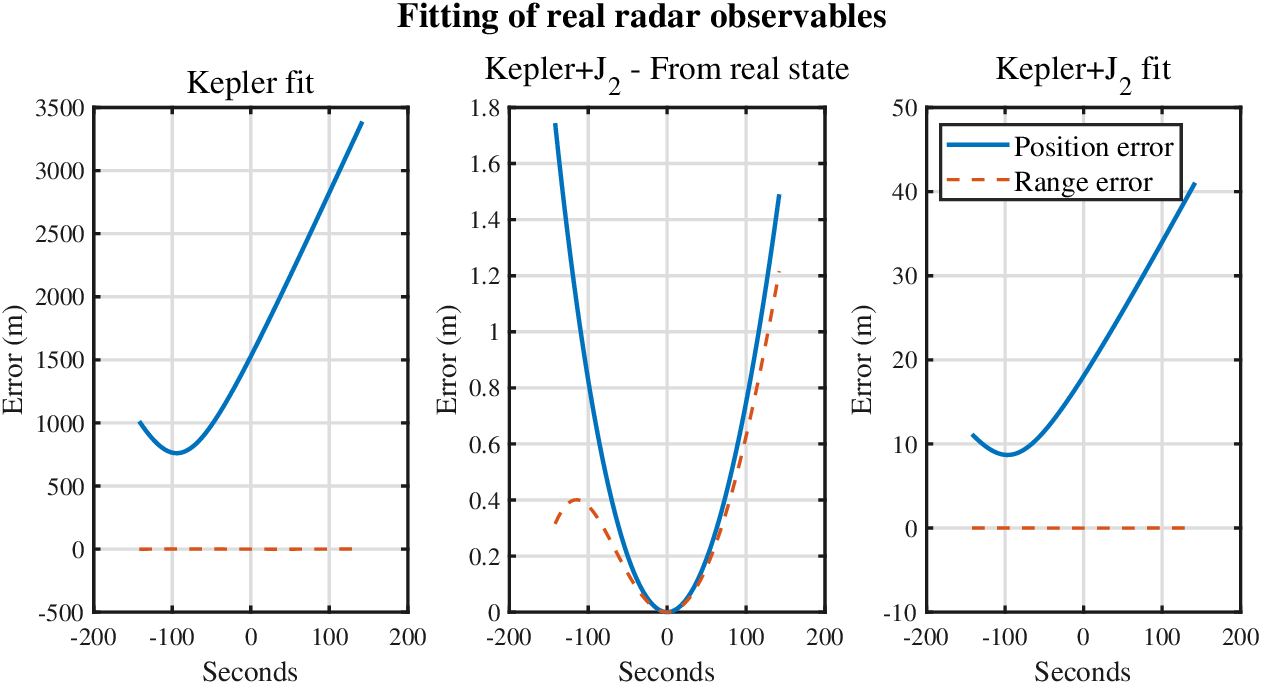}
    \caption{Position and range errors using \emph{real} radar measurements directly, including range-rate, in the same sample track as in Figure \ref{fig:kepler_fit}. The left plot shows the errors when the fitting is done using Kepler dynamics, see Section \ref{sec:kep_fit_radar_observables}. The central plot shows the errors when the analytical $\text{J}_2$ propagator is used from the true state at the epoch ($t_0$). The right plot is the result of applying the analytical fitting with $\text{J}_2$ developed in Section \ref{sec_j2_fit_observables}.}
    \label{fig:kj2_fit}
\end{figure}

This problem is made worse when the actual radar observables are fitted directly with a Kepler model\footnote{The derivatives of the predicted state are computed with the expansion of the Kepler functions up to the terms present in Equation (\ref{eq:fg_expansion}), and the convergence speed is comparable to the GTDS method.}, see Figure \ref{fig:kj2_fit} (Left), where now the epoch position error is $1.5$ Km. This is due to the range being now even more relevant to the weighting, but also because a strong range-rate constraint is added to the algorithm. Kepler dynamics deviate even more from the real trajectory when both residuals are so relevant. 

\begin{table}[ht]
    \caption{Constants used for the analytical $\text{J}_2$ propagator developed in Section \ref{sec:j2_expansion_theory}.}
   \label{tab:J2_constants}
        \centering 
   \begin{tabular}{c | c | c} 
      \hline 
      $\mu$ ($\text{Km}^3/\text{s}^2$) & $R_e$ (Km) & J$_2$  \\
      \hline 
      $398600.4418$  & $6378.137$ & $1.082626683553 \cdot 10^{-3}$   \\
      \hline
   \end{tabular}
\end{table}

\begin{figure}
    \centering
    \begin{subfigure}{0.474\textwidth} 
      \centering
      \includegraphics[width=\textwidth]{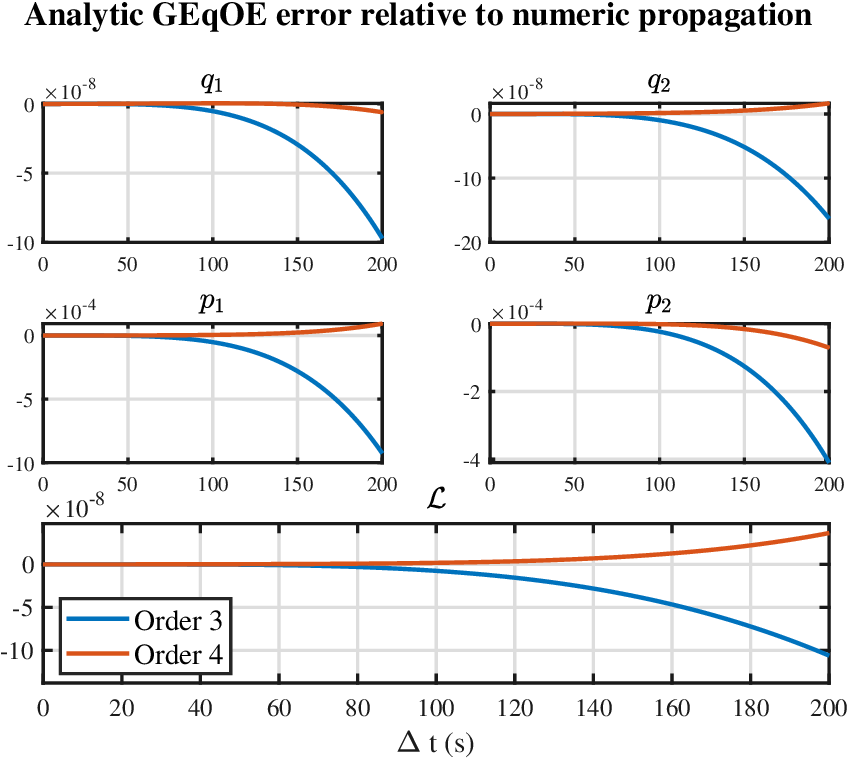}
      \caption{}
      \label{fig:j2_geqoe_rel_error}
    \end{subfigure}
    \begin{subfigure}{0.5190\textwidth}
      \centering
      \includegraphics[width=\textwidth]{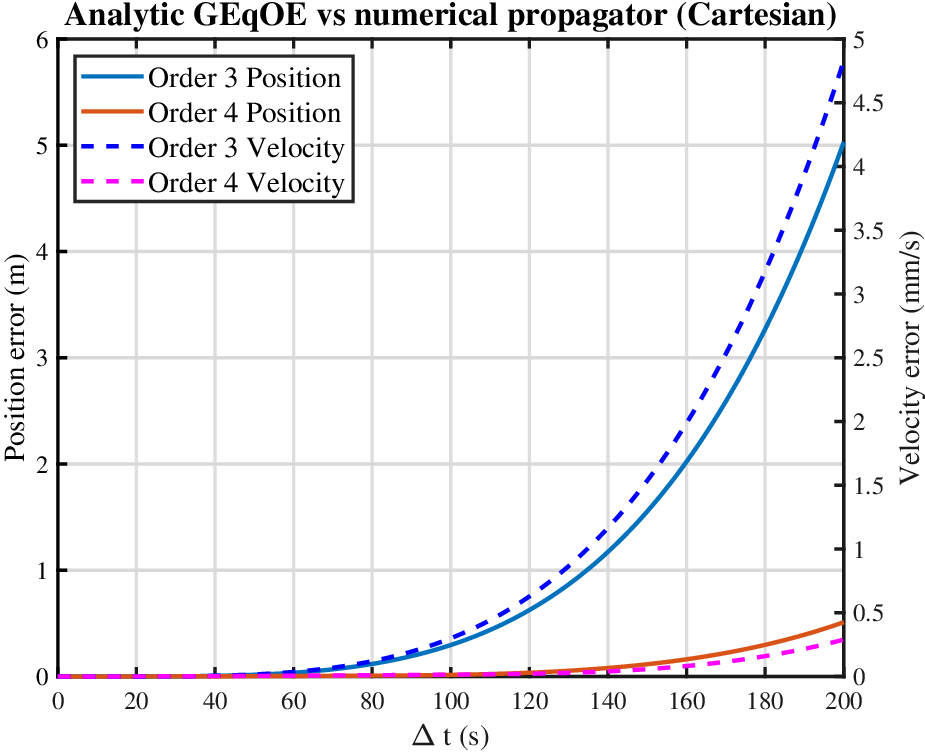}
      \caption{}
      \label{fig:j2_geqoe_pos_error}
    \end{subfigure}
    \caption{The relative error of the GEqOE Taylor expansion improves by an order of magnitude with each added term to the series for the first 200 seconds. The normalizing value is the output of the numerical propagation with a tolerance of $10^{-14}$ (\ref{fig:j2_geqoe_rel_error}). After 100 seconds of propagation, the Cartesian error (using Algorithm \ref{alg:geqoe2eci}) is on the order of $10^{-2}$ m in position and $10^{-2}$ mm/s in velocity for the fourth order approximation (\ref{fig:j2_geqoe_pos_error}).}
    \label{fig:J2_expansion_error}
\end{figure}

The approximated $\text{J}_2$ propagator in Section \ref{sec:j2_expansion_theory} is now tested up to the fourth term of the Taylor expansion (see Table \ref{tab:J2_constants} for the constants used). The absolute position errors compared to a numerical propagation are on the order of the centimeter for a $100$ seconds propagation, see Figure \ref{fig:J2_expansion_error}. Figure \ref{fig:kj2_fit} (Center) shows the errors when the true state is propagated from the epoch using the $\text{J}_2$ propagator, with much better results now that the range residuals are barely over $1$ meter at worst. As a consequence the radar observables fitting is much more similar to the original trajectory, see Figure \ref{fig:kj2_fit} (Right), which indicates this might be a superior method to the GTDS if only position errors are considered. 

\begin{figure}[h]
    \centering
    \includegraphics[width=0.75\textwidth]{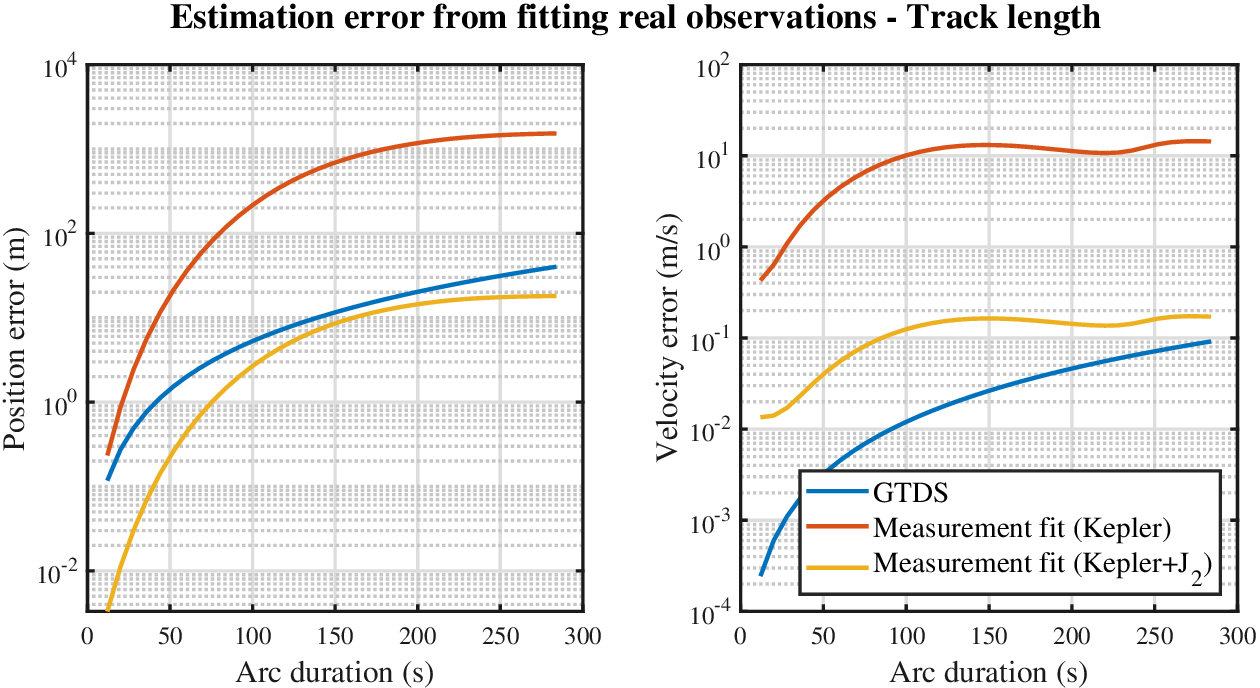}
    \caption{Fitting errors (at $t_0$) for a radar track of decreasing duration using the three main IOD algorithms developed in Section \ref{sec:estimation}. By leveraging the same radar track as in Figures \ref{fig:kepler_fit} and \ref{fig:kj2_fit}, and taking off two measurements at a time from each side, the effect of ever shorter tracks in the estimation accuracy is studied here. In all fittings the measurements are the real values (either position for GTDS or radar observables in the other cases). }
    \label{fig:fit_with_Arc_length}
\end{figure}

Before including noise into the discussion, Figure \ref{fig:fit_with_Arc_length} is presented for a complete comparison between the methods for different track lengths when measurements are noiseless. This is a representation of the error at the epoch (middle of the track) when the fitted trajectory gets shorter as measurements are extracted in pairs from the sides of the test track. The Keplerian fit of measurements is by far the worst of the three for the reasons already discussed, but very short tracks do not suffer so much in terms of position error. The observables fit with $\text{J}_2$ perturbation seems to be the best in this regard, but GTDS has a small advantage in velocity errors. This can only occur under the condition of flawless fitted position measurements. Since the GTDS method assigns equal weight to all residuals, errors stemming from the basic dynamics are effectively averaged out.

\subsubsection{Fitting of noisy measurements}

\begin{figure}[h]
    \centering
    \includegraphics[width=0.75\textwidth]{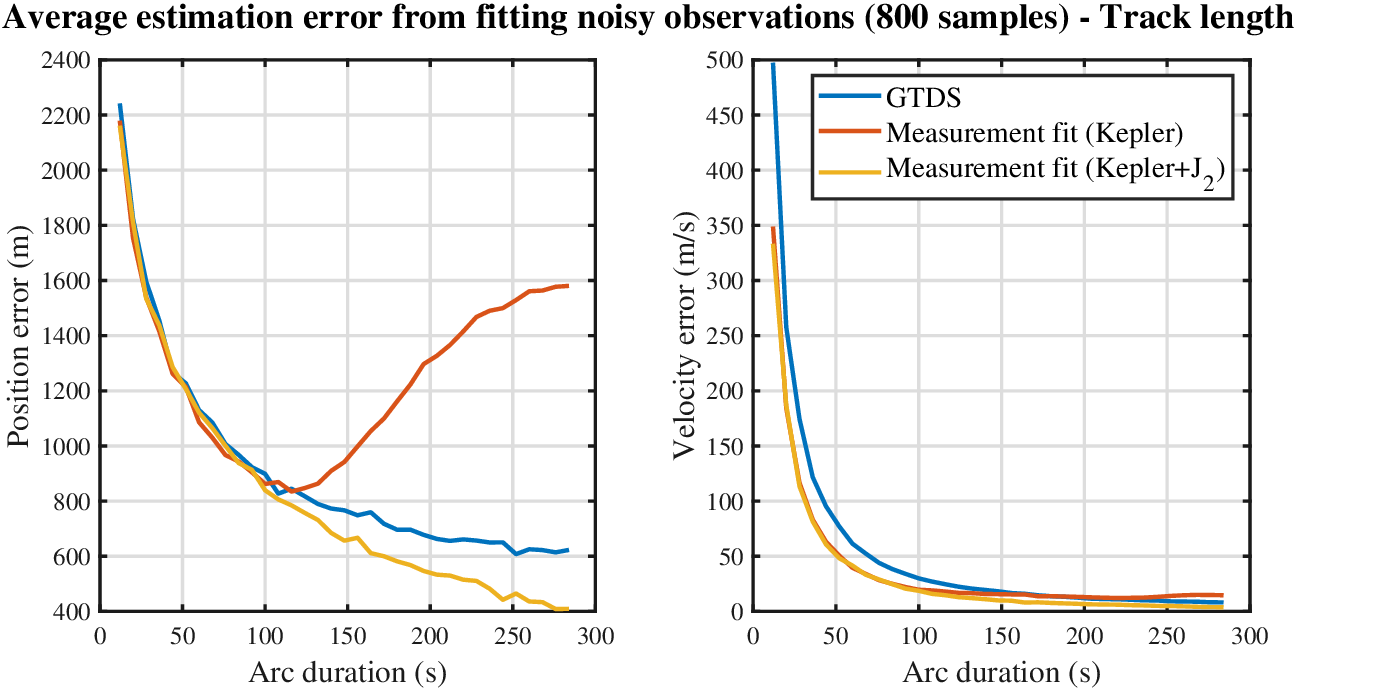}
    \caption{Fitting errors for a radar track of decreasing duration using the three main IOD algorithms developed in Section \ref{sec:estimation}, with the addition of noise and averaged over a number of samples. In all fittings the measurements (either position for GTDS or radar observables in the other cases) are noisy values sampled in measurement space .}
    \label{fig:fit_with_Arc_length_noise}
\end{figure}

Noise is now introduced, see Figure \ref{fig:fit_with_Arc_length_noise}, and results are averaged over a number of fittings to make them comparable. Despite shorter tracks being better represented with simplified dynamics, the uncertainty and measurement errors dominate the problem. In this instance there is not much difference between the methods for shorter than $100$ seconds tracks, at least in terms of average position error. There is a slight disadvantage for GTDS in terms of velocity errors throughout the first half of the interval, which can be attributed to the lack of range-rate information (very precise for this radar). As the track gets longer though, the Keplerian fitting of the measurements shows its fundamental flaw. It is interesting to see how adding information can deteriorate the estimation, and it is only due to the pairing of non-homogeneous data with an unsuitable model for the dynamics. The GTDS algorithm suffers also from the Kepler approximation, but much less so thanks to the lack of weighting, which enables the averaging of errors by default. The approximated $\text{J}_2$ method is the best in general, particularly for the longest radar tracks.

\begin{figure}[h]
    \centering
    \includegraphics[width=0.75\textwidth]{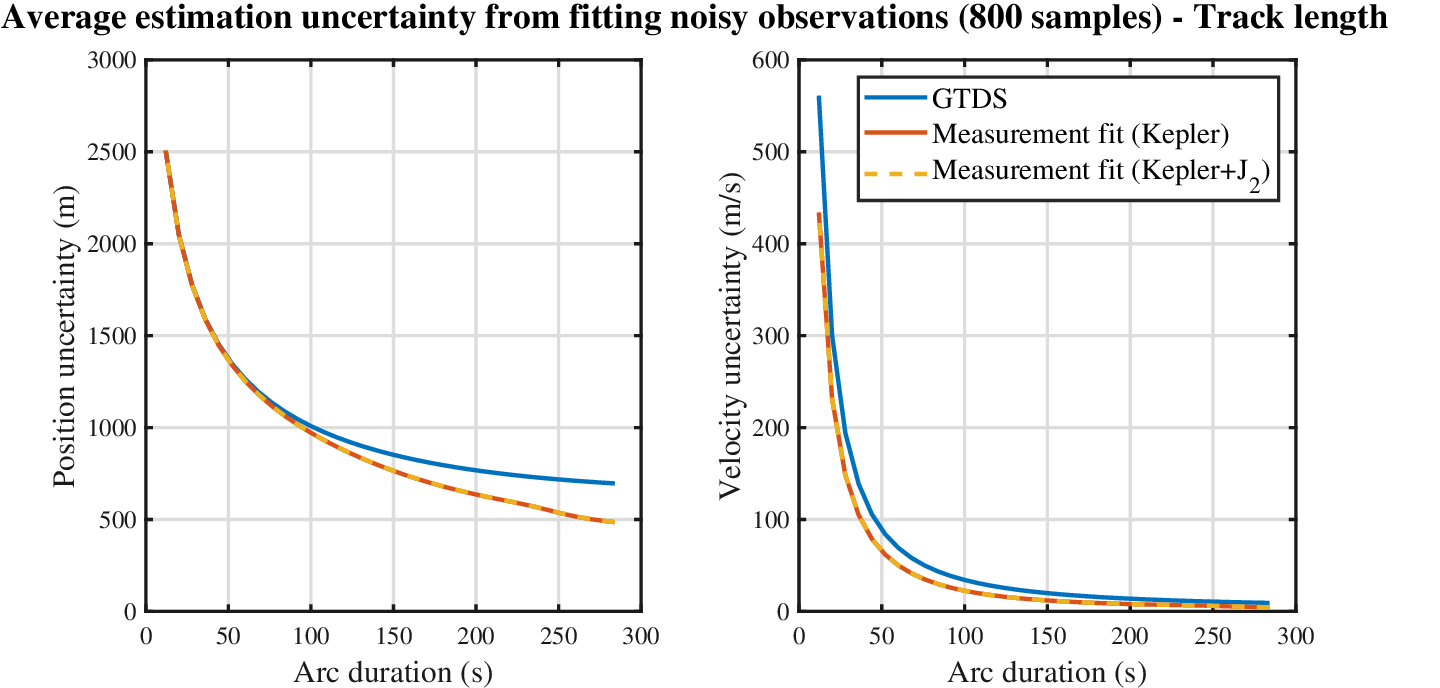}
    \caption{Estimation uncertainty using the three algorithms developed in Section \ref{sec:estimation}, with the addition of noise and averaged over a number of samples. The value of uncertainty displayed corresponds to the square root of the sum of eigenvalues for the position and velocity blocks of the uncertainty matrix $\text{C}_y$, so this is a conservative representation that does not consider correlations.}
    \label{fig:fit_uncertainty_with_Arc_length_noise}
\end{figure}

These results are incomplete without taking a look at the evolution of the uncertainty for different track lengths, see Figure \ref{fig:fit_uncertainty_with_Arc_length_noise}. Due in part to the absence of range-rate data, the GTDS is more conservative in the uncertainty aspect of the estimation, but it is also a consequence of the way uncertainty of measurements is used (indirectly), see Section \ref{sec_gtds_fitting}. The information shown in Figure \ref{fig:fit_uncertainty_with_Arc_length_noise} does not consider correlations, so this is an incomplete picture of the covariance. In any case, the evolution is compatible with the errors, so this estimation method should behave correctly. The Keplerian fit of radar observables on the other hand suffers from overconfidence very soon for longer than $100$ seconds tracks, as the uncertainty keeps getting lower with more measurements added, but the error does not evolve accordingly, see Figure \ref{fig:fit_with_Arc_length_noise}. A more comprehensive exploration of the uncertainty realism is conducted in Section \ref{sec:stats_IOD_radar_only} using a broader dataset.

\subsection{Preliminary testing of OPOD}\label{sec:prel_testing_plane_astror}

The IOD method that includes information of inclination ($i$) and right ascension of the ascending node ($\Omega$), see Section \ref{sec:orb_plane_fitting}, is tested here. First a check is performed on the degree of predictability of the orbital plane and the expected uncertainty for two different LEO. The presence of maneuvers is also considered. Then the fitting methodology is tested when this information is included with different levels of virtual measurement uncertainty on the same test track used in Section \ref{sec:prel_testing_non_plane_alg} (and for different track lengths).

\subsubsection{Orbital plane predictability check}\label{sec:orb_plane_pred_check}

\begin{figure}
    \centering
    \begin{subfigure}{0.45\textwidth} 
      \centering
      \includegraphics[width=\textwidth]{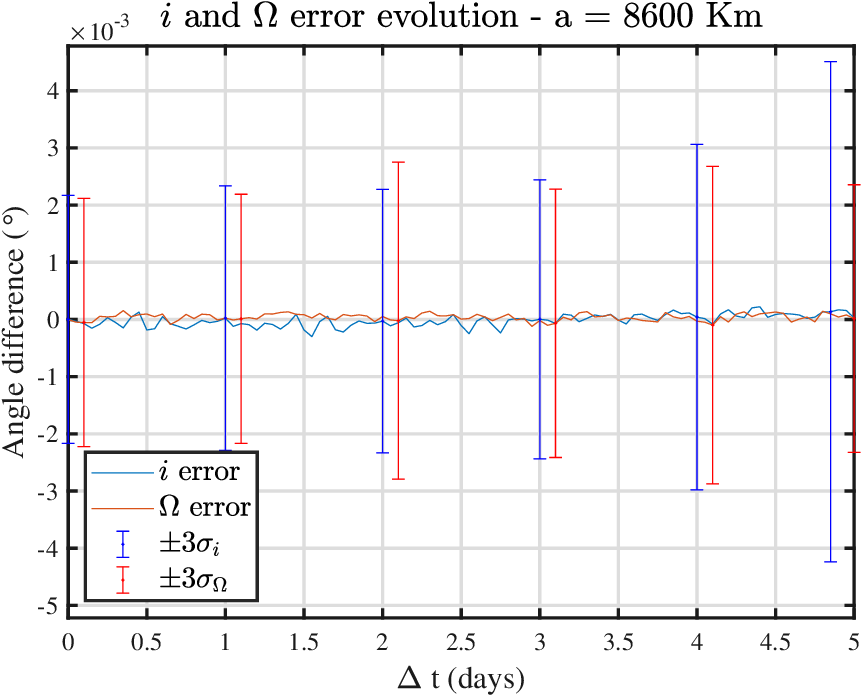}
      \caption{}
      \label{fig:orb_plane_error_8600}
    \end{subfigure}
    \begin{subfigure}{0.45\textwidth}
      \centering
      \includegraphics[width=\textwidth]{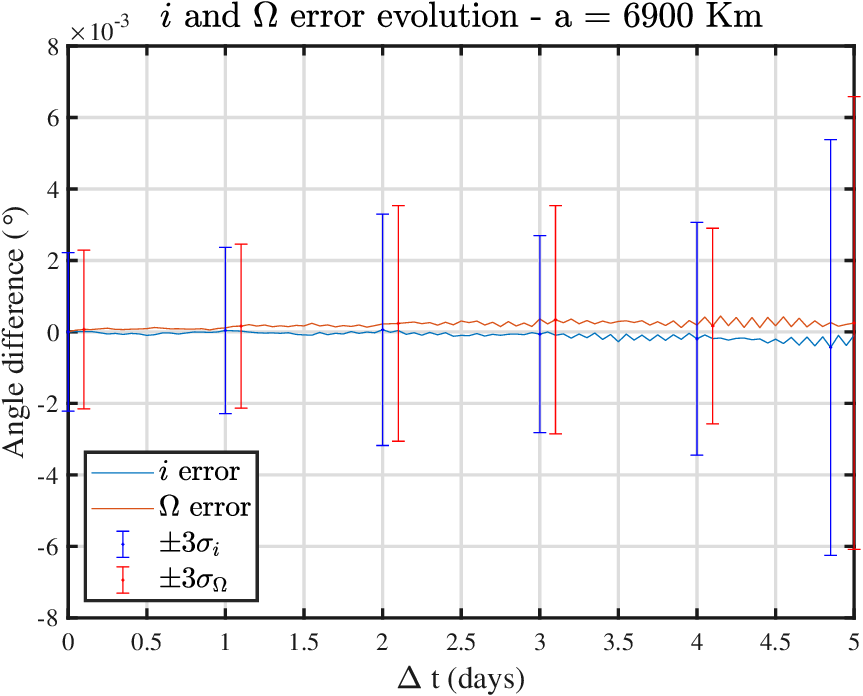}
      \caption{}
      \label{fig:orb_plane_error_6900}
    \end{subfigure}
    \caption{The prediction error against a high fidelity propagator is low for both inclination and RAAN. The model only includes earth harmonics of order and degree 4 and luni-solar perturbation (no drag). These perturbations suffice to accurately predict the orbital plane for an orbit with $a = 8600$ Km (\ref{fig:orb_plane_error_8600}). For the example of $a=6900$ Km a better model is required due to the atmosphere and earth shape being more relevant. Harmonics are of order and degree 20, and a simple Harris-Priester atmosphere model is used. This is enough to get good predictions of the orbital plane for the first 5 days, although the position error goes to $33$ Km (\ref{fig:orb_plane_error_6900}).}
    \label{fig:orb_plane_error}
\end{figure}

The orbital plane predictability is tested first with a circular LEO of $a=8600$ Km semi-major axis and $i=60^{\circ}$. A high fidelity propagation is compared to a very simple one in Figure \ref{fig:orb_plane_error_8600}, and the prediction error of $i$ and $\Omega$ is shown. The high fidelity propagator uses the Real dynamics (see Table \ref{tab:dynmodels}) with the Sentinel-1A HF model. The exact same initial conditions are also propagated with very simple dynamics (and the LF model of Sentinel-1A). No atmosphere is included in this case, and only the luni-solar perturbation and harmonics of degree and order 4 are used . In order to consider state uncertainty a Monte-Carlo approach has been used for simplicity here, sampling position and velocity with standard deviations on the order of $\sigma_r \approx 30 $ m and $\sigma_v \approx 0.2$ m/s. The result in this first test is that the orbital plane is actually very predictable, despite the low computational effort required to run the prediction model. The angular errors for both $i$ and $\Omega$ are on the order of $10^{-4}$ degrees at most, and the $\sigma$ values go up to around $10^{-3}$. This is in contrast to the position error, not included in the graph, which evolves up to $13$ Km after 5 days. This example is of low difficulty because the atmosphere plays almost no role at those altitudes. The second example, see Figure \ref{fig:orb_plane_error_6900}, uses the same initial conditions except for the semi-major axis, now at $6900$ Km, and eccentricity, which is $e=10^{-4}$. The closeness to the earth requires increased fidelity of its shape, so the harmonics used now are of degree and order 20. The atmosphere has to be included, but it is an static model (Modified Harris-Priester included in Orekit, see \cite{hatten2017smooth}), and the drag is modeled with a constant area and coefficient. This model is enough to get similar values of errors for the orbital plane, but now the predicted position has an error of $33$ Km.

\begin{figure}
    \centering
    \includegraphics[width=0.65\textwidth]{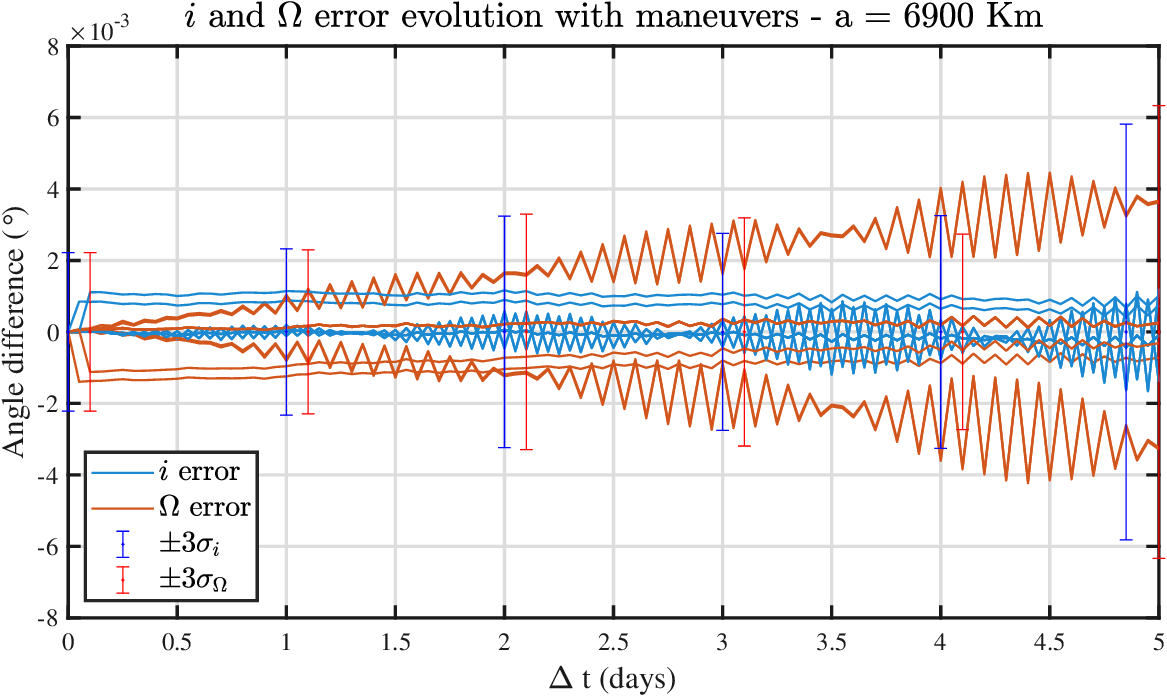}
    \caption{For 10 different maneuvered cases the predicted errors of $i$ and $\Omega$ can become greater than in the non-maneuvered case, see Figure \ref{fig:orb_plane_error_6900}. Despite that, in the worst cases the order of magnitude of these errors does not evolve much over $5\cdot10^{-3}$ degrees after 5 days.}
    \label{fig:orb_plane_error_maneuver}
\end{figure}

These findings indicate that in the absence of maneuvers, even a low-fidelity propagator possesses the capability to accurately predict the orbital plane. The next step is to ascertain the possible effects of maneuvers in general over this part of the prediction. This has been done for a series of maneuvered cases, where variations in the maneuver instant and direction have been considered, for a total of 5 directions and 2 different instants after $t_0$. The only impulsive value used is $0.2$ m/s, which from previous works \citep{montilla2023conference} is considered a rather high value (easy to detect in general). The evolution of errors is shown in Figure \ref{fig:orb_plane_error_maneuver} for all cases. One of the directions considered is the \emph{out-of-plane} maneuver, and those cases, for two different instants, are noticeable as an immediate change in inclination and RAAN. The divergent $\Omega$ cases correspond to the \emph{prograde} and \emph{regrograde} maneuvers, which are also the most efficient ones at changing the semi-major axis of the orbit, and thus have a greater effect over the $\text{J}_2$ secular rate on $\Omega$. These are good examples of what is the expected magnitude of the errors in orbital plane prediction, even when there have been maneuvers.

With the previous discussion in mind, the proposed method in this section is to use both the predicted $i$ and $\Omega$ at the epoch of the estimation as if they were extra measurements. Considering the possible effects of a maneuver in this prediction, it is important to use reasonable values for the uncertainties of these virtual measurements. Failing to do so would impose a constraint over the state incompatible with the measurements of a possibly maneuvered spacecraft, and induce error on the estimation. These values ought to be on the order of $5\cdot10^{-3}$ degrees, but the validity and effect of the chosen value over the estimation is discussed next.

\subsubsection{Effect of the $i$ and $\Omega$ virtual measurements on the estimation error}


This section tests the OPOD method and the effect of the chosen orbital plane uncertainty on the estimation error. Given the initial conditions S1A-1 (see Table \ref{tab:sto}), the HF model of Sentinel-1A is simulated with 4 different maneuvered cases. A very long track from Radar 3 that occurs $3.6$ days after $t_0$ is sampled (for each case) and used to derive average results when the track duration is modified. The predicted orbital plane is calculated using the LF model, and the obtained $i$ and $\Omega$ values are the same for all the tests (the same initial conditions are always used). The test is exclusively with the $\text{J}_2$ fitting method.

\begin{figure}[h]
    \centering
    \includegraphics[width=0.85\textwidth]{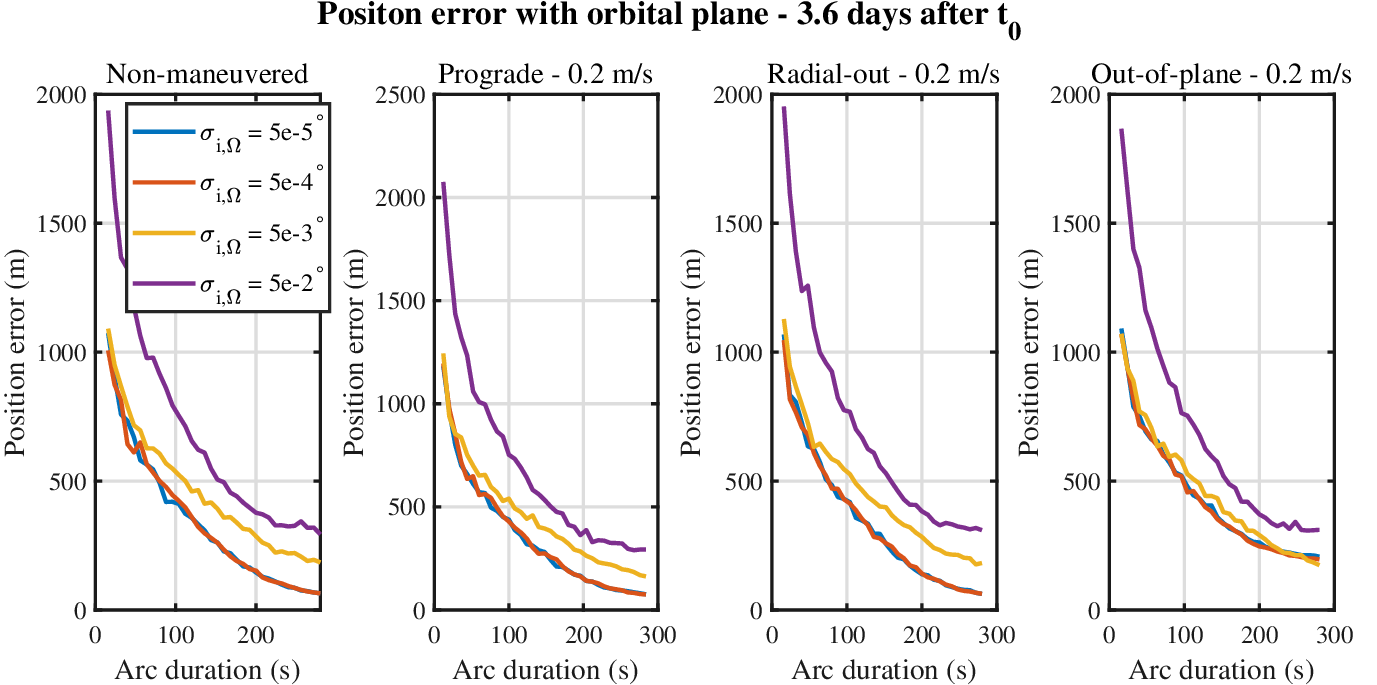}
    \caption{Error evolution when the information of the orbital plane is added using the $\text{J}_2$ analytical estimation method. This is done for a single radar track that happens $3.6$ days after $t_0$, and sampling measurements 500 times. Different maneuvered cases are compared. The effect of the uncertainty in the $i$ and $\Omega$ information is the focus of this test.}
    \label{fig:error_orbital_plane}
\end{figure}

Figure \ref{fig:error_orbital_plane} is a summary of the improvements that are to be expected for different magnitudes of uncertainty considered. The orbit used here is higher than in the test case of Figure \ref{fig:orb_plane_error_maneuver}, on top of that the prediction model is of much higher fidelity for this experiment (see Table \ref{tab:dynmodels}). Accounting for both factors prediction errors ought to be of lower magnitude (the secular effect of $\text{J}_2$ is less dominant, about $10\%$ weaker). 


The non-maneuvered case is the reference, as the errors here behave in a somewhat expected manner when the predicted $i$ and $\Omega$ are very precise. Adding this information always improves the estimation, but when the uncertainty is big, as in $5 \cdot 10^{-2}$ degrees for both angles, then the effect is considerably reduced. It can be compared to Figure \ref{fig:fit_with_Arc_length_noise} (Left) to notice the slight reduction of estimation error at all lengths, but more noticeable in shorter tracks. Then, as the uncertainty is reduced one order of magnitude, there is a great reduction of error, particularly for very short tracks (1 Km reduction in position error). Decreasing it even more only affects longer tracks, but the improvement is smaller and even non-existent when the uncertainty gets unreasonably low.

Maneuvers can have an effect on the accuracy of the estimation now, and it can be negative. Notice how for the \emph{Out-of-plane} maneuver, which has an immediate effect on the orbital plane, the overconfidence in the extra information increases the prediction error for longer tracks. This is in contrast to the uncertainty of the estimation (not showed here), that is in fact decreasing when the $\sigma_{i,\Omega}$ is reduced, so that the estimation becomes overconfident in this case. This alone shows how using values of uncertainty smaller than $\sigma_{i,\Omega}=5 \cdot 10^{-3}$ degrees is not recommended for the magnitudes of maneuvers considered in the context of this work (a different value should be considered otherwise). Another example of this is the \emph{Prograde} maneuver, which also sees a small increase in error when the orbital plane information is assumed to be too precise. The effect of this maneuver is primarily in $\Omega$ because of the $\text{J}_2$ perturbation, but not enough on this orbit to be more noticeable in the induced error. Finally, the \emph{Radial-out} case is not affected enough to see an increase in error. 

Thus, with maneuvers up to $0.2$ m/s, it is safe to assume the predicted orbital plane error has a standard deviation of $\sigma_{i,\Omega}=5 \cdot 10^{-3}$ degrees for the purpose of estimating the state. The effect on the estimation statistics is studied for this augmented method separately from the others, see Section \ref{sec:results_orbital_plane}.


\subsection{Synthetic scenarios}\label{sec:scenarios_k2_results}

The goal of the testing presented here is to establish realistic expectations for the estimation algorithms with varying track lengths and data density. In each scenario, the satellite\textquotesingle s true state at $t_0$ sets the real orbit, which remains invariant. As mentioned before, testing the OPOD method requires to compute predicted orbital plane as well. In order to introduce variability on the prediction accuracy, a non-maneuvered case and various combinations of impulse value ($N_{imp}$), maneuver direction ($N_{dir}$), and maneuver instant after $t_0$ ($N_{t}$) are part of the scenario configuration. Each of these $N_{imp}N_{dir}N_{t}+1$ cases (which spawn a real trajectory) generate a battery of tracks ($N_{int}$) in a particular radar station within a given integration window. All the generated tracks are considered individually, and are then sampled $N_{ms}$ times (measurement noise) to generate statistics of the estimate distribution (which should be represented by $\text{C}_y$). This is done by computing the Mahalanobis Distance squared of the estimation error for all samples, denoted as $k^2$, and comparing its distribution to the theoretical one.

If the fitting methodology does not require information outside of the radar observables, all the scenario variability only serves to generate a greater amount of slightly different tracks. Notice how scenarios STARL-1 and STARL-2 do not consider maneuvers, as these scenarios have been included to increase the track length variety in the testing of the 3 main IOD algorithms. Thus, these two scenario do not require LF satellite model or uncertainty information over the initial state. The uncertainty aspect is relevant when doing the fitting that includes the predicted $i$ and $\Omega$, which in turns comes from a sampled initial state around the real one. A total of $N_{sp}$ different initial samples are propagated individually. One of the $N_{sp}$ predictions is randomly chosen for the computation of the estimation.

In particular, 5 different scenarios are used for the generation of statistics on the methods under study. Table \ref{tab:scenarios} contains all defining aspects of these scenarios. The 2 maneuver instants considered are $0.1$ and $0.5$ days after $t_0$. The uncertainty (\emph{Low}) is locally defined and has values of $5$, $30.4$ and $5$ meters in position, and $0.002$, $0.01$ and $0.006$ m/s in velocity (diagonal covariance in the LVLH frame). Finally, the variation in integration window $N_{int}$ comes from the established maximum of $4$ days ($10$ for STARL-1 and 2), and from the combination of initial conditions and the radar station. During the maximum window, $N_{int}$ different radar tracks, are found and considered independently of each other. This is, non of them is used in conjunction with the others to increase the information available, but only by themselves. 




\begin{table}[ht]
    \caption{The six impulse directions used in all scenarios are defined in the LVLH frame (or QSW in OREKIT), so $+x$ is the direction of the position vector and $+z$ is in the orbit angular momentum. From 1 to 6 these are (approximately) prograde, retrograde, radial-in, radial-out, in-plane ($45^{\circ}$ from $-x$ and $+y$ in LVLH) and out-of-plane (at $+z$).}
   \label{tab:dir}
    \centering 
   \begin{tabular}{c | c | c | c } 
      \hline 
      Direction & $d_x$ & $d_y$ & $d_z$  \\
      \hline 
      1     & 0 & 0.9397     & 0.3420  \\
      2     & 0 & -0.9397     & 0.3420  \\
      3     & -0.9397 & 0     & 0.3420  \\
      4     & 0.9397 & 0     & 0.3420  \\
      5     & -0.7071 & 0.7071    & 0  \\
      6     & 0 & 0   & 1  \\
      \hline
   \end{tabular}
\end{table}

\begin{table}[ht]
	\fontsize{8}{8}\selectfont
    \caption{Scenarios used for the synthetic data generation and metric testing. All of them use the same $6$ different maneuver directions, in Table \ref{tab:dir}. The 5 impulse values used are $20$, $10$, $8$, $5$, $2$ (cm/s), and the number of individual radar observations considered comes from the maximum integration window, 4 days, and the combination of I.C and radar station.}
   \label{tab:scenarios}
        \centering 
   \begin{tabular}{c | c | c | c | c | c | c | c | c | c | c | c} 
      \hline 
      SC Name & HF Satellite & LF Satellite & Radar Station & I.C. & I.Uncertainty & $N_{sp}$ & $N_{imp}$ & $N_{dir}$ & $N_{t}$ & $N_{int}$ & $N_{ms}$ \\
      \hline 
      SEN-1A-1  &  Sentinel-1A  & Sentinel-1A$_{1}$ & Radar 1 & S1A-1 & Low     & 50 & 5 & 6 & 2 & 4    & 600 \\
      SEN-1A-2   & Sentinel-1A  & Sentinel-1A$_{1}$ & Radar 2 & S1A-1 & Low     & 35 & 5 & 6 & 2 & 15   & 600 \\
      SW-C-1   & Swarm-C        & Swarm-C$_{1}$     & Radar 1 & SWC-1 & Low    & 50 & 5 & 6 & 2 & 4    & 600 \\
      STARL-1   & Starlink-1        & -  & Radar 1 & SRL-1 & -    & - & 0 & - & - & 13    & 600 \\
      STARL-2   & Starlink-2        & -  & Radar 1 & SRL-2 & -    & - & 0 & - & - & 16    & 600 \\
      \hline
   \end{tabular}
\end{table}

\subsection{$k^2$ metric statistics} \label{sec:k2_metric_def}

This section presents an in depth commentary on the performance of the estimation methods. A battery of tests are performed to check on the accuracy and uncertainty realism of the methods for the different radar tracks of the established scenarios. Before this, a single radar pass was studied, see Sections \ref{sec:prel_testing_non_plane_alg} and \ref{sec:prel_testing_plane_astror}, and with no consideration for the correlation information. The current approach involves utilizing all radar tracks generated from the scenarios listed in Table \ref{tab:scenarios}. Following the methodology outlined in \cite{reihs2021application}, an isolated $k^2$ metric (independent of predictions) is employed to assess the accurate characterization of estimation uncertainty. This metric is $k^2=\bm d^T \text{C}_y \bm d$, with $\bm d$ being the vector of the difference between the fitted state and the ground truth.


When the fitting model is adequate and the relation between the estimated state and the measurements is close to linear the statistics of $k^2$ for a set of samples in a particular track should be that of the chi-square distribution of the corresponding degrees of freedom $p$. The consequence is that the mean value of those $k^2$ should meet $\mu=p$, the variance should be $\sigma^2=2 p$, and $10\%$ of the distribution should be over $\chi_{inv}^2(0.9,p)$. With $\chi_{inv}^2(\eta,p)$ the chi-square inverse cumulative distribution function of $p$ degrees of freedom evaluated at the percentage $\eta$ of the distribution. 

\subsubsection{$k^2$ statistics on the IOD methods} \label{sec:stats_IOD_radar_only}

\begin{figure}[h]
    \centering
    \includegraphics[width=0.9\textwidth]{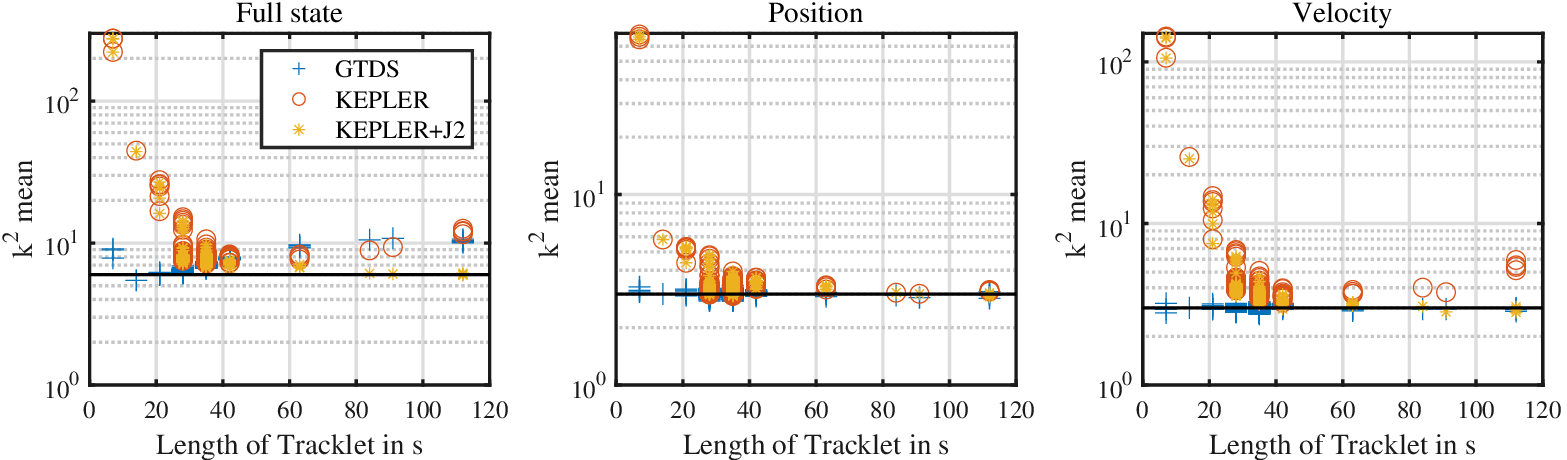}
    \caption{Estimation only $k^2$ mean values for Radar 1 ($r_t=7$ seconds). Each radar track in the scenarios SEN-1A-1, SW-C-1, STARL-1 and STARL-2 is sampled 600 times and the $k^2$ statistics of the corresponding method are computed.}
    \label{fig:k2_stats_mean_s3t}
\end{figure}

\begin{figure}[h]
    \centering
    \includegraphics[width=0.9\textwidth]{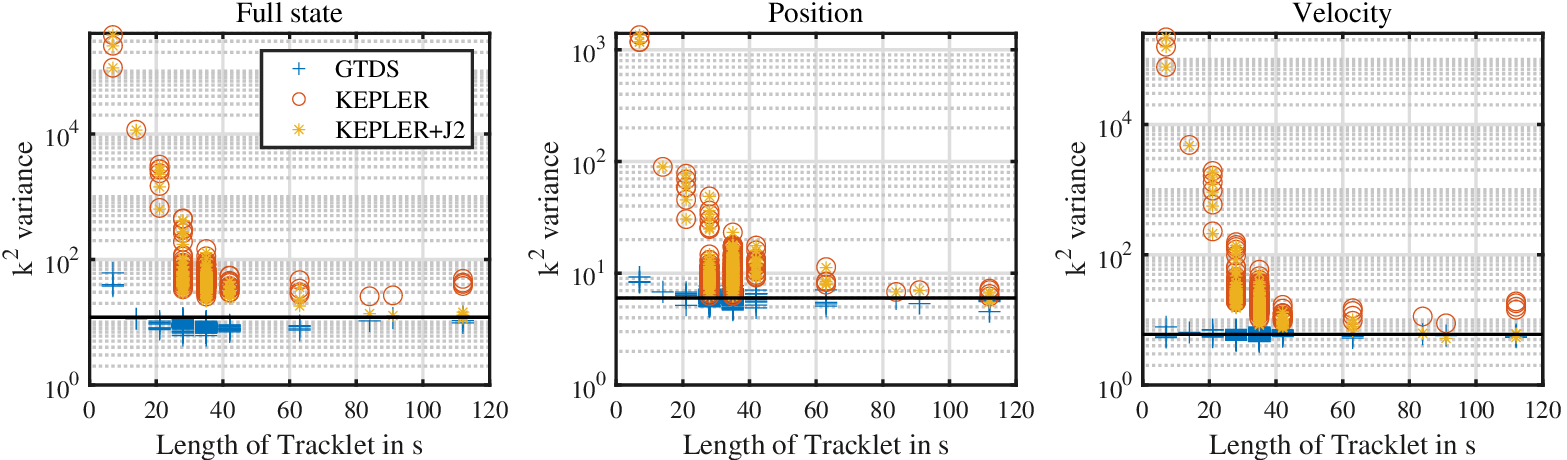}
    \caption{Estimation only $k^2$ variance values for Radar 1 ($r_t=7$ seconds). Each radar track in the scenarios SEN-1A-1, SW-C-1, STARL-1 and STARL-2 is sampled 600 times and the $k^2$ statistics of the corresponding method are computed.}
    \label{fig:k2_stats_var_s3t}
\end{figure}

\begin{figure}[h]
    \centering
    \includegraphics[width=0.9\textwidth]{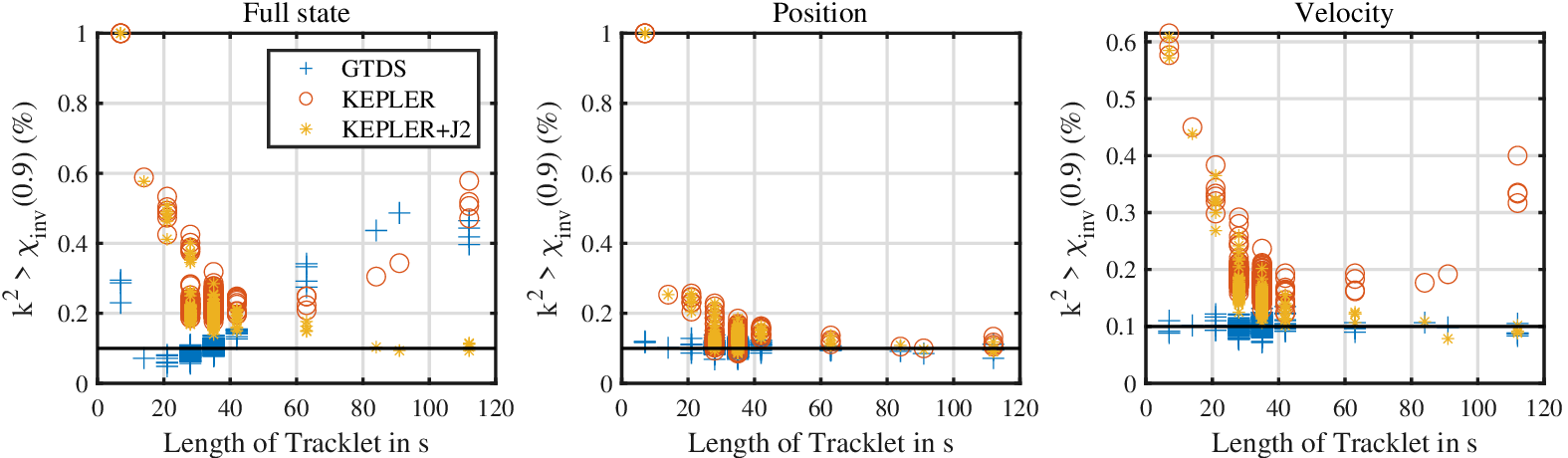}
    \caption{Estimation only $k^2$ percentage of anomalous estimations for Radar 1 ($r_t=7$ seconds). Each radar track in the scenarios SEN-1A-1, SW-C-1, STARL-1 and STARL-2 is sampled 600 times and the $k^2$ statistics of the corresponding method are computed.}
    \label{fig:k2_stats_fn_s3t}
\end{figure}

Considering only the subset of simulations for Radar 1 in Table \ref{tab:scenarios}, the results of \cref{fig:k2_stats_mean_s3t,fig:k2_stats_var_s3t,fig:k2_stats_fn_s3t} compile these statistics, where it has been divided between the full state and the position and velocity parts (which are 3 degrees of freedom distributions in the ideal case). The horizontal black line indicates the expected value. Differentiating between the results for Radar 1 and 2 is a consequence of the different rates of measurement, which affect the density of plots for a given track length and thus the behavior of the fitting.

\begin{figure}[h]
    \centering
    \includegraphics[width=0.85\textwidth]{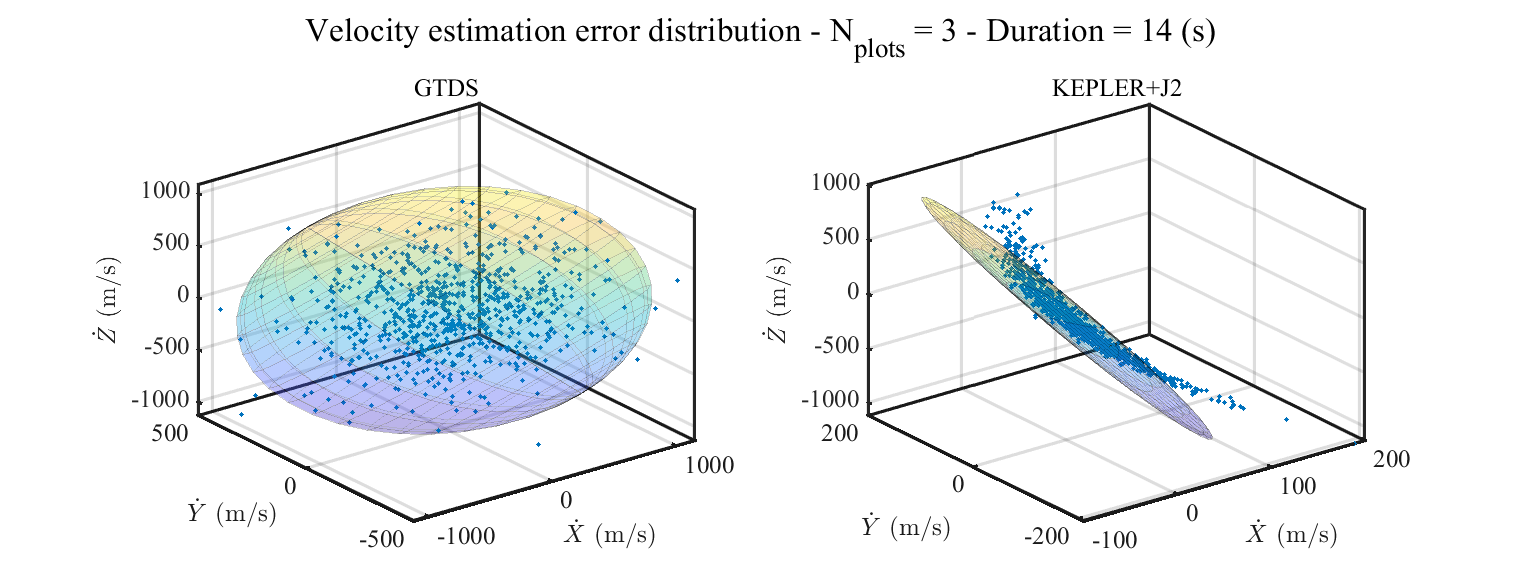}
    \caption{Distribution of the estimated velocity errors (against the real state) for a very short track. Both measurement fits (KEP and KEP+$\text{J}_2$) show the same behavior when the fitted track is too short (2 to 4 measurements in Radar 1), where the real distribution of the velocity estimation differs from the analytical covariance of the linear least-squares method. This makes the estimated covariance not very realistic.}
    \label{fig:error_VEL_short}
\end{figure}

The initial and perhaps most surprising finding is that both measurement fittings perform poorly in very short tracks (less than 21 seconds or 4 measurements for this radar station). This is due to a combination of factors that, in order to better understand them, require a closer inspection to the errors distribution. Figure \ref{fig:error_VEL_short} shows the velocity errors for a very short track that has been fitted with GTDS and KEP+$\text{J}_2$ methods (sampling the track 600 times). The distribution is not that of the analytical covariance in the case of the measurement fit. The reason comes down to the linear approximation that is done for the iterative least-squares algorithm, which happens to be a bad approximation in this case. The position part seems to have the same problem, but looking to the corresponding error distribution (not shown here) it is apparent that the problem is of a different kind. In this case there is a slight bias in the estimation that happens to be perpendicular to the direction of smaller variance. This penalizes the $k^2$ considerably. The interplay of both phenomena leads to an unfavorable behavior in the full state metric, exhibiting abnormally high percentages of values exceeding $\chi_{inv}^2(0.9,p)$. This signals a potential reliability issue for a maneuver detection metric that operates on any of the radar observables fittings, particularly in the context of very short tracks. The GTDS method on the other hand behaves as expected in this range of track lengths.


For longer than 5 measurements (in Radar 1) the situation improves considerably in both KEP and KEP+$\text{J}_2$, the non-linearity of the velocity estimation ceases and the position bias also goes down. In the case of the $\text{J}_2$ fitting the consequence is a full state estimation that behaves in the way the analytical covariance is indicating, even when the track length goes to the higher values of the scenarios (these are for Starlink-2, which has the lower inclination and gets the lengthier tracks). The KEP method starts to degrade in the velocity estimation part, due to a bias, from 50 seconds and upwards. Position does not seem to be affected in these scenarios.

Despite not showing any visible degradation in the position and velocity estimations separately, the GTDS method has increased rates of $k^2$ values with increasing track lengths, see Figure \ref{fig:k2_stats_fn_s3t}. This can only be due to unrealistic correlations of the full covariance, which makes the use of the full state estimation not recommendable, at least for tracks longer than 50 seconds and up to 120 seconds.

From these example scenarios (with a measurement rate of 7 seconds) it seems that short track state estimation is better done with an unweighted position fit (GTDS) but longer tracks can benefit from doing a direct fitting of the observables that includes the range-rate. This last statement is done under the premise of using the estimations as they are obtained from the methods explained in Section \ref{sec:estimation}. That said, notice how the velocity errors in the X and Y directions are of much lower magnitude in the KEP+$\text{J}_2$ fitting of Figure \ref{fig:error_VEL_short} compared to the GTDS. If the covariance of the estimation is modified to account for the non-linear behavior of the fit in short tracks, it might be a better alternative still, but this is out of the scope of this work. The same could be done with the position covariance, by inflating it slightly in the direction of the bias, to obtain more precise estimations.

\begin{figure}[h]
    \centering
    \includegraphics[width=0.9\textwidth]{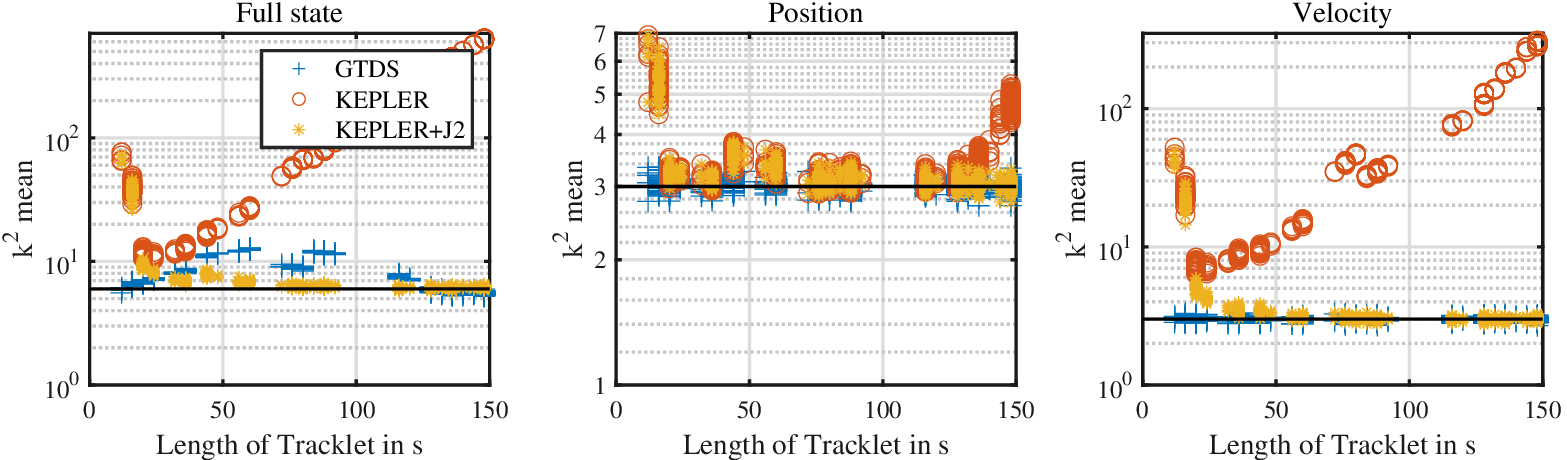}
    \caption{Estimation only $k^2$ mean values for Radar 2 ($r_t=4$ seconds).}
    \label{fig:k2_stats_mean_pol}
\end{figure}

\begin{figure}[h]
    \centering
    \includegraphics[width=0.9\textwidth]{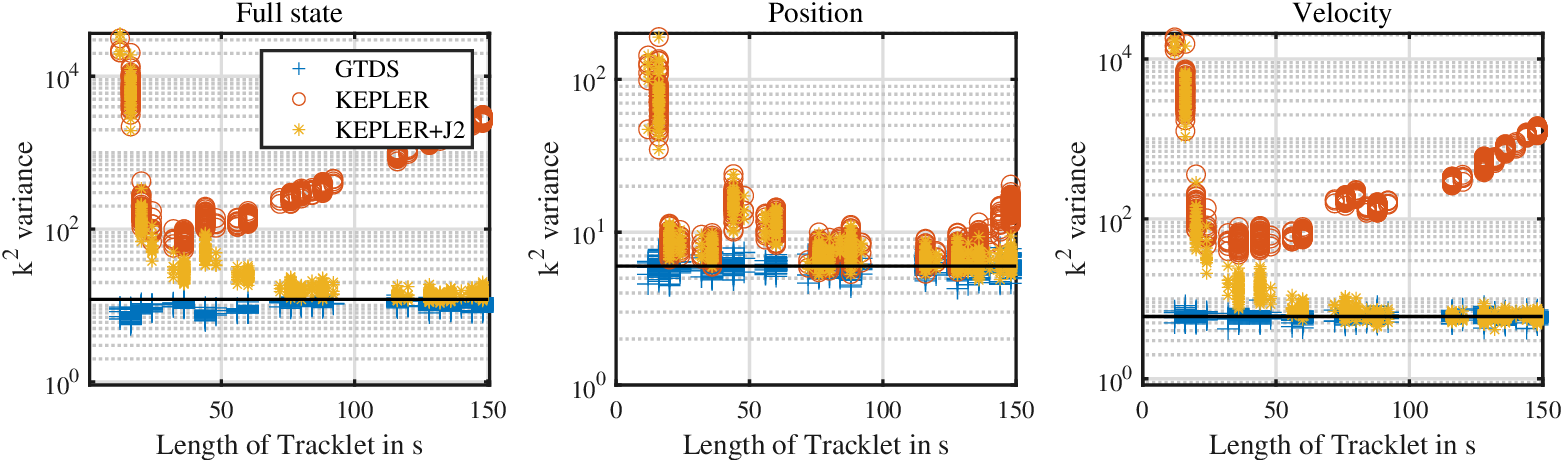}
    \caption{Estimation only $k^2$ variance values for Radar 2 ($r_t=4$ seconds).}
    \label{fig:k2_stats_var_pol}
\end{figure}

\begin{figure}[h]
    \centering
    \includegraphics[width=0.9\textwidth]{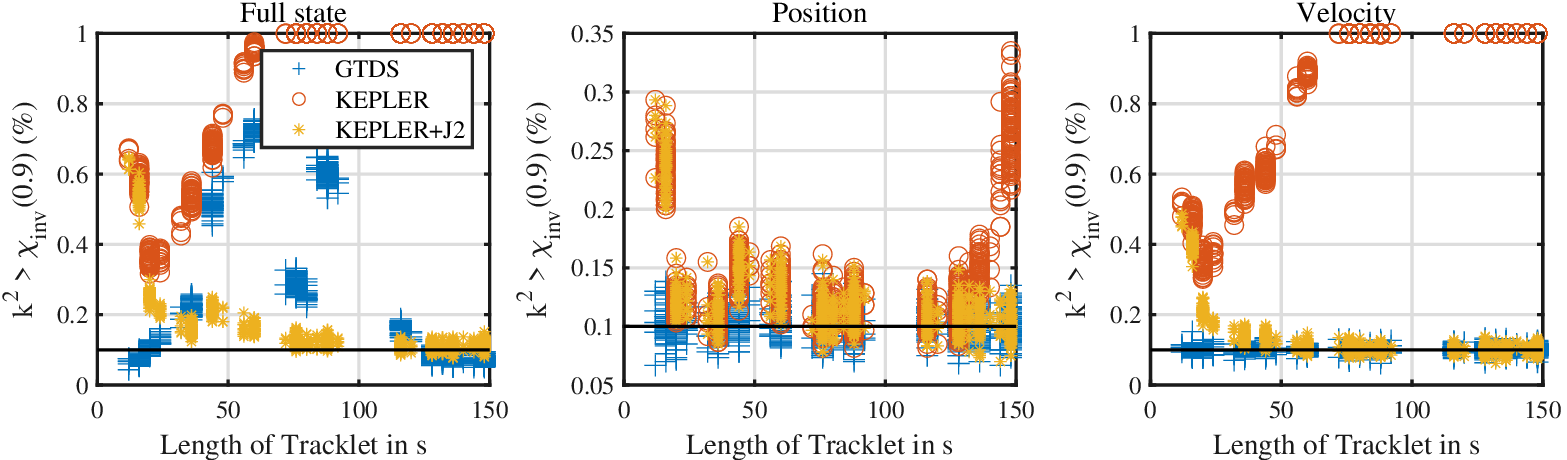}
    \caption{Estimation only $k^2$ percentage of anomalous estimations for Radar 2 ($r_t=4$ seconds).}
    \label{fig:k2_stats_fn_pol}
\end{figure}

The results for Radar 2 (measurement rate of 4 seconds) are slightly modified, see \cref{fig:k2_stats_mean_pol,fig:k2_stats_var_pol,fig:k2_stats_fn_pol}. The scenario considered here (SEN-1A-2) includes longer tracks than before, so it shows an extended range of the metric performance as well. The main difference is apparent with the KEP method, which degrades sooner in the velocity aspect, and starts to deviate from the ideal distribution also in position for longer than 120 tracks (same behavior as was shown in Figure \ref{fig:fit_with_Arc_length_noise} for an example track of Radar 3). The combination of all these problems makes the KEP method almost unusable for the whole range of track lengths, at least if the full state metric is considered, see Figure \ref{fig:k2_stats_fn_pol}. The GTDS shows similar behavior, although the degradation of the metric is exacerbated for tracks of around 90 seconds. Surprisingly, there is a recovery for tracks longer than 120 seconds, which were not present before. This indicates the position-velocity correlations are compatible with the error distribution for very long tracks. Despite that, generalizing this behavior for long single track estimation using GTDS is not a good idea. A separate test with Radar 3 has been performed, showing that this method might not always provide adequate correlations in the full state covariance with very long radar tracks.

With these results it is important to highlight the excellent performance of the KEP+$\text{J}_2$ throughout (almost) the complete range, only failing in the very short tracks as before, see Figure \ref{fig:k2_stats_fn_pol}. This method presents itself as the most interesting. It is not only a good estimator in terms of realism of covariance representation, but also with increased accuracy (due to the use of measurement uncertainty in the fitting and the inclusion of range-rate).

\subsubsection{$k^2$ statistics on OPOD}\label{sec:results_orbital_plane} 

The scenario SEN-1A-2 has been used again to test the performance of the estimation algorithms, but now including the predicted orbital plane information. In each sampled radar track the information of a propagated initial condition (randomly chosen from the sampling of the I.C) is used to include predicted inclination and RAAN to the fitting. This is only done with the radar observables fitting methods, as the GTDS is unweighted and does not allow for other types of measurement (other than position).

\begin{figure}[h]
    \centering
    \includegraphics[width=0.9\textwidth]{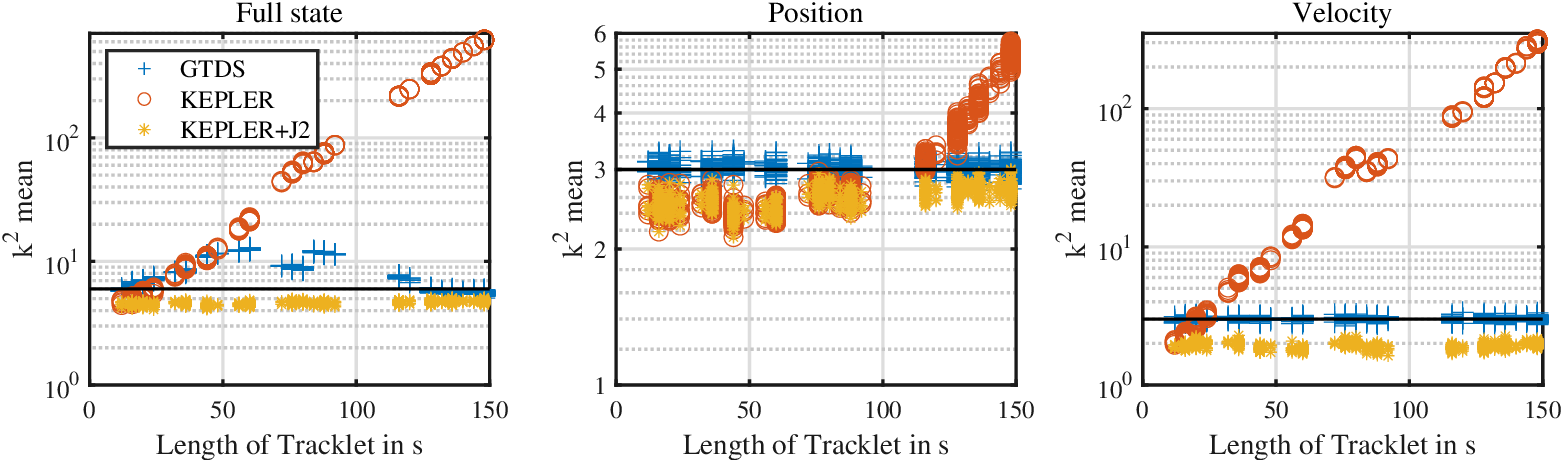}
    \caption{Estimation only $k^2$ mean values for Radar 2 ($r_t=4$ seconds), including the information of the orbital plane with $\sigma_{i,\Omega}=5e-3^{\circ}$ in the KEP and KEP+$\text{J}_2$ methods.}
    \label{fig:k2_stats_mean_pol_op}
\end{figure}

\begin{figure}[h]
    \centering
    \includegraphics[width=0.9\textwidth]{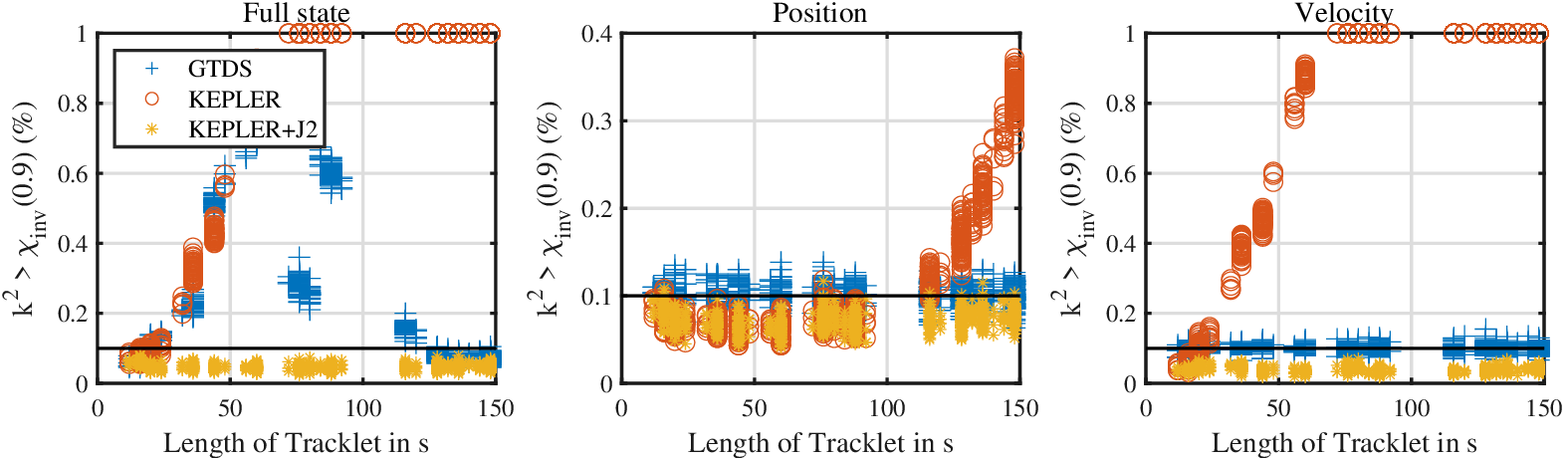}
    \caption{Estimation only $k^2$ percentage of anomalous estimations for Radar 2 ($r_t=4$ seconds), including the information of the orbital plane with $\sigma_{i,\Omega}=5e-3^{\circ}$ in the KEP and KEP+$\text{J}_2$ methods.}
    \label{fig:k2_stats_fn_pol_op}
\end{figure}

By looking at Figure \ref{fig:k2_stats_mean_pol_op} it is apparent that including the orbital plane information has little effect with the KEP fittings. Only the problematic with very short tracks is solved, but the degradation for longer ones is still present. On the other hand, the KEP+$\text{J}_2$ method sees its only drawback fixed, and it now behaves more appropriately for all track lengths. It is true that in this case the $k^2$ values obtained are lower than expected, which becomes obvious in Figure \ref{fig:k2_stats_fn_pol_op}, but a conservative covariance is in general more desirable to avoid false correlations. This is due to the virtual measurements being more consistent than what is indicated by the used value of uncertainty ($\sigma_{i,\Omega}=5e-3^{\circ}$ in all cases), signaling there is room for reducing the estimation uncertainty. As mentioned before the GTDS method results are unchanged (no extra information is added to the fit). One other aspect that has considerably improved and is not present in the $k^2$ statistics is the accuracy of the estimations when compared to the GTDS method. This is not surprising, as the augmented methodology uses range-rate and also orbital plane data to generate a better estimation in all aspects (uncertainty realism included).

\begin{figure}[h]
    \centering
    \includegraphics[width=0.9\textwidth]{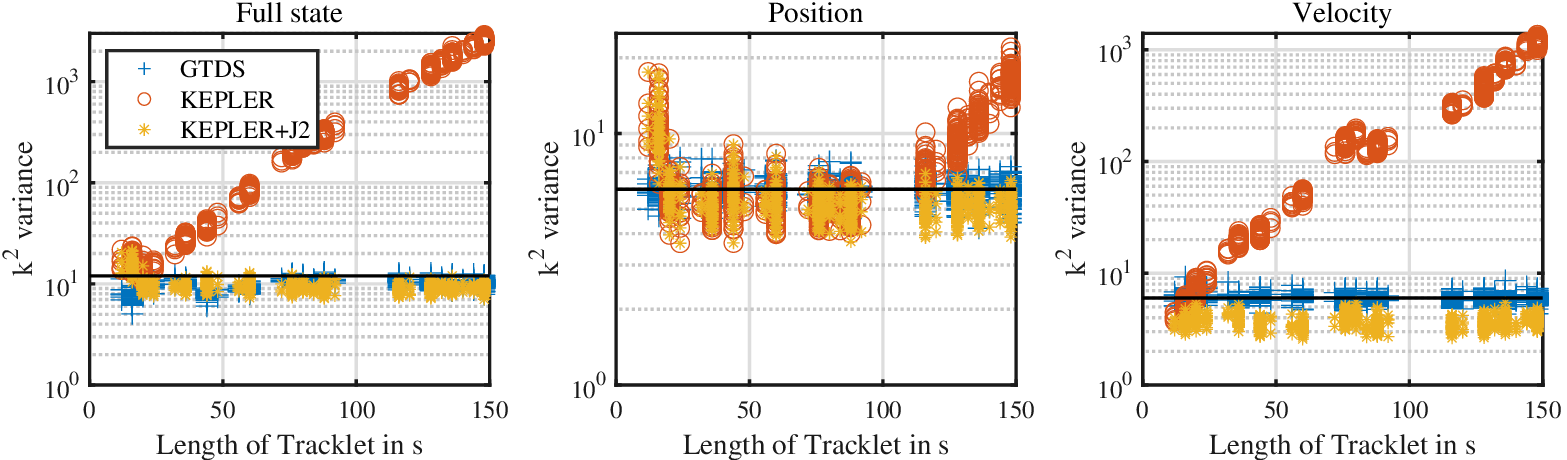}
    \caption{Estimation only $k^2$ variance values for Radar 2 ($r_t=4$ seconds), including the information of the orbital plane with $\sigma_{i,\Omega}=5e-3^{\circ}$ in the KEP and KEP+$\text{J}_2$ methods.}
    \label{fig:k2_stats_var_pol_op}
\end{figure}

\section{Conclusions}\label{sec:conlusions}




This work has introduced a novel approach to initial orbit determination, utilizing data sourced from a single radar track and combining it with the predicted orbital plane of the object. Emphasizing the imperative need for rapid algorithms in operational scenarios, this study focuses on the development of a least-squares fitting procedure. The procedure incorporates an analytically formulated approximation of dynamics under the $\text{J}_2$ perturbation, specifically designed for short-term propagation. Noteworthy for its inclusion of range-rate observables, the algorithm distinguishes itself from similar methods. A comprehensive evaluation is conducted through a series of synthetic tests, comparing its performance against a classical range and angles fitting method (GTDS). This investigation explores the impact of track length and measurement density on full state estimation.

When using data from a single radar, doing a proper full state estimation does not work as intended if the fitting is done using the radar observables directly, at least if Keplerian dynamics are used. The magnitude of the errors of this simple model is in fact high enough that precise range measurements become incompatible. A trajectory that meets the uncertainty imposed by the range and range-rate measurements tends to deviate from the real one when the the perpendicular constrains are much more imprecise, at least if the track is long enough. Solving this is a matter of improving the fidelity of the fitting dynamics by just including the $\text{J}_2$ perturbation. This has been accomplished while maintaining the fully analytical aspect of the methodology. The result is a very fast IOD method applicabe to maneuver detection, or simply find correlations between cataloged objects and track as an extended, more general application. 

It has been shown that the proposed method behaves as expected for tracks longer than 40 seconds, especially when the density of data is high. There is still a problem for shorter arcs, where the non-linear relation between the estimate and the radar observables seems to affect the velocity part of the estimation in a non-negligible way. Some kind of covariance inflation post-process could be applied to the uncertainty output in order to leverage the estimation. While not pursued in this study, this aspect is currently under investigation.




Although it has more limited applications, the method developed here admits the inclusion of chosen parts of the predicted state as virtual measurements. Given that the orbital plane is less susceptible to changes in LEO (high $\Delta V$ cost) and integration windows are short enough, it has been shown that the level of precision for the predicted $i$ and $\Omega$ justify including them in the estimation process. The resulting method, denoted as OPOD, shows improved accuracy and final reliability for all track lengths. This method in particular shows promise for its application to maneuver detection metrics, which is the next step of the process and will be the subject of a future publication.

The analytical time derivatives of the GEqOE equations of motion developed for this work could be employed in an enhanced numerical propagator that makes use of higher order derivatives. This could be leveraged to improve its accuracy at each integration step, thus reducing the total number of steps needed for a given propagation length and tolerance. This is certainly an interesting line of research that is being pursued at the moment as an extended application of the work presented here.




\section*{Acknowledgments}

Jose M. Montilla acknowledges the support from Universidad de Sevilla under VI PPIT-US. Special thanks are due to Guillermo Escribano for his help in developing the ideas and methods in this work.

During the preparation of this work the author(s) used ChatGPT in order to revise and correct grammar and refine the language. After using this tool/service, the authors reviewed and edited the content as needed and take full responsibility for the content of the publication.

\bibliographystyle{jasr-model5-names}
\biboptions{authoryear}
\bibliography{refs}

\appendix
\section{Auxiliary functions}\label{sec:appendix}

\subsection{Radar measurement function derivatives} \label{appendix:meas_derivatives}

The measurements provided by the radar in LEO are range ($\rho$), azimuth ($Az$), elevation ($el$) and range-rate ($\dot{\rho}$). The observation function $\left[ \rho , {Az}, {el}, \dot{\rho} \right] = h(\bm y,t)$ of these observables is in Equation (\ref{eq:rad_measurements}). The fitting algorithms in Sections \ref{sec:kep_fit_radar_observables} and \ref{sec_j2_fit_observables} require the derivatives of these measurements with respect to position ($\bm r$) and velocity ($\bm v$), denoted as $\partial h / \partial \bm y$, see Equation (\ref{eq:obs_der_app}). 
\begin{align} \label{eq:obs_der_app}
    \frac{\partial h}{\partial \bm y} = \begin{bmatrix}
        \rho_{\bm r} & \rho_{\bm v} \\
        Az_{\bm r} & Az_{\bm v} \\
        {el}_{\bm r} & {el}_{\bm v} \\
        \dot{\rho}_{\bm r} & \dot{\rho}_{\bm v}
    \end{bmatrix}
\end{align}
The explicit functions of these derivatives are in Equation (\ref{eq:radar_obs_der}), which requires the knowledge of radar inertial position ($\bm{P}_R$), velocity ($\bm{V}_R$) and orientation ($\text{T}_{I}^{R}$). Note that $\bm \rho = \bm r - \bm{P}_R$ and $\bm u = \bm v - \bm{V}_R$.
\begin{equation}\label{eq:radar_obs_der}
\begin{aligned}
    \rho_{\bm r} & = \frac{\bm \rho^\intercal}{\rho}, & \rho_{\bm v} & = \bm 0^\intercal, \\
    Az_{\bm r} & = \frac{1}{1+\tan^2(Az)} \left[ \bm{e}_1 ^\intercal  -\tan{(Az)}\bm{e}_2^\intercal \right]\frac{\text{T}_{I}^{R}}{\bm{e}_2^\intercal \bm \rho \rfloor_{R}}, & Az_{\bm v} & = \bm 0 ^\intercal, \\
    {el}_{\bm r} & = \frac{1}{\sqrt{1-\sin^2(el)}} \left[ \bm{e}_3^\intercal \text{T}_{I}^{R} - \sin{(el)} \rho_{\bm r} \right]\frac{1}{\rho}, & {el}_{\bm v} & = \bm 0 ^\intercal, \\
    \dot{\rho}_{\bm r} & = (\bm u - \dot{\rho}\rho_{\bm r} ) \frac{1}{\rho}, & \dot{\rho}_{\bm v} & =\frac{\bm \rho}{\rho}. \\
\end{aligned}
\end{equation}

\subsection{Time derivatives of the inverse function}\label{appendix:inverse_time_der}

In Section \ref{sec_j2_fit_observables} the linear least-squares algorithm is particularized to a dynamical model that includes the J$_2$ perturbation. This is done by utilizing the GEqOE from \cite{giulio2021generalization}, for which a Taylor expansion has been developed in order to maintain the analytical aspect of the fitting method. The expansion\textquotesingle s coefficients are fully developed in the \ref{appendix:j2_derivatives}, where there is an intensive use of the inverse function $\mathscr{f}_s= \frac{1}{s}$ as well as its time derivatives, included here for completeness.
\begin{equation} \label{eq:fs_td}
\begin{split}
    \dot{\mathscr{f}}_s = & -\frac{\dot{s}}{s^2}, \\ 
    \ddot{\mathscr{f}}_s = & \, 2\frac{\dot{s}^2}{s^3} - \frac{\ddot{s}}{s^2}, \\
    \overset{\dots}{\mathscr{f}}_s  = & \, 6\frac{\ddot{s}\dot{s}}{s^3} - \frac{\dddot{s}}{s^2} - 6 \frac{\dot{s}^3}{s^4}, \\
    \overset{\scriptscriptstyle(4)}{\mathscr{f}}_s  = & \, 8 \frac{\overset{\scriptscriptstyle(3)}{s}\dot{s}}{s^3} + 6 \frac{\ddot{s}^2}{s^3} -36\frac{\dot{s}^2 \ddot{s}}{s^4} -\frac{\overset{\scriptscriptstyle(4)}{s}}{s^2} - 24 \frac{\dot{s}^4}{s^5}.
\end{split}
\end{equation}

\subsection{Time derivatives of the function of $s$ times $\dot{s}$}\label{appendix:s_dots_fun}

As with the inverse function in the \ref{appendix:inverse_time_der}, the function that multiplies $s$ by its time derivative $\dot{s}$, $\mathscr{g}_{s} = s \dot{s}$, is repeatedly used in the derivation of the J$_2$ Taylor propagator, see the \ref{appendix:j2_derivatives}. The time derivatives of $\mathscr{g}_{s}$ have been included here.
\begin{equation}
\begin{split}
    \dot{\mathscr{g}}_{s} = & \, \dot{s}^2 + s\ddot{s}, \\
    \ddot{\mathscr{g}}_{s} = & \, 3\dot{s}\ddot{s} + s\dddot{s} \\
    \overset{\dots}{\mathscr{g}}_s = & \, 3\ddot{s}^2 + 4 \dot{s} \overset{\scriptscriptstyle(3)}{s} + s \overset{\scriptscriptstyle(4)}{s}, \\ 
    \overset{\scriptscriptstyle(4)}{\mathscr{g}}_s = & \, 10 \ddot{s} \overset{\scriptscriptstyle(3)}{s} + 5 \dot{s} \overset{\scriptscriptstyle(4)}{s} + s \overset{\scriptscriptstyle(5)}{s}.
\end{split}
\end{equation}

\subsection{Time derivatives of the $\text{J}_2$ GEqOE equations of motion} \label{appendix:j2_derivatives}

In order to obtain an efficient propagator with J$_2$ perturbation this work develops a Taylor expansion of the solution expressed in terms of GEqOE \citep{giulio2021generalization}. This is done in a systematic way by first simplifying the expressions of the equations of motion (first order derivatives) in a more compact form, see Equation (\ref{eq:J2_geqoe_simplified}). The resulting equations only have multiplying elements thanks to the use of the inverse function, see the \ref{appendix:inverse_time_der}. This section includes the time derivatives of Eq. (\ref{eq:J2_geqoe_simplified}) up to fourth order. All the simplifications that have been applied are first defined in the \ref{appendix:j2_second_order}, allowing for a systematic application of the chain rule in the derivation. From \ref{appendix:j2_second_order} to \ref{appendix:j2_fourth_order} all the steps for the computation of the Taylor J$_2$ propagator coefficients are included.

\subsubsection{Second order derivatives} \label{appendix:j2_second_order}
The second order time derivatives of the GEqOE can be computed with Equation (\ref{eq_geqoe_2o}).
\begin{equation} \label{eq_geqoe_2o}
   \begin{aligned}
    \ddot{\nu} & = 0, & \ddot{q}_1 & = -\dot{I} \mathscr{s}_L -I \dot{\mathscr{s}}_L, \\
    \ddot{p}_1 & = \dot{p}_2 \left( \mathscr{d} - w_h \right) + p_2 \left( \dot{\mathscr{d}} - \dot{w}_h \right) - \left(\dot{\mathscr{f}}_{c} \xi_1 + \mathscr{f}_{c} \dot{\xi}_1 \right) \mathscr{U} - \mathscr{f}_{c} \xi_1 \dot{\mathscr{U}}, & \ddot{q}_2 & = -\dot{I} \mathscr{c}_L -I \dot{\mathscr{c}}_L, \\
    \ddot{p}_2 & = \dot{p}_1 \left( w_h - \mathscr{d}  \right) + p_1 \left( \dot{w}_h - \dot{\mathscr{d}}  \right) + \left(\dot{\mathscr{f}}_{c} \xi_2 + \mathscr{f}_{c} \dot{\xi}_2 \right) \mathscr{U} + \mathscr{f}_{c} \xi_1 \dot{\mathscr{U}}, & \ddot{\mathcal{L}} & = \dot{\mathscr{d}} - \dot{w}_h - \left(\dot{\mathscr{f}}_{c} \Gamma + \mathscr{f}_{c} \dot{\Gamma} \right) \mathscr{U} - \mathscr{f}_{c} \Gamma \dot{\mathscr{U}}.
\end{aligned} 
\end{equation}
All the elements that appear in Eq. (\ref{eq_geqoe_2o}) are computed with the following formulation. Equation (\ref{eq:z_adim_dot}) computes the first time derivative of $\hat{z}$, where $\dot{\mathscr{f}}_{\mathcal{D}}$ is calculated from $\dot{\mathcal{D}}$.
\begin{equation} \label{eq:z_adim_dot}
    \begin{aligned} 
        \hat{z} & = \frac{ 2 \left(Y q_2 - X q_1 \right) }{r (1+q_1^2+q_2^2)} = \frac{ \mathcal{C} }{ \mathcal{D} } = \mathcal{C}\mathscr{f}_{\mathcal{D}},  & \dot{\hat{z}} & = \dot{\mathcal{C}}\mathscr{f}_{\mathcal{D}} + \mathcal{C}\dot{\mathscr{f}}_{\mathcal{D}}, \\
        \mathcal{C} & = 2 \left(Y q_2 - X q_1 \right), & \dot{\mathcal{C}} &  = 2 \left(\dot{Y} q_2 + Y \dot{q}_2 - \dot{X} q_1 - X \dot{q}_1\right), \\
        \mathcal{D} & = r (1+q_1^2+q_2^2) = r q_s, & \dot{\mathcal{D}} & = \dot{r} q_s + r \dot{q}_s \\
        q_s & = (1+q_1^2+q_2^2), & \dot{q}_s & = 2 \left( q_1 \dot{q}_1 + q_2 \dot{q}_2 \right).
    \end{aligned}
\end{equation}
Equation (\ref{eq:Uz_td}) computes the first time derivative of $\mathscr{U}$ and $U_z$.
\begin{equation}\label{eq:Uz_td}
\begin{aligned}
    \mathscr{U} & = - \frac{A}{r^3} (1-3 \hat{z}^2) = -A U_z \mathscr{f}_{r^3}, & \dot{\mathscr{U}} & = -A \left( \dot{U}_z \mathscr{f}_{r^3} + U_z \dot{\mathscr{f}}_{r^3} \right), \\ 
    U_z & = (1-3 \hat{z}^2), & \dot{U}_z & = -6 \hat{z} \dot{\hat{z}}. \\
\end{aligned}
\end{equation}
The value $\dot{\mathscr{f}}_{r^3}$ in Eq. (\ref{eq:Uz_td}) is computed from the derivative of $r^3$ in Equation (\ref{eq:rc_dot})
\begin{equation} \label{eq:rc_dot}
     \frac{d (r^3)}{d t} = \dot{r}_c  = 3 r^2 \dot{r}. \\
\end{equation}
Equation (\ref{eq:gamma_dot}) includes the derivative of $\Gamma$ and $\beta$, which in turn allows to compute $\dot{\mathscr{f}}_{1+\beta}$, $\dot{\mathscr{f}}_{\beta}$ and $\dot{\mathscr{f}}_{\alpha}$.
\begin{equation}\label{eq:gamma_dot}
    \begin{aligned}
        \Gamma & = \mathscr{f}_{\alpha} + \alpha \left(1 - r/a \right), & \dot{\Gamma} & = \dot{\mathscr{f}}_{\alpha} + \dot{\alpha} \left(1 - \frac{r}{a} \right) - \alpha \frac{\dot{r}}{a}, \\
        \alpha & = \frac{1}{1 + \beta} = \mathscr{f}_{1 + \beta}, & \dot{\alpha} & = \dot{\mathscr{f}}_{1 + \beta}, \\
        \beta  & = \sqrt{1 - p1^2 - p_2^2}, & \dot{\beta} & = \frac{-\dot{p}_s}{2 \beta} = - \frac{1}{2}\dot{p}_s \mathscr{f}_{\beta}, \\
        p_s & = 1 + p_1^2 + p_2^2, & \dot{p}_s & = 2 \left( p_1 \dot{p}_1 + p_2 \dot{p}_2 \right). \\
    \end{aligned}
\end{equation}
Equation (\ref{eq:I_dot}) computes the first time derivative of $I$. This includes the computation of $\dot{\mathscr{f}}_{\mathscr{h}}$ from $\dot{\mathscr{h}}$, and $\dot{\mathscr{f}}_{h}$ is computed from $\dot{h}$ for later use. The same goes for $\dot{\mathscr{f}}_{c}$ and $\mathscr{g}_{c}$, calculated from $\dot{c}$.
\begin{equation}\label{eq:I_dot}
    \begin{aligned}
        I & = \frac{3 A}{h r^3} \hat{z} (1-q_1^2-q_2^2) = 3A \frac{ \hat{z} \delta }{ \mathscr{h} } = 3A  \, \hat{z} \,  \delta \, \mathscr{f}_{\mathscr{h}},   & \dot{I} & = 3 A \left[ \left( \dot{\hat{z}}\delta + \hat{z}\dot{\delta} \right) \mathscr{f}_{\mathscr{h}} + \hat{z} \delta \dot{\mathscr{f}}_{\mathscr{h}}  \right], \\
    \delta & = 1-q_1^2-q_2^2, & \dot{\delta} & = -\dot{q}_s, \\
    \mathscr{h} & = h r^3 = h r_c, & \dot{\mathscr{h}} & = \dot{h} r_c + h \dot{r}_c, \\
    h & = \sqrt{c^2 - 2 r^2 \mathscr{U}}, & \dot{h} &  = \left( \mathscr{g}_c -2\mathscr{g}_r\mathscr{U} - r^2 \dot{\mathscr{U}}  \right) \mathscr{f}_h,  \\
    c & = \left( \frac{\mu^2}{\nu} \right)^{1/3} \sqrt{1 - p_1^2 - p_2^2} = \left( \frac{\mu^2}{\nu} \right)^{1/3}  \beta, & \dot{c}  & = \left( \frac{\mu^2}{\nu} \right)^{1/3} \dot{\beta}.
    \end{aligned}
\end{equation}
Equation (\ref{eq:d_td}) computes the first time derivative of $\mathscr{d}$, where $\dot{\mathscr{f}}_{r^2}$ is computed from the derivation of $r^2$.
\begin{equation}\label{eq:d_td}
    \begin{aligned}
        \mathscr{d} & = \frac{h-c}{r^2} = \left( h - c \right) \mathscr{f}_{r^2}, \\ 
        \dot{\mathscr{d}} & = \left( \dot{h} - \dot{c} \right)\mathscr{f}_{r^2} + \left( h - c \right)\dot{\mathscr{f}}_{r^2}, \\ 
        \frac{d (r^2)}{d t} & = \dot{r}_s  = 2 r \dot{r} = 2 \mathscr{g}_r.
    \end{aligned}
\end{equation}
Finally, the time derivatives of $w_h$, $\xi_1$, $\xi_2$, $\mathscr{s}_L$ and $\mathscr{c}_L$ are in Equation (\ref{eq:extra_td}).
\begin{equation}\label{eq:extra_td}
    \begin{aligned}
        w_h & = I \hat{z}, & \dot{w}_h & = \dot{I} \hat{z} + I \dot{\hat{z}}, \\
    \xi_1 & = \frac{X}{a} + 2 p_2, & \dot{\xi}_1 & = \frac{\dot{X}}{a} + 2 \dot{p}_2, \\
    \xi_2 & = \frac{Y}{a} + 2 p_1, & \dot{\xi}_2 & = \frac{\dot{Y}}{a} + 2 \dot{p}_1, \\
    \mathscr{s}_L & = \frac{Y}{r} = Y \mathscr{f}_r, & \dot{\mathscr{s}}_L & = \dot{Y} \mathscr{f}_r + Y \dot{\mathscr{f}}_r, \\
    \mathscr{c}_L & = \frac{X}{r} = X \mathscr{f}_r, & \dot{\mathscr{c}}_L & = \dot{X} \mathscr{f}_r + X \dot{\mathscr{f}}_r.
    \end{aligned}
\end{equation}

    In this section the computation of every $\mathscr{f}_s$ time derivative has been explicitly noted. In the following this comments are omitted for the sake of brevity, and it will be assumed all instances can be computed with the corresponding derivatives of $s$ (to the needed order).
\subsubsection{Third order derivatives} \label{appendix:j2_third_order}
The third order time derivatives of the GEqOE can be computed with Equation (\ref{eq_geqoe_3o}).
\begin{equation}\label{eq_geqoe_3o}
    \begin{split}
        \dddot{\nu} & = 0, \\
        \dddot{q}_1 & = -\ddot{I} \mathscr{s}_L - 2\dot{I}\dot{\mathscr{s}}_L -I \ddot{\mathscr{s}}_L, \\
        \dddot{q}_2 & = -\ddot{I} \mathscr{c}_L - 2\dot{I}\dot{\mathscr{c}}_L -I \ddot{\mathscr{c}}_L, \\
        \dddot{p}_1 & = \ddot{p}_2 \left( \mathscr{d} - w_h \right) + 2 \dot{p}_2 \left( \dot{\mathscr{d}} - \dot{w}_h \right) + p_2 \left( \ddot{\mathscr{d}} - \ddot{w}_h \right) - \left(\ddot{\mathscr{f}}_{c} \xi_1 + 2\dot{\mathscr{f}}_{c}\dot{\xi}_1 + \mathscr{f}_{c} \ddot{\xi}_1 \right) \mathscr{U} - 2\left(\dot{\mathscr{f}}_{c} \xi_1 + \mathscr{f}_{c} \dot{\xi}_1 \right)\dot{\mathscr{U}} - \mathscr{f}_{c} \xi_1 \ddot{\mathscr{U}}, \\
        \dddot{p}_2 & = \ddot{p}_1 \left( w_h - \mathscr{d} \right) + 2 \dot{p}_1 \left( \dot{w}_h - \dot{\mathscr{d}}  \right) + p_1 \left( \ddot{w}_h - \ddot{\mathscr{d}} \right) + \left(\ddot{\mathscr{f}}_{c} \xi_2 + 2\dot{\mathscr{f}}_{c}\dot{\xi}_2 + \mathscr{f}_{c} \ddot{\xi}_2 \right) \mathscr{U} + 2\left(\dot{\mathscr{f}}_{c} \xi_2 + \mathscr{f}_{c} \dot{\xi}_2 \right)\dot{\mathscr{U}} + \mathscr{f}_{c} \xi_2 \ddot{\mathscr{U}}, \\
        \dddot{\mathcal{L}} & = \ddot{\mathscr{d}} - \ddot{w}_h - \left(\ddot{\mathscr{f}}_{c} \Gamma + 2\dot{\mathscr{f}}_{c}\dot{\Gamma} + \mathscr{f}_{c} \ddot{\Gamma} \right) \mathscr{U} - 2\left(\dot{\mathscr{f}}_{c} \Gamma + \mathscr{f}_{c} \dot{\Gamma} \right) \dot{\mathscr{U}} - \mathscr{f}_{c} \Gamma \ddot{\mathscr{U}}.        
    \end{split}
\end{equation}
All the elements that appear in Eq. (\ref{eq_geqoe_3o}) are computed with the following formulation. Equation (\ref{eq:z_adim_ddot}) computes the second time derivative of $\hat{z}$.
\begin{equation}\label{eq:z_adim_ddot}
    \begin{aligned}
        \ddot{\hat{z}} & = \ddot{\mathcal{C}}\mathscr{f}_{\mathcal{D}} + 2\dot{\mathcal{C}}\dot{\mathscr{f}}_{\mathcal{D}} + \mathcal{C}\ddot{\mathscr{f}}_{\mathcal{D}}, & \ddot{\mathcal{C}} &  = 2 \left( \ddot{Y} q_2 + 2\dot{Y}\dot{q}_2 + Y \dot{q}_2 - \ddot{X} q_1 - 2\dot{X} \dot{q}_1 - X \ddot{q}_1 \right), \\
    \ddot{\mathcal{D}} & = \ddot{r} q_s + 2\dot{r}\dot{q}_s + r \ddot{q}_s \rightarrow  \ddot{\mathscr{f}}_{\mathcal{D}}, & \ddot{q}_s & = 2 \left( \dot{q}_1^2 + q_1 \ddot{q}_1 +\dot{q}_2^2 + q_2 \ddot{q}_2 \right).
    \end{aligned}
\end{equation}
 The second order time derivatives of $X$ and $Y$ are computed in Equation (\ref{eq:X_Y_ddot}).
\begin{equation}\label{eq:X_Y_ddot}
    \begin{aligned}
        \dot{X} & = \dot{r} \cos{L} - \frac{h}{r} \sin{L} = \dot{r} \mathscr{c}_L - \mathscr{w} \mathscr{s}_L, & \ddot{X} & = \ddot{r} \mathscr{c}_L + \dot{r} \dot{\mathscr{c}}_L - \dot{\mathscr{w}} \mathscr{s}_L -\mathscr{w} \dot{\mathscr{s}}_L, \\
    \dot{Y} & = \dot{r} \sin{L} + \frac{h}{r} \cos{L} = \dot{r} \mathscr{s}_L + \mathscr{w} \mathscr{c}_L, & \ddot{Y} & = \ddot{r} \mathscr{s}_L + \dot{r} \dot{\mathscr{s}}_L + \dot{\mathscr{w}} \mathscr{c}_L + \mathscr{w} \dot{\mathscr{c}}_L, \\
    \mathscr{w} & = \frac{h}{r} = h \mathscr{f}_r, & \dot{\mathscr{w}} & =  \dot{h} \mathscr{f}_r + h \dot{\mathscr{f}}_r, \\
    \dot{r} & = \frac{\mu}{c} \left( p_2 \sin{L} - p_1 \cos{L} \right) = \mu \mathscr{f}_c r_{pl}, & \ddot{r} & = \mu \left(\dot{\mathscr{f}}_c r_{pl} + \mathscr{f}_c \dot{r}_{pl} \right),  \\
    r_{pl} & = p_2 \mathscr{s}_L - p_1 \mathscr{c}_L, &  \dot{r}_{pl} & = \dot{p}_2 \mathscr{s}_L + p_2 \dot{\mathscr{s}}_L - \dot{p}_1 \mathscr{c}_L - p_1 \dot{\mathscr{c}}_L. \\
    \end{aligned}
\end{equation}
Equation (\ref{eq:Uz_tdd}) computes the second time derivative of $\mathscr{U}$ and $U_z$.
\begin{equation}\label{eq:Uz_tdd}
    \begin{aligned}
        \ddot{\mathscr{U}} & = -A \left( \ddot{U}_z \mathscr{f}_{r^3} + 2\dot{U}_z\dot{\mathscr{f}}_{r^3} + U_z \ddot{\mathscr{f}}_{r^3} \right), & \ddot{U}_z & = -6 \left( \dot{\hat{z}}^2 + \hat{z} \ddot{\hat{z}}\right), & \frac{d^2 (r^3)}{d t^2} = \ddot{r}_c  = 3 \left( 2r \dot{r}^2 + r^2 \ddot{r} \right).
    \end{aligned}
\end{equation}
Equation (\ref{eq:gamma_ddot}) includes the second order derivatives of $\Gamma$ and $\beta$.
\begin{equation}\label{eq:gamma_ddot}
    \begin{aligned}
        \ddot{\Gamma} & = \ddot{\mathscr{f}}_{\alpha} + \ddot{\alpha} \left(1 - \frac{r}{a} \right) - 2 \dot{\alpha}\frac{\dot{r}}{a} - \alpha \frac{\ddot{r}}{a}, & \ddot{\alpha} & = \ddot{\mathscr{f}}_{1 + \beta}, \\
    \ddot{\beta} & = - \frac{1}{2} \left( \ddot{p}_s \mathscr{f}_{\beta} + \dot{p}_s \dot{\mathscr{f}}_{\beta} \right) \rightarrow \ddot{\mathscr{f}}_{1+\beta},\ddot{\mathscr{f}}_{\beta}, &  \ddot{p}_s & = 2 \left( \dot{p}_1^2 + p_1 \ddot{p}_1 +\dot{p}_2^2 + p_2 \ddot{p}_2 \right). \\
    \end{aligned}
\end{equation}
Equation (\ref{eq:I_ddot}) computes the second time derivative of $I$.
\begin{equation}\label{eq:I_ddot}
    \begin{aligned}
        \ddot{I} & = 3 A \left[ \left( \ddot{\hat{z}}\delta + 2\dot{\hat{z}}\dot{\delta} + \hat{z}\ddot{\delta} \right) \mathscr{f}_{\mathscr{h}} + 2\left( \dot{\hat{z}}\delta + \hat{z}\dot{\delta} \right)\dot{\mathscr{f}}_{\mathscr{h}} + \hat{z} \delta \ddot{\mathscr{f}}_{\mathscr{h}}  \right], \\
        \ddot{\mathscr{h}} & = \ddot{h} r_c  +  2\dot{h}\dot{r}_c + h \ddot{r}_c, \\
        \ddot{h} &  = \left( \dot{\mathscr{g}}_c -2\dot{\mathscr{g}}_r\mathscr{U} - 4\mathscr{g}_r\dot{\mathscr{U}} - r^2 \ddot{\mathscr{U}}  \right) \mathscr{f}_h   +    \left( \mathscr{g}_c -2\mathscr{g}_r\mathscr{U} - r^2 \dot{\mathscr{U}}  \right) \dot{\mathscr{f}}_h, \\
        \ddot{c}  & = \left( \frac{\mu^2}{\nu} \right)^{1/3} \ddot{\beta}, \\
        \ddot{\delta} & = -\ddot{q}_s. \\
    \end{aligned}
\end{equation}
Equation (\ref{eq:d_tdd}) computes the second time derivative of $\mathscr{d}$.
\begin{equation}\label{eq:d_tdd}
    \begin{aligned}
        \ddot{\mathscr{d}} & = \left( \ddot{h} - \ddot{c} \right)\mathscr{f}_{r^2} +  2\left( \dot{h} - \dot{c} \right)\dot{\mathscr{f}}_{r^2} + \left( h - c \right)\ddot{\mathscr{f}}_{r^2}, \\
        \frac{d^2 (r^2)}{d t^2} & = \ddot{r}_s = 2 \dot{\mathscr{g}}_r. \\
    \end{aligned}
\end{equation}
The second order time derivatives of $w_h$, $\xi_1$, $\xi_2$, $\mathscr{s}_L$ and $\mathscr{c}_L$ are in Equation (\ref{eq:extra_tdd}).
\begin{equation}\label{eq:extra_tdd}
    \begin{aligned}
        \ddot{w}_h & = \ddot{I}\hat{z} + 2\dot{I}\dot{\hat{z}}  + I\ddot{\hat{z}}, \\
        \ddot{\xi}_1 & = \frac{\ddot{X}}{a} + 2 \ddot{p}_2, \\
        \ddot{\xi}_2 & = \frac{\ddot{Y}}{a} + 2 \ddot{p}_1, \\
        \ddot{\mathscr{s}}_L & =  \ddot{Y} \mathscr{f}_r + 2\dot{Y}\dot{\mathscr{f}}_r + Y \ddot{\mathscr{f}}_r, \\
        \ddot{\mathscr{c}}_L & =  \ddot{X} \mathscr{f}_r + 2\dot{X}\dot{\mathscr{f}}_r + X \ddot{\mathscr{f}}_r. \\
    \end{aligned}
\end{equation}

\subsubsection{Fourth order derivatives} \label{appendix:j2_fourth_order}
The third order time derivatives of the GEqOE can be computed with Equation (\ref{eq_geqoe_4o}).
\begin{equation}\label{eq_geqoe_4o}
    \begin{aligned}
        \overset{\scriptscriptstyle(4)}{\nu}  = & 0, \\
        \overset{\scriptscriptstyle(4)}{q}_1  = & -\dddot{I} \mathscr{s}_L - 3 \ddot{I}\dot{\mathscr{s}}_L - 3\dot{I}\ddot{\mathscr{s}}_L -I \dddot{\mathscr{s}}_L, \\
        \overset{\scriptscriptstyle(4)}{q}_2  = & -\dddot{I} \mathscr{c}_L - 3 \ddot{I}\dot{\mathscr{c}}_L - 3\dot{I}\ddot{\mathscr{c}}_L -I \dddot{\mathscr{c}}_L, \\
         \overset{\scriptscriptstyle(4)}{p}_1 = & \dddot{p}_2 \left( \mathscr{d} - w_h \right) + 3 \ddot{p}_2\left( \dot{\mathscr{d}} - \dot{w}_h \right) + 3\dot{p}_2 \left( \ddot{\mathscr{d}} - \ddot{w}_h \right) + p_2 \left( \dddot{\mathscr{d}} - \dddot{w}_h \right) \\ & - \left(\dddot{\mathscr{f}}_{c} \xi_1 + 3\ddot{\mathscr{f}}_{c}\dot{\xi}_1 + 3 \dot{\mathscr{f}}_{c}\ddot{\xi}_1  + \mathscr{f}_{c} \dddot{\xi}_1 \right) \mathscr{U} -3 \left(\ddot{\mathscr{f}}_{c} \xi_1 + 2\dot{\mathscr{f}}_{c}\dot{\xi}_1 + \mathscr{f}_{c} \ddot{\xi}_1 \right) \dot{\mathscr{U}} - 3 \left(\dot{\mathscr{f}}_{c} \xi_1 + \mathscr{f}_{c} \dot{\xi}_1 \right) \ddot{\mathscr{U}}  - \mathscr{f}_{c} \xi_1 \dddot{\mathscr{U}}, \\
         \overset{\scriptscriptstyle(4)}{p}_2  = & \dddot{p}_1 \left( w_h - \mathscr{d}  \right) + 3 \ddot{p}_1\left(  \dot{w}_h - \dot{\mathscr{d}}  \right) + 3\dot{p}_1 \left(  \ddot{w}_h - \ddot{\mathscr{d}}  \right) + p_1 \left(  \dddot{w}_h - \dddot{\mathscr{d}}  \right) \\ & + \left(\dddot{\mathscr{f}}_{c} \xi_2 + 3\ddot{\mathscr{f}}_{c}\dot{\xi}_2 + 3 \dot{\mathscr{f}}_{c}\ddot{\xi}_2  + \mathscr{f}_{c} \dddot{\xi}_2 \right) \mathscr{U} +3 \left(\ddot{\mathscr{f}}_{c} \xi_2 + 2\dot{\mathscr{f}}_{c}\dot{\xi}_2 + \mathscr{f}_{c} \ddot{\xi}_2 \right) \dot{\mathscr{U}} + 3 \left(\dot{\mathscr{f}}_{c} \xi_2 + \mathscr{f}_{c} \dot{\xi}_2 \right) \ddot{\mathscr{U}}  + \mathscr{f}_{c} \xi_2 \dddot{\mathscr{U}}, \\
         \overset{\scriptscriptstyle(4)}{\mathcal{L}}  = & \dddot{\mathscr{d}} - \dddot{w}_h - \left(\dddot{\mathscr{f}}_{c} \Gamma + 3\ddot{\mathscr{f}}_{c}\dot{\Gamma} + 3\dot{\mathscr{f}}_{c}\ddot{\Gamma}  + \mathscr{f}_{c} \dddot{\Gamma} \right) \mathscr{U} - 3 \left(\ddot{\mathscr{f}}_{c} \Gamma + 2\dot{\mathscr{f}}_{c}\dot{\Gamma} + \mathscr{f}_{c} \ddot{\Gamma} \right)\dot{\mathscr{U}} - 3\left(\dot{\mathscr{f}}_{c} \Gamma + \mathscr{f}_{c} \dot{\Gamma} \right) \ddot{\mathscr{U}} - \mathscr{f}_{c} \Gamma \dddot{\mathscr{U}}.
    \end{aligned}
\end{equation}

All the elements that appear in Eq. (\ref{eq_geqoe_4o}) are computed with the following formulation. Equation (\ref{eq:z_adim_dddot}) computes the second time derivative of $\hat{z}$.
\begin{equation}\label{eq:z_adim_dddot}
    \begin{aligned}
        \dddot{\hat{z}} & = \dddot{\mathcal{C}}\mathscr{f}_{\mathcal{D}} + 3\ddot{\mathcal{C}}\dot{\mathscr{f}}_{\mathcal{D}} +  3\dot{\mathcal{C}}\ddot{\mathscr{f}}_{\mathcal{D}} + \mathcal{C}\dddot{\mathscr{f}}_{\mathcal{D}}, & \dddot{\mathcal{C}} &  = 2 \left( \dddot{Y} q_2  + 3\ddot{Y}\dot{q}_2 + 3\dot{Y}\ddot{q}_2 + Y \dddot{q}_2 - \dddot{X} q_1  - 3\ddot{X}\dot{q}_1 - 3\dot{X}\ddot{q}_1 - X \dddot{q}_1 \right), \\
    \dddot{\mathcal{D}} & = \dddot{r} q_s + 3\ddot{r}\dot{q}_s + 3\dot{r}\ddot{q}_s + r \dddot{q}_s, & \dddot{q}_s & = 2 \left( 3 \dot{q}_1 \ddot{q}_1 + q_1 \dddot{q}_1 +3\dot{q}_2\ddot{q}_2 + q_2 \dddot{q}_2 \right).
    \end{aligned}
\end{equation}
The third order time derivatives of $X$ and $Y$ are computed in Equation (\ref{eq:X_Y_dddot}).
\begin{equation}\label{eq:X_Y_dddot}
    \begin{aligned}
        \dddot{X} & = \dddot{r}\mathscr{c}_L + 2\ddot{r}\dot{\mathscr{c}}_L + \dot{r}\ddot{\mathscr{c}}_L - \ddot{\mathscr{w}} \mathscr{s}_L - 2\dot{\mathscr{w}}\dot{\mathscr{s}}_L - \mathscr{w} \ddot{\mathscr{s}}_L, \\
    \dddot{Y} & = \dddot{r}\mathscr{s}_L + 2\ddot{r}\dot{\mathscr{s}}_L + \dot{r}\ddot{\mathscr{s}}_L + \ddot{\mathscr{w}} \mathscr{c}_L + 2\dot{\mathscr{w}}\dot{\mathscr{c}}_L + \mathscr{w} \ddot{\mathscr{c}}_L, \\
    \ddot{\mathscr{w}} & =  \ddot{h} \mathscr{f}_r + 2 \dot{h}\dot{\mathscr{f}}_r  + h \ddot{\mathscr{f}}_r, \\
    \dddot{r} & = \mu \left(\ddot{\mathscr{f}}_c r_{pl} + 2 \dot{\mathscr{f}}_c \dot{r}_{pl} + \mathscr{f}_c \ddot{r}_{pl} \right), \\
    \ddot{r}_{pl} & = \ddot{p}_2 \mathscr{s}_L+ 2 \dot{p}_2\dot{\mathscr{s}}_L  + p_2 \ddot{\mathscr{s}}_L - \ddot{p}_1 \mathscr{c}_L - 2\dot{p}_1 \dot{\mathscr{c}}_L - p_1 \ddot{\mathscr{c}}_L. \\
    \end{aligned}
\end{equation}
Equation (\ref{eq:Uz_tddd}) computes the third order time derivative of $\mathscr{U}$ and $U_z$.
\begin{equation}\label{eq:Uz_tddd}
    \begin{aligned}
        \dddot{\mathscr{U}} & = -A \left( \dddot{U}_z \mathscr{f}_{r^3} + 3  \ddot{U}_z \dot{\mathscr{f}}_{r^3} + 3\dot{U}_z\ddot{\mathscr{f}}_{r^3} + U_z \dddot{\mathscr{f}}_{r^3} \right), \\
        \dddot{U}_z & = -6 \left( 3\dot{\hat{z}}\ddot{\hat{z}} + \hat{z} \dddot{\hat{z}}\right), \\
        \frac{d^3 (r^3)}{d t^3}  & = \dddot{r}_c  = 3 \left( 2 \dot{r}^3 + 6 r \dot{r} \ddot{r} + r^2 \dddot{r} \right).
    \end{aligned}
\end{equation}
Equation (\ref{eq:gamma_dddot}) includes the third order derivatives of $\Gamma$ and $\beta$.
\begin{equation}\label{eq:gamma_dddot}
    \begin{aligned}
        \dddot{\Gamma} & = \dddot{\mathscr{f}}_{\alpha} + \ddot{\alpha} \left(1 - \frac{r}{a} \right) - 3\ddot{\alpha}\frac{\dot{r}}{a} - 3 \dot{\alpha}\frac{\ddot{r}}{a} - \alpha \frac{\dddot{r}}{a}, & \dddot{\alpha} & = \dddot{\mathscr{f}}_{1 + \beta}, \\
    \dddot{\beta} & = - \frac{1}{2} \left( \dddot{p}_s \mathscr{f}_{\beta} + 2 \ddot{p}_s \dot{\mathscr{f}}_{\beta} + \dot{p}_s \ddot{\mathscr{f}}_{\beta} \right),\dddot{\mathscr{f}}_{\beta}, &  \dddot{p}_s & = 2 \left( 3 \dot{p}_1 \ddot{p}_1 + p_1 \dddot{p}_1 +3\dot{p}_2\ddot{p}_2 + p_2 \dddot{p}_2 \right). \\
    \end{aligned}
\end{equation}
Equation (\ref{eq:I_dddot}) computes the third order time derivative of $I$.
\begin{equation}\label{eq:I_dddot}
    \begin{aligned}
        \dddot{I} & = 3 A \left[ \left( \dddot{\hat{z}}\delta + 3\ddot{\hat{z}}\dot{\delta} + 3\dot{\hat{z}}\ddot{\delta} + \hat{z}\dddot{\delta} \right) \mathscr{f}_{\mathscr{h}} + 3 \left( \ddot{\hat{z}}\delta + 2\dot{\hat{z}}\dot{\delta} + \hat{z}\ddot{\delta} \right)\dot{\mathscr{f}}_{\mathscr{h}}  + 3\left( \dot{\hat{z}}\delta + \hat{z}\dot{\delta} \right)\ddot{\mathscr{f}}_{\mathscr{h}} + \hat{z} \delta \dddot{\mathscr{f}}_{\mathscr{h}}  \right], \\
         \dddot{\mathscr{h}} & = \dddot{h} r_c + 3\ddot{h}\dot{r}_c  +  3 \dot{h}\ddot{r}_c + h \dddot{r}_c, \\
         \dddot{h} &  = \left( \ddot{\mathscr{g}}_c - 2 \ddot{\mathscr{g}}_r \mathscr{U} - 6 \dot{\mathscr{g}}_r \dot{\mathscr{U}}  - 6\mathscr{g}_r\ddot{\mathscr{U}} - r^2 \dddot{\mathscr{U}}  \right) \mathscr{f}_h + 2 \left( \dot{\mathscr{g}}_c -2\dot{\mathscr{g}}_r\mathscr{U} - 4\mathscr{g}_r\dot{\mathscr{U}} - r^2 \ddot{\mathscr{U}}  \right) \dot{\mathscr{f}}_h  +    \left( \mathscr{g}_c -2\mathscr{g}_r\mathscr{U} - r^2 \dot{\mathscr{U}}  \right) \ddot{\mathscr{f}}_h, \\
         \dddot{c}  & = \left( \frac{\mu^2}{\nu} \right)^{1/3} \dddot{\beta}, \\
         \dddot{\delta} & = -\dddot{q}_s.
    \end{aligned}
\end{equation}
Equation (\ref{eq:d_tddd}) computes the third time derivative of $\mathscr{d}$.
\begin{equation}\label{eq:d_tddd}
    \begin{aligned}
        \dddot{\mathscr{d}} & = \left( \dddot{h} - \dddot{c} \right)\mathscr{f}_{r^2} + 3\left( \ddot{h} - \ddot{c} \right)\dot{\mathscr{f}}_{r^2} +  3\left( \dot{h} - \dot{c} \right)\ddot{\mathscr{f}}_{r^2} + \left( h - c \right)\dddot{\mathscr{f}}_{r^2}, & \frac{d^3 (r^2)}{d t^3} = \dddot{r}_s = 2 \ddot{\mathscr{g}}_r. \\
    \end{aligned}
\end{equation}
The third order time derivatives of $w_h$, $\xi_1$, $\xi_2$, $\mathscr{s}_L$ and $\mathscr{c}_L$ are in Equation (\ref{eq:extra_tddd}).
\begin{equation}\label{eq:extra_tddd}
    \begin{aligned}
        \dddot{w}_h & = \dddot{I}\hat{z} + 3\ddot{I}\dot{\hat{z}} + 3\dot{I}\ddot{\hat{z}}  + I\dddot{\hat{z}}, \\
    \dddot{\xi}_1 & = \frac{\dddot{X}}{a} + 2 \dddot{p}_2, \\
    \dddot{\xi}_2 & = \frac{\dddot{Y}}{a} + 2 \dddot{p}_1, \\
    \dddot{\mathscr{s}}_L & =  \dddot{Y} \mathscr{f}_r + 3\ddot{Y}\dot{\mathscr{f}}_r + 3\dot{Y}\ddot{\mathscr{f}}_r + Y \dddot{\mathscr{f}}_r, \\
    \dddot{\mathscr{c}}_L & =  \dddot{X} \mathscr{f}_r + 3 \ddot{X}\dot{\mathscr{f}}_r  + 3\dot{X}\ddot{\mathscr{f}}_r + X \dddot{\mathscr{f}}_r. \\
    \end{aligned}
\end{equation}
    
\subsection{Derivatives of the $\mathscr{f}_s$ and $\mathscr{g}_s$ time derivatives with respect to the element $x$} \label{appendix:f_g_derivatives}

When $\overset{\scriptscriptstyle(k)}{\mathscr{f}}_s$ is derived with respect to $x$, then all the derivatives up to $\overset{\scriptscriptstyle(k)}{s}$ with respect to $x$ (with standard notation $\overset{\scriptscriptstyle(k)}{s}_x$) are needed (or up to $\overset{\scriptscriptstyle(k+1)}{s}$ in the case of $\overset{\scriptscriptstyle(k)}{\mathscr{g}}_s$). The notation for the cross derivative used here meets $\partial \overset{\scriptscriptstyle(k)}{\mathscr{f}_s} / \partial x = \overset{\scriptscriptstyle(k),x}{\mathscr{f}_s}$, chose on purpose for the compact expressions developed in this appendix section.

Thus, the function that computes $\overset{\scriptscriptstyle{(k),x}}{\mathscr{f}}_s$ needs not only the vector $\left[s,\dot{s} ,\cdots, \overset{\scriptscriptstyle(k)}{s} \right]$, but also $\left[s_x,\dot{s}_x ,\cdots, \overset{\scriptscriptstyle(k)}{s}_x \right]$ as input. Here are the expressions of the function that implements these derivatives:


\begin{equation}
    \begin{aligned}
        \frac{\partial \mathscr{f}_s}{\partial x} = \overset{\scriptscriptstyle{x}}{\mathscr{f}}_s = &  -\frac{s_x}{s^2}, \\ \frac{\partial \dot{\mathscr{f}}_s}{\partial x} = \overset{\scriptscriptstyle{(1),x}}{\mathscr{f}}_s  = & - \frac{\dot{s}_x}{s^2} + 2 \frac{\dot{s} s_x}{s^3},  \\
        \frac{\partial \ddot{\mathscr{f}}_s}{\partial x} = \overset{\scriptscriptstyle{(2),x}}{\mathscr{f}}_s  = & - \frac{\ddot{s}_x}{s^2} + 2 \frac{ \ddot{s} s_x + 2 \dot{s} \dot{s}_x }{s^3} - 6 \frac{ \dot{s}^2 s_x }{s^4}, \\
        \frac{\partial \dddot{\mathscr{f}}_s}{\partial x} = \overset{\scriptscriptstyle{(3),x}}{\mathscr{f}}_s  = &  - \frac{\dddot{s}_x}{s^2} + 2 \frac{ \dddot{s} s_x + 3 \left(  \ddot{s}_x \dot{s}  + \ddot{s} \dot{s}_x \right) }{s^3} - 18 \frac{ \ddot{s} \dot{s} s_x + \dot{s}^2 \dot{s}_x  }{s^4} + 24 \frac{\dot{s}^3 s_x}{s^5}, \\
        \frac{\partial \overset{\scriptscriptstyle{(4)}}{\mathscr{f}}_s}{\partial x} = \overset{\scriptscriptstyle{(4),x}}{\mathscr{f}}_s  = &  - \frac{\overset{\scriptscriptstyle(4)}{s}_x}{s^2} + 2 \frac{ \overset{\scriptscriptstyle(4)}{s} s_x + 6 \ddot{s} \ddot{s}_x + 4 \left(  \dddot{s}_x \dot{s}  + \overset{\scriptscriptstyle(3)}{s} \dot{s}_x \right) }{s^3} - 6 \frac{ 4 \overset{\scriptscriptstyle(3)}{s} \dot{s} s_x + 6 \left( 2 \dot{s}\dot{s}_x \ddot{s} + \dot{s}^2\ddot{s}_x \right) + 3 \ddot{s}^2 s_x  }{s^4} \\ & + 12 \frac{12 \dot{s}^2 \ddot{s} s_x - 8 \dot{s}^3\dot{s}_x }{s^5} + 120 \frac{\dot{s}^4 s_x}{s^6}.
    \end{aligned}
\end{equation}

And similarly for $\overset{\scriptscriptstyle{(k),x}}{\mathscr{g}}_s$ the derivatives are:
\begin{equation}
    \begin{aligned}
        \frac{\partial \mathscr{g}_s}{\partial x} = \overset{\scriptscriptstyle{x}}{\mathscr{g}}_s = & s_x \dot{s} + s \dot{s}_x, \\
        \frac{\partial \dot{\mathscr{g}}_s}{\partial x} = \overset{\scriptscriptstyle{(1),x}}{\mathscr{g}}_s  = & 2 \dot{s} \dot{s}_x + s_x \ddot{s} + s \ddot{s}_x, \\
        \frac{\partial \ddot{\mathscr{g}}_s}{\partial x} = \overset{\scriptscriptstyle{(2),x}}{\mathscr{g}}_s  = & 3 \dot{s}_x \ddot{s} + 3 \dot{s} \ddot{s}_x + s_x \dddot{s} + s \dddot{s}_x, \\
        \frac{\partial \dddot{\mathscr{g}}_s}{\partial x} = \overset{\scriptscriptstyle{(3),x}}{\mathscr{g}}_s  = & 6 \ddot{s} \ddot{s}_x + 4 \dot{s}_x \overset{\scriptscriptstyle(3)}{s} + 4 \dot{s} \overset{\scriptscriptstyle(3)}{s}_x + s_x \overset{\scriptscriptstyle(4)}{s} + s \overset{\scriptscriptstyle(4)}{s}_x, \\
    \frac{\partial \overset{\scriptscriptstyle{(4)}}{\mathscr{g}}_s}{\partial x} = \overset{\scriptscriptstyle{(4),x}}{\mathscr{g}}_s  = & 10 \ddot{s}_x \overset{\scriptscriptstyle(3)}{s} + 10 \ddot{s} \overset{\scriptscriptstyle(3)}{s}_x + 5 \dot{s}_x \overset{\scriptscriptstyle(4)}{s} + 5 \dot{s} \overset{\scriptscriptstyle(4)}{s}_x + s_x \overset{\scriptscriptstyle(5)}{s} + s \overset{\scriptscriptstyle(5)}{s}_x.
    \end{aligned}
\end{equation}

\subsection{Generalized equinoctial orbital elements auxiliary functions} \label{appendix:eci2geqoe}
Algorithms \ref{alg:eci2geqoe} and \ref{alg:geqoe2eci} are basic conversion functions needed to work with the generalized equinoctial orbital elements. The analytical jacobians of these conversions are given explicitly in \cite{giulio2021generalization}.
\begin{algorithm}[ht] \caption{$\: \bm \chi = \left[\nu \: p_1 \: p_2 \: q_1 \: q_2 \: \mathcal{L} \right]^\intercal  = \text{RV2GEQOE}_{fun}( \bm r\rfloor_{ECI}, \bm v\rfloor_{ECI},\mu,J_2,R_{\oplus} )$}\label{alg:eci2geqoe}
    \begin{algorithmic}[1]
        \State $r = \lVert \bm r \lVert$, \: $\mathscr{v} = \lVert \bm v \lVert$
        \State $\bm h = \bm r \times \bm v, \: h = \lVert \bm h \lVert $, $ \dot{r} = \frac{\bm r \cdot \bm v}{ r }$
        \State $ \hat{z} = \bm r(3) / r$
        \State $ A = \frac{\mu \text{J}_2 R_{\oplus}^2}{2}$, $ \mathscr{U} = - \frac{A}{r^3} (1-3 \hat{z}^2) $, $ \mathscr{U}_{\text{eff}} =  \frac{h^2}{2 r^2} + \mathscr{U} $
        \State $ \bm e_{r} = \bm r / r$, $ \bm e_{h} = \bm h / h$, $ \bm e_{f} = \bm e_{h} \times \bm e_{r}$
        \State $ \bm e_{x} = \left[1,0,0\right]^{\intercal}$, $ \bm e_{y} = \left[0,1,0\right]^{\intercal}$, $ \bm e_{z} = \left[0,0,1\right]^{\intercal}$
        \State \textbf{Total energy: }$ \varepsilon = \varepsilon_K + \mathscr{U} = \frac{\mathscr{v}^2}{2} -\frac{\mu}{r} + \mathscr{U} $
        \State \textbf{Generalized mean motion: }$ \nu = \frac{1}{\mu} \left( -2 \varepsilon \right)^{3/2} $
        \State $ q_1 = \frac{ e_{h} \dot e_{x} }{1 + e_{h} \dot e_{z} } $, $ q_2 = \frac{ - e_{h} \dot e_{x} }{1 + e_{h} \dot e_{z} } $
        \State $\bm e_{X} = \frac{1}{1+ q_1^2 + q_2^2}\left[1 - q_1^2 + q_2^2,2 q_1 q_2 , -2 q_1 \right]^{\intercal}$
        \State $\bm e_{Y} = \frac{1}{1+ q_1^2 + q_2^2}\left[2 q_1 q_2, 1 + q_1^2 - q_2^2 , 2 q_2 \right]^{\intercal}$
        \State \textbf{Generalized angular momentum: }$ c = \sqrt{2 r^2 \mathscr{U}_{\text{eff}}}$
        \State \textbf{Generalized velocity vector: }$ \bm \upsilon = \dot{r} \bm e_{r} + \frac{c}{r} \bm e_{f}$
        \State \textbf{Generalized eccentricity vector: }$ \bm g = \frac{1}{\mu} \bm \upsilon \left( \bm r \times \bm \upsilon \right) - \bm e_{r}$
        \State $ p_1 = \bm g \cdot \bm e_{Y} $, $ p_2 = \bm g \cdot \bm e_{X} $
        \State $ X = \bm r \cdot \bm e_{X}$, $ Y = \bm r \cdot \bm e_{Y}$
        \State $ \beta = \sqrt{1 - p1^2 - p_2^2}$, $\alpha = \frac{1}{1 + \beta}$ \label{alg_geqoe_alpha}
        \State \textbf{Generalized semi-major axis: }$ a = - \frac{\mu}{2 \varepsilon} = \left( \frac{\mu }{\nu^2}\right)^{1/3}$ \label{alg_geqoe_a}
        \State $\cos{\mathcal{K}} = p_2 + \frac{1}{a \beta} \left[ \left( 1 - \alpha p_2^2 \right) X - \alpha p_1 p_2 Y \right]$
        \State $\sin{\mathcal{K}} = p_1 + \frac{1}{a \beta} \left[ \left( 1 - \alpha p_1^2 \right) Y - \alpha p_1 p_2 X \right]$
        \State $ \mathcal{L} = \text{atan2}(\sin{\mathcal{K}},\cos{\mathcal{K}}) + \frac{1}{a \beta} \left( X p_1 - Y p_2 \right)$
    \end{algorithmic}
\end{algorithm}

\begin{algorithm}[ht] \caption{$\: \bm X\rfloor_{ECI} = \left[ \bm r^\intercal, \bm v^\intercal \right]^\intercal = \text{GEQOE2RV}_{fun}(\nu, p_1,p_2,q_1,q_2,\mathcal{L},\mu,\text{J}_2,R_{\oplus} )$}\label{alg:geqoe2eci}
    \begin{algorithmic}[1]
        \State \textbf{Numerically solve Kepler\textquotesingle  s equation: }$ \mathcal{K} \gets \mathcal{L} = \mathcal{K} + p_1 \cos{\mathcal{K}} - p_2 \sin{\mathcal{K}}$
        \State $ \bm e_{X} = \frac{1}{1+ q_1^2 + q_2^2}\left[1 - q_1^2 + q_2^2,2 q_1 q_2 , -2 q_1 \right]^{\intercal}$
        \State $\bm e_{Y} = \frac{1}{1+ q_1^2 + q_2^2}\left[2 q_1 q_2, 1 + q_1^2 - q_2^2 , 2 q_2 \right]^{\intercal}$
        \State $ \beta = \sqrt{1 - p1^2 - p_2^2}$, $\alpha = \frac{1}{1 + \beta}$
        \State $ a = \left( \frac{\mu }{\nu^2}\right)^{1/3}$
        \State $ X = a  \left[ \alpha p_1 p_2 \sin{\mathcal{K}} + (1-\alpha p_1^2)\cos{\mathcal{K}}  - p_2 \right]$
        \State $ Y = a  \left[ \alpha p_1 p_2 \cos{\mathcal{K}} + (1-\alpha p_2^2)\sin{\mathcal{K}}  - p_1 \right]$
        \State $ \bm r = X \bm e_{X} + Y \bm e_{Y}$
        \State $  r = a \left( 1 - p_1 \sin{\mathcal{K}} - p_2 \cos{\mathcal{K}} \right) $
        \State $ \hat{z} = \bm r(3) / r$
        \State $ A = \frac{\mu \text{J}_2 R_{\oplus}^2}{2}$, $ \mathscr{U} = - \frac{A}{r^3} (1-3 \hat{z}^2) $
        \State $ \dot{r} = \frac{\sqrt{\mu a}}{r} \left( p_2 \sin{\mathcal{K}} - p_1 \cos{\mathcal{K}} \right) $
        \State $ \cos{L} = \frac{X}{r} $, $ \sin{L} = \frac{Y}{r} $
        \State $ c = \left( \frac{\mu^2}{\nu} \right)^{1/3} \sqrt{1 - q_1^2 - q_2^2} $, $ h = \sqrt{c^2 - 2 r^2 \mathscr{U}} $
        \State $ \dot{X} = \dot{r} \cos{L} - \frac{h}{r} \sin{L} $
        \State $  \dot{Y} = \dot{r} \sin{L} + \frac{h}{r} \cos{L} $
        \State $ \bm v = \dot{X} \bm e_{X} + \dot{Y} \bm e_{Y}$
    \end{algorithmic}
\end{algorithm}

\end{document}